\documentclass[preprints,article,accept,moreauthors,pdftex]{Definitions/mdpi}
\usepackage[utf8]{inputenc}

\usepackage{braket}
\usepackage{amssymb}

\firstpage{1} 
\pubvolume{xx}
\issuenum{1}
\articlenumber{5}
\pubyear{2019}
\copyrightyear{2019}
\history{Received: date; Accepted: date; Published: date}

\Title{Quantum Black Holes in the Sky}

\Author{Jahed Abedi $^{1,2, \dagger}$, Niayesh Afshordi $^{3,4,5,\dagger}$*, Naritaka Oshita $^{5, \dagger}$ and Qingwen Wang $^{3,4,5, \dagger}$}

\AuthorNames{Firstname Lastname, Firstname Lastname and Firstname Lastname}

\address{%
$^{1}$ \quad Max-Planck-Institut f\"ur Gravitationsphysik, D-30167 Hannover, Germany\\
$^{2}$ \quad Leibniz Universit\"at Hannover, D-30167 Hannover, Germany\\
$^{3}$ \quad Department of Physics and Astronomy, University of Waterloo, Waterloo, ON N2L 3G1, Canada\\
$^{4}$ \quad Waterloo Centre for Astrophysics, University of Waterloo, Waterloo, ON N2L 3G1, Canada\\
$^{5}$ \quad Perimeter Institute For Theoretical Physics, 31 Caroline St N, Waterloo, ON N2L 2Y5, Canada
}

\corres{Correspondence: nafshordi@pitp.ca}

\firstnote{All authors have contributed equally to this work. The order of authors is alphabetical.}

\abstract{Black Holes are possibly the most enigmatic objects in our Universe. From their detection in gravitational waves upon their mergers, to their snapshot eating at the centres of galaxies, black hole astrophysics has undergone an observational renaissance in the past 4 years. Nevertheless, they remain active playgrounds for strong gravity and quantum effects, where novel aspects of the elusive theory of quantum gravity may be hard at work. In this review article, we provide an overview of the strong motivations for why ``Quantum Black Holes'' may be radically different from their classical counterparts in Einstein's General Relativity. We then discuss the observational signatures of quantum black holes, focusing on gravitational wave echoes as smoking guns for quantum horizons (or exotic compact objects), which have led to significant recent excitement and activity. We review the theoretical underpinning of gravitational wave echoes and critically examine the seemingly contradictory observational claims regarding their (non-)existence. Finally, we discuss the future theoretical and observational landscape for unraveling the ``Quantum Black Holes in the Sky''.  
}


\begin{document}
\tableofcontents

\section{Introduction}
Black holes (BHs) are very interesting ``stars'' in the Universe where both strong gravity and macroscopic quantum behavior are expected to coexist. Classical BHs in General Relativity (GR) have been thought to have only three hairs, i.e., mass, angular momentum, and charge, making observational predictions for BHs relatively easy \cite{1968CMaPh...8..245I,1971PhRvL..26..331C} (compared to other astrophysical compact objects). For astrophysical BHs, due to the effect of ambient plasma, this charge is vanishingly small, leaving us with effectively two hairs for isolated black holes, with small accretion rates. In other words, finding conclusive deviations from {\textit standard} predictions of these 2-parameter models, may be interpreted as fingerprints of a quantum theory of gravity or other possible deviations from GR. For example, the quasinormal modes (QNMs) of spinning BHs, which  have been widely-studied over the past few decades (a subject often referred to as BH spectroscopy), only depend on the mass and spin of the Kerr BH (e.g., \cite{Kokkotas:1999bd}).  The ringdown of the perturbations of the BH is regarded as a superposition of these QNMs, and thus can be used to test the accuracy of GR predictions and no-hair theorem (e.g., see \cite{Isi:2019aib}). As a result, precise detection of QNMs from the ringdown phase (from BH mergers or formation) in gravitational wave (GW) observations may enable us to test the classical and quantum modifications to GR (e.g., \cite{Bhagwat:2019dtm}). 

A concrete path towards this goal is paved through the study of ``GW echoes'', a smoking gun for near-horizon modifications of GR which are motivated from the resolutions of the proposed resolutions to the BH information paradox and dark energy problems \cite{Almheiri:2013hfa,PrescodWeinstein:2009mp}. The list of these models include wormholes \cite{Cardoso:2016rao}, gravastars \cite{Mazur:2004fk}, fuzzballs \cite{Lunin:2001jy}, 2-2 holes \cite{Holdom:2016nek}, Aether Holes \cite{PrescodWeinstein:2009mp}, Firewalls \cite{Almheiri:2013hfa} and the Planckian correction in the dispersion relation of gravitational field \cite{Oshita:2018fqu,Oshita:2019sat}.

The possibility of observing GW echoes was first proposed shortly after the first detection of GWs by LIGO \cite{Cardoso:2016rao,Cardoso:2016oxy,Cardoso:2019rvt}, which has led to several observational searches \cite{Abedi:2016hgu,Uchikata:2019frs, Conklin:2017lwb, Westerweck:2017hus,Nielsen:2018lkf, Abedi:2018npz,Salemi:2019uea, Holdom:2019bdv, Ashton:2016xff, Abedi:2017isz, Abedi:2018pst}. Tentative evidence for and/or detection of these echoes can be seen in the results reported by different groups \cite{Abedi:2016hgu, Conklin:2017lwb, Westerweck:2017hus,Nielsen:2018lkf, Abedi:2018npz,Salemi:2019uea, Uchikata:2019frs, Holdom:2019bdv} from O1 and O2 LIGO observations of binary BH and neutron star mergers, but the origin and the statistical significance of these signals remain controversial \cite{Westerweck:2017hus, Ashton:2016xff, Abedi:2017isz, Abedi:2018pst,Salemi:2019uea}, motivating further investigation. 

Given their uncertain theoretical and observational status, GW echoes are gathering much attention from those who are interested in the observational signatures of quantum gravity, and the field remains full of excitement, controversy and confusion. In this review article, we aim to bring some clarity to this situation, from its background, to its current status, and into its future outlook. 

The review article is organized as follows: In the next section, we provide builds the motivation to investigate the quantum signatures from BHs. In Sec. \ref{sec:QBHs}, we discuss theoretical models of quantum BHs, starting from the BH information loss paradox, and then its proposed physical resolutions that lead to observable signatures. In Sec. \ref{sec:echo_predictions}, we review how to predict the GW echoes from spinning BHs based on the Chandrasekhar-Detweiler (CD) equation, and also review the Boltzmann reflectivity model \cite{Oshita:2019sat,Wang:2019rcf} for quantum black holes. Sec. \ref{sec:echo_searches} is devoted to the echo searches, where we summarize positive, negative, and mixed reported outcomes, and attempt to provide a balanced and unified census. In Sec. \ref{sec:future_prospects}, we discuss the future prospects for advancement in theoretical and observational studies of quantum black holes, while Sec. \ref{sec:final_words} concludes the review article. 

Throughout the article, we use the following notations:
\begin{center}
\begin{tabular}{c c} 
\hline
Symbol & Description\\
\hline \hline
$a$ & spin parameter\\ 
\hline
$\bar{a}$ & non-dimensional spin parameter ($a/(GM)$)\\
\hline
$c$ & speed of light\\
\hline
$\hbar$ & Planck constant\\
\hline
$k_{\rm B}$ & Boltzmann constant\\
\hline
$G$ & gravitational constant\\
\hline
$M_{\text{Pl}}$ & Planck mass\\
\hline
$E_{\text{Pl}}$ & Planck energy\\
\hline
$l_{\text{Pl}}$ & Planck length\\
\hline
$M$ & mass of a balck hole or exotic compact object\\
\hline
$M_{\odot}$ & solar mass ($1.988 \times 10^{30}$ kg)\\
\hline
$r_g$ & Schwarzschild radius\\
\hline
$T_{\rm H}$ & Hawking temperature\\
\hline
\end{tabular}
\end{center}

Furthermore,  unless noted otherwise, we use the natural Planck units with $\hbar = c = 1= G=1$.

\section{Invitation}
\label{sec:invitation}
\subsection{Classical BHs}

Schwarzschild and Kerr spacetimes are solutions of general relativity giving the spacetime configurations of BHs, which are the most dense objects in our universe. In some regions, matter accumulates and attracts more matter with its gravity which is classically always attractive. At the end, the force is so strong that even the light cannot escape from those regions, where then BHs form. 
The first and most important feature (the definition of BHs) is the formation of horizon. Inside the (event) horizon, all the light cones are directed into the singularity, and nothing can escape, unless it could travel faster than the speed of light. Therefore, horizons stand as the causal boundaries of BHs in Einstein's theory of Relativity. 

Realistic BHs in the sky have different hairs (mass, spin and charge), and their dynamics share more complicated structure, thus, have different kinds of horizons. To list some of them, event horizons are defined as the boundaries where no light can escape to the infinite future. However, for a dynamically evolving BH, event horizons are teleological, i.e.  we cannot predict them until we have the entire history of  the spacetime. Apparent horizons, however, are predictable at a specific time without knowing the future. Any surface has two null normal vectors and if expansion of both of them are negative, the surface is called ``trapped''. Apparent horizons are the outermost of all the trapped surfaces, which is why they are also known as the ``marginally outer trapped surface''. 

Here is a simple example to distinguish these two horizons --- we start with a Schwarzschild BH at time $t_1$, now the event and apparent horizons coincide at the Schwarzschild radius. We throw a spherical null shell into the BH and let it cool down at $t_2$. This process is perfectly described by Vaidya metric \cite{Poisson:2009pwt}. The apparent horizon changes immediately when the shell falls into the BH but the event horizon starts to expand earlier, even before the shell reaches it. It is because that after throwing the shell, the gravity of BH increases. Thus it is harder for light to escape from the BH to infinity. In other words, particles might be doomed to fall into a singularity, even before they had a chance to meet the infalling gravitating matter that is responsible for their fate. Therefore, the event horizon is modified earlier than the apparent horizon. While this result is counter-intuitive, it is a result of the formal definition of the event horizons, which requires the information about the entire history of spacetime, in particular, the future! 

Beyond the horizons, another intriguing trait of BHs is the curvature singularity, which sits at the centre of the BHs. Horizons can also be singular, but usually only coordinate singularities and (in classical General Relativity) removable by changing to a proper coordinate system. However, the singularities inside the BHs are where the general relativity breaks down and so far we do not have any good physics to describe them. We cannot chase the information lost into these singularities (using standard physics), which leads to the information paradox (more on this later).

Back in November 1784, John Michell, an English clergyman, advanced the idea that light might not be able to escape from a very massive object (at a fixed density). For example, light cannot escape from the surface of a star with the density of the sun, if it was 500 times bigger than the sun. Albert Einstein, later in 1915, developed general relativity. Soon after this, Karl Schwarzschild solved the Einstein vacuum field equation under spherical symmetry with a singular mass at the center, which was the first solution for BHs, the Schwarzschild metric. 

While 20th century saw a golden age of general relativity with blooming of dozens of different BH solutions, the existence of BHs was not directly confirmed until one century later in 2015. LIGO-Virgo collaboration reported unprecedented detection of GWs from the binary BH merger events \cite{TheLIGOScientific:2016agk, TheLIGOScientific:2016pea, Abbott:2016blz, Abbott:2016nmj, Abbott:2017vtc, Abbott:2017oio, TheLIGOScientific:2017qsa, Abbott:2017gyy}. Numerical relativity is consistent with LIGO data at least up to quite near the horizon range. But the detection has not confirmed the existence of the horizons. We will discuss in this article how the detection opens a window for searching for quantum nature of the BHs beyond the general relativity.

\subsubsection{Schwarzschild spacetime}

The Schwarzschild spacetime was the first exact solution in the Einstein theory of general relativity. It models a static gravitational field outside a mass which has spherical symmetry, zero charge and rotation. Karl Schwarzschild found this solution in 1915, and four months later, Johannes Droste published a more concrete study on this independently. The metric in the Schwarzschild coordinate is:
\begin{equation}
    ds^2= -\left(1-\frac{2M}{r} \right)dt^2+\left(1-\frac{2M}{r} \right)^{-1} dr^2 + r^2 d \Omega^2,
\end{equation}
where $M$ is the mass of the centre object, $2M$ is Schwarzschild radius and $d \Omega ^2 =d \theta^2 + \sin^2\theta d\phi^2$ is the metric on a 2-sphere. The metric describes gravitational field outside any spherical object without charges. If the radius of the central object is smaller than the Schwarzschild radius, the object is then too dense to be stable, and will go through a gravitational collapse and form the Schwarzschild BH.

Later in 1923, G.D.Birkhoff proved that any spherically symmetric solution of the vacuum Einstein field equation must be static and asymptotically flat. Hence, Schwarzschild metric is the only solution in that case. For any static solution, the event horizon always coincides with the apparent horizon at $r=2M$. In general relativity, Schwarzschild metric is singular at the horizon but , as stated above, this is only a coordinate artifact. That is to say, a free falling observer feels no drama going through the horizon. It takes the observer a finite amount of proper time but infinite coordinate time. Particularly, we can remove the singularity by a proper coordinate transformation. 
In contrast, the origin $r=0$ is the intrinsic curvature singularity. The scalar curvature is infinite and the general relativity is no longer valid at this point.

\subsubsection{Kerr spacetime}

The Kerr spacetime \cite{PhysRevLett.11.237}, discovered by Roy Kerr, is a realistic generalization of the Schwarzschild spacetime. It describes the gravitational field of an empty spacetime outside a rotating object. The spacetime is stationary and has axial symmetry. The metric in the Boyer-Lindquist coordinate is:

\begin{eqnarray}
 ds^2 & =-\left(1-\frac{2Mr}{\rho ^2}\right) dt^2 + \frac{\rho^2}{\Delta} dr^2 + \rho^2 d\theta^2 +\left(r^2+a^2+\frac{2Mra^2}{\rho^2} \sin^2{\theta} \right)\sin^2{\theta} d \phi^2 - \frac{4Mra\sin^2{\theta}}{\rho^2}dt d\phi, \\
 & = -\frac{\rho^2\Delta}{\Sigma} dt^2 + \frac{\Sigma}{\rho^2}\sin^2{\theta}(d\phi - \omega dt)^2 +\frac{\rho^2}{\Delta}dr^2 + \rho^2 d\theta^2,
\end{eqnarray}
where $a=J/M$, $\rho^2=r^2+a^2\cos^2{\theta}$, $\Delta=r^2-2Mr+a^2$, $\Sigma=(r^2+a^2)^2-a^2 \Delta \sin^2{\theta}$ and $\omega=-\frac{g_{t\phi}}{g_{\phi\phi}}=\frac{2Mar}{\Sigma}$. 
The Cartesian coordinates can be defined as \begin{equation}
    x=\sqrt{r^2+a^2}\sin{\theta}\cos{\phi}, \quad
    y=\sqrt{r^2+a^2}\sin{\theta}\sin{\phi}, \quad
    z=r\cos{\theta}.
\end{equation}
There are two singularities easily reading from the coordinate where the $g^{rr}$ and $g_{tt}$ vanish. The first one gives $r_{\pm} = M \pm \sqrt{M^2-a^2}$ corresponding to the horizon analog to the Schwarzschild metric. The larger root $r_+ = M + \sqrt{M^2-a^2}$ is the event horizon, while the other root is inner apparent horizon. 

The second singularity is related to an interesting effect in the Kerr spacetime called frame-dragging effect: When reaching close to the Kerr BHs, the observers even with zero angular momentum (ZAMOs) will co-rotate with the BHs because of the swirling of spacetime from the rotating body. We assume that $u^{\alpha}$ is the four-velocity of ZAMOs, and from the conservation of angular momentum $g_{\phi t} \dot{t} +g_{\phi \phi}\dot{\phi} =0 $, where an overdot is differentiation with respect to the proper time of the observers $\tau$. Thus, $\frac{d\phi}{dt}=-\frac{g_{t\phi}}{g_{\phi\phi}}$. Because of this frame-dragging effect, there is a region of spacetime where static observers cannot exist, no matter how much external force is applied. This region is known as the  ``ergosphere'' $r \leq M+\sqrt{M^2-a^2\cos^2{\theta}}$. The rotation also leads to another interesting feature, called ``superradiance''. That is, we can extract energy from scattering waves off the Kerr BHs. The exact formalism of superradiance is defined and discussed in Sec. \ref{super}.

Finally, the Kerr spacetime also possesses a curvature singularity at the origin $\rho^2=r^2+a^2 \cos^2{\theta}$. However, in contrast to Schwarzschild case, this singularity can be avoided since it is a ring at r=0 and $\theta = \pi/2$, where z=0 and $x^2+y^2=a^2$. In principle, observers can go through the ring without hitting the singularity. However, it is widely believed that the inner horizon, $r_-$ in Kerr spacetime is subject to an instability which would dim the analytic extension of Kerr metric beyond $r_-$ unphysical \cite{Poisson:1989zz}. 

\subsubsection{Blue-shift near horizon}

As shown in the metric, different observers have different proper time. Hence, in the general relativity, the clocks at a gravitational field tick in a different speed in a different spacetime point. This is the blue(red)-shift effect, and it is extremely strong close to the dense object, especially near horizon.

Assuming static clocks in the Schwarzschild spacetime $ds^2=-d\tau^2=-(1-2M/r_o)dt^2$, where $\tau$ is the proper (clock) time of an observer at distance $r_o$. Hence, $t$ is the proper time of an observer at infinity. The shifted wavelength $\lambda_{o}$ measured by observers at $r_o$  compared to observers at infinite is 
\begin{equation}
    \frac{ \lambda_{o}}{ \lambda_{\infty}} =\frac{d\tau}{dt}=\left(1-\frac{2M}{r_o} \right)^{1/2}.
\end{equation}{}

\subsubsection{Thermodynamics of Semi-classical BH}

Jacob Bekenstein and Stephen Hawking first proposed that the entropy of BHs is related to the area of their event horizons divided by the Planck area \cite{Bekenstein:1972tm,Bekenstein:1973ur,Bekenstein:1974ax,Gibbons:1976ue,Hawking:1978jz}. Furthermore, in 1974, Stephen Hawking showed that rather than being totally black, BHs emit thermal radiation at the Hawking temperature, $T_{\rm H} = \frac{\kappa}{2\pi}$, where $\kappa$ is the surface gravity at the horizon \cite{Hawking:1974rv,Hawking:1974sw,PhysRevD.13.2188}. This then lead to the celebrated Bekenstein-Hawking entropy formula $S_{\rm BH} = \frac{A}{ 4 }$ \cite{Gibbons:1976ue,Hawking:1978jz}, where $A$ is the area of the event horizon.
However, the nature of microstates of BHs that are enumerated by this entropy remains so far unknown. String theory associates it with higher dimensional fuzzball solutions, as discussed later in Sec. \ref{fuzzball}. Loop quantum gravity relates the quantum geometries of the horizon to the microstates \cite{Rovelli:1996dv}. Both these approaches can give the right Bekenstein-Hawking entropy, given specific assumptions and idealizations. 

Interestingly, not only the entropy exists for the BHs, but also Brandon Carter, Stephen Hawking and James Bardeen \cite{Bardeen:1973gs} discovered the four laws of BH thermality analogous to the four laws of thermodynamics. The latter is presented in the parentheses.

\begin{itemize}
    \item \textbf{The zeroth law}: A stationary BH has constant surface gravity $\kappa$. (A thermal equilibrium system has a constant temperature $T_{\rm H}$.)
     \item \textbf{The first law}: A small change of mass for a stationary BH is related to the changes in the horizon area A, the angular momentum J, and the electric charge Q:  $dM = \frac{\kappa}{8\pi} dA + \Omega d J + \Phi d Q$, where $\Omega$ is the angular velocity and $\Phi$ is the electrostatic potential (Energy conservation: $dE=TdS-PdV-\mu dN$).
      \item \textbf{The second law}: The area of event horizon $A$  never decreases in general relativity (The entropy of isolated systems never decreases).
       \item \textbf{The third law}: BHs with a zero surface gravity cannot be achieved (Matter in a zero temperature cannot be reached).
\end{itemize}{}{}

\subsection{Membrane Paradigm}\label{sec:membrane}

As mentioned above, in classical general relativity, freely falling observers experience no drama as they cross the event BH horizons, at least not until they reach the singularity inside the BH.  However, to a distant and static observer outside a BH, any infalling objects are frozen at the horizons due to the blue-shift effect. Hence, the BH interior can be regarded as an irrelevant region for the static observers. Based on this complementary picture near horizon, in 1986, Kip S. Thorne, Richard H. Price and Douglas A. Macdonald published the idea of \textit{membrane paradigm} \cite{Thorne:1986iy}. They use a classically radiating membrane to model the BHs, which is motivated as a useful tool to study physics outside BHs without involving any obscure behavior within BH interior.

Let us introduce a spherical membrane located infinitesimally outside the Schwarzschild radius, a.k.a. the {\it stretched horizon}. When the membrane is sufficiently thin, one can use the Israel junction condition to nicely embed the membrane in the Schwarzschild spacetime. The condition is
\begin{equation}
(K^{(+)} f_{ab} -K^{(+)}_{ab}) - (K^{(-)} f_{ab} -K^{(-)}_{ab}) = 8 \pi T_{ab},
\end{equation}
where $f_{ab}$ is the induced metric of the membrane, $K^{(\pm)}_{ab}$ is the extrinsic curvatures on its two sides, and $T_{ab}$ is its stress tensor. The infalling observer will cross the horizon and enter the BH interior without possibility of seeing the membrane. However the static observer outside the BH can remove irrelevant interior region from the remaining spacetime with a membrane. Assuming reflection symmetry $K^+_{ab} = -K^-_{ab},$ the Israel junction condition on the membrane becomes
\begin{equation}
K^{(+)} f_{ab} -K^{(+)}_{ab} = 4 \pi T_{ab},
\label{membrane_junc_out}
\end{equation}
where the stress tensor $T_{ab}$ is no longer zero but has contribution from the extrinsic curvature on the membrane. Rewriting the left hand side of (\ref{membrane_junc_out}), one can obtain the following relation
\begin{equation}
T^a_b = \frac{1}{4 \pi} \left( -\sigma^a_b + \delta^a_b \left( \frac{\theta}{2} + \kappa \right) \right),
\end{equation}
where $\sigma^a_b$ is the shear, $\theta$ is the expansion, and $\kappa$ is the surface gravity at the horizon. From the analogy with the energy momentum tensor of the 2-dimensional compressible fluid, one can read that the shear viscosity $\eta$ and bulk viscosity $\zeta$ are given by $\eta = 1/(16 \pi)$ and $\zeta = -1/(16 \pi)$, respectively. The negativity of the bulk viscosity implies gravitational instability for the expansion or compression at the horizon.
The viscosity at the membrane lead to the thermal dissipation of infalling gravitational waves into the horizon and in this sense, the causality of a classical BH is dual to the viscosity of a 2-dimensional fluid at the stretched horizon. This was the earliest example of fluid-gravity correspondence, which is now active area of research. 

Modifying Einstein gravity which revises the structure of BHs can provide a modified structure of the thin-shell membrane. For example, by adapting the transport properties of the membrane fluid, we can investigate various models of quantum BHs. As an example, we provide a simple idea \cite{Oshita:2019sat} which relates the reflectivity of the ``horizon'' of quantum BHs to viscosity in the context of membrane paradigm. We start by perturbing the Schwarzschild spacetime, whose metric is $g_{\mu \nu}^{\rm Sch}$. Within Regge-Wheeler formalism \cite{Regge:1957td}, the axial axisymmetric perturbation $g_{\mu\nu} = g^{\rm Sch}_{\mu\nu}(r) +\delta g_{\mu\nu}(r,\theta,t)$ take teh form:
\begin{align}
\delta g_{t \phi}&= \epsilon e^{- i \omega t} h_0(r) y(\theta),\label{htphi} \\
\delta g_{r \phi}&= \epsilon e^{- i \omega t} h_1(r) y(\theta),
\end{align}
where other $\delta g_{\mu \nu}$ components vanish, and $\epsilon \ll 1$ controls the order of perturbation. The membrane stands at $r=r_0+ \epsilon R(t,\theta)$, where $r_0$ is its unperturbed position. We apply the Israel junction conditions $K_{ab}-K f_{ab} = - 4 \pi T_{ab}$ to Brown-York stress tensor as defined in \cite{Jacobson:2011dz}. The indexes $\mu, \nu$ run over $(t,r,\theta,\phi)$ in the 4d spacetime, while $a, b$ run over $(t,\theta,\phi)$ on the 3d membrane. We further assume that $T_{ab}$ is the energy stress tensor of a viscous fluid:
\begin{align}
     T_{ab}= [\rho_0+ \epsilon \rho_1(t,\theta) ] u_{a} u_{b}+ \nonumber\\ [ p_0+ \epsilon p_1(t,\theta)-\zeta \Theta] \gamma_{ab} -2 \eta \sigma_{ab},\\
     \sigma_{ab}=\frac{1}{2}(u_{a;c} \gamma^c_b+u_{b;c}\gamma^c_a- \Theta \gamma_{ab}), \\
     \gamma_{ab} \equiv h_{ab}+u_{a}u_{b}, \quad \Theta \equiv u^a_{;a},\label{theta}
\end{align}
where $\rho_0$ and $p_0$ ($\rho_1$ and $p_1$) are background (perturbation on) membrane density and pressure, and $u_a$, $\eta$ and $\zeta$ are fluid velocity, shear viscosity, and bulk viscosity, respectively.
Plugging Eqs. (\ref{htphi}-\ref{theta}) into the the Israel junction condition, we find in the zeroth order in $\epsilon$:
\begin{align}
     \rho_0(r_0)&= -\frac{\sqrt{f(r_0)}}{4 \pi r_0 },\\
     p_0(r_0)&= \frac{\sqrt{f(r_0)} (g(r_0)+r_0 g'(r_0))}{8 \pi r_0 g(r_0)},
\end{align}
where $g(r_0) = (1-2M/r_0)^{1/2}$ and $f(r_0) = 1-2M/r_0$. Assuming $u_{\phi}=0$,  equation of ${\theta \phi}$ component gives in next order of $\epsilon$:
 \begin{equation}
     \omega h_1(r) =-8 i \pi \eta [h_1(r) +(r-r_g)h'_1(r)]. 
     \label{boundary}
 \end{equation}
We can further use $\psi_{\omega} = \frac{1}{r} \left(1-\frac{2M}{r}\right) h_1(r)$ and the tortoise coordinate $x=r+2M\log[r/(2M) -1]$ to rewrite Eq. (\ref{boundary}) as
\begin{equation}
    \omega\psi_{\omega} = 16 i \pi \eta \frac{\partial \psi_{\omega}}{ \partial x}.
    \label{boundaryrw}
\end{equation}
For the classical BHs with a purely ingoing boundary condition $\psi_{\omega} \propto e^{-i\omega x}$ at the horizon, Eq. (\ref{boundaryrw}) gives $\eta =\frac{1}{16 \pi}$, which is consistent with the standard membrane paradigm. If instead we assume there is no longer horizon but a reflective surface with $\psi_{\omega} = A_{out} e^{i\omega x} + A_{in} e^{-i\omega x} $,  Eq. (\ref{boundaryrw}) gives:
\begin{equation}
    \frac{A_{\rm out}}{A_{\rm in}} =\frac{1-16 \pi \eta}{1+ 16 \pi \eta} e^{-2i\omega x}.
\end{equation}
Which relates the reflectivity of the membrane to the viscosity of the surface fluid.

\subsection{Dawn of Gravitational Wave Astronomy}

From 2015 onwards, the LIGO/Virgo collaboration reported unprecedented GW observations from binary BH merger events \cite{TheLIGOScientific:2016agk, TheLIGOScientific:2016pea, Abbott:2016blz, Abbott:2016nmj, Abbott:2017vtc, Abbott:2017oio, TheLIGOScientific:2017qsa, Abbott:2017gyy}. It is the first time that humankind can detect GWs after one century of the Einstein's general theory of gravity. In 2017, Rainer Weiss, Kip Thorne and Barry C. Barish won the Nobel Prize in Physics ``for decisive contributions to the LIGO detector and the observation of Gravitational Waves''. 

The first and most prominent binary BH merger signal seen by LIGO, GW150914, matches well with predictions of numerical relativity simulations that settle into Kerr metric, but contrary to original claims, it could not confirm the existence of the event horizons \cite{Cardoso:2016rao}. However, it opened a new front to test general relativity in strong gravity regime and Kerr-like spacetimes (e.g., Quantum BHs) from modified gravity, which is the main topic of this review article.   

This is the dawn of GW astronomy, and we stand at the threshold of a new age. We are detecting even more compact binary merger events with a better sensitivity from the O3 run of LIGO/Virgo. Future experiments such as Einstein Telescope, Cosmic Explorer, and LISA are expected to improve this by orders of magnitude. More studies on the echo-emission mechanism as well as observational strategies will be crucial for taking advantage of these new observations, to shed light on the nature of quantum BHs. It is our point of view that the best bet is on a sustained synergy between theory and observation, relying on well-motivated theoretical models (such as the Boltzmann reflectivity, aether holes, 2-2 holes, or fuzzballs, discussed in this review) to provide concrete templates for data analysis, which in turn could be used to pin down the correct theory underlying quantum BHs. With some luck, this has the potential to revolutionize our understanding of fundamental physics and quantum gravity. 

\subsection{Quantum Gravity and Equivalence Principle}

The Einstein's general theory of relativity is classical. However, in the Einstein field equation $G_{\mu \nu} = 8 \pi G T_{\mu \nu}$, the classical spacetime geometry is related to stress energy tensor of quantum matter. For decades, scientist have tried to reconcile this inconsistency by embedding general relativity (or its generalizations) within some quantum mechanical framework, i.e. quantum gravity. 

Conventional approach to quantizing Einstein gravity fails because it is not renormalizable. This implies that making predictions for observables, such as scattering cross-sections, requires knowledge of infinitely many parameters at high energies, leading to loss of {\it predictivity}. In the modern language, general relativity could at best be an effective field theory, and requires UV-completion beyond a cutoff near (or below) Planck energy (e.g., \cite{Donoghue:1994dn}). 

Most proposals for this UV-completion involve replacing spacetime geometry with a more fundamental degree of freedom, such as strings (string theory) \cite{Polchinski:1998rq}, discrete spins (loop quantum gravity) \cite{Ashtekar:2004eh}, spacetime atoms (causal sets) \cite{Bombelli:1987aa}, or tetra-hydra (causal dynamical triangulation) \cite{Ambjorn:2004qm}. More exotic possibilities include Asymptotic Safety \cite{Niedermaier:2006wt}, Quadratic Gravity \cite{Holdom:2015kbf}, and Fakeon approach \cite{Anselmi:2017ygm} that introduce a non-perturbative or non-traditional quantization schemes for 4d geometry. Yet another possibility is to modify the symmetry structure of General Relativity in the UV, as is proposed in Lorentz-violating (or Horava-Lifshitz) quantum gravity \cite{Horava:2009uw}. 

While proponents of these various proposals (with varying degrees of popularity) have claimed limited success in empirical explanations of some natural phenomena, it should be fair to say that none can objectively pass muster of {\it predicitivity}. As such, for now, the greatest successes of these proposals remain in the realm of Mathematics.   


Due to this lack of concrete predictivity, the EFT estimates (discussed above) are instead commonly used to argue that the quantum gravitational effects should only show up at Planck scale $ \sim 10^{-35} meter$ or $10^{28}$ eV, which is far from anything accessible by current experiments. However, such arguments miss the possibilities of non-perturbative effects (such as phase transitions) which depend on a more comprehensive understanding of the full phase space of the specific quantum gravity proposal.

For example, it has been shown that the non-perturbative quantum gravitational effects may lead to Planck-scale modifications of the classical BH horizons \cite{Mathur:1997wb}. Proposed models like gravastars \cite{Mazur:2004fk}, fuzzballs \cite{Lunin:2001jy, Lunin:2002qf, Mathur:2005zp, Mathur:2008nj, Mathur:2012jk}, aether BHs \cite{PrescodWeinstein:2009mp}, and firewalls \cite{Braunstein:2009my, Almheiri:2012rt} amongst others \cite{Barcelo:2015noa, Kawai:2017txu,Giddings:2016tla} all drastically alter the standard structure of the BH stretched horizons with a non-classical surface. Soon after the first reported detection of gravitational waves, \cite{Cardoso:2016rao} discerned that Planck-length structure modification around horizons leads to a similar waveform as in classical GR, but followed by later repeating signal --- \textit{echoes} --- in the ringdown from the reflective surface that replaces the classical horizon. This discovery equals a new road leading to Rome --- quantum nature of gravity --- and has sparked off a novel area of modeling and searching for signatures from Quantum BHs. The next section will discuss the quantum theories of BH models and possible road maps to probe them, inspired by the detection of binary BH merger events in gravitational waves.

\section{Quantum BHs}

\label{sec:QBHs}
\subsection{Evaporation of BHs and the Information Paradox}
It was already recognized by Stephen Hawking in the 1970s that the evaporation of a BH leads to an apparent breakdown of the unitarity of quantum mechanics. Here, we will briefly review this problem, which is known as the BH information loss paradox \cite{Hawking:1976ra}.
In the context of quantum field theory in curved spacetime, the energy flux out of a BH horizon is obtained by specifying a proper vacuum state and fixing the (classical) background spacetime. However, a radiating BH must lose its mass in time, and so fixing the background is valid only for a much shorter timescale than the evaporation timescale. One can roughly estimate the lifetime of a BH as follows: The energy expectation value of a Hawking particle is of the order of the Hawking temperature $T_H \equiv (8 \pi M)^{-1}$, which would be emitted over the timescale of $t \sim M$. Then we can estimate the luminosity of the BH as
\begin{equation}
\frac{d M}{dt} \sim \frac{- T_H}{M} \sim - ( M^2)^{-1},
\end{equation}
and this gives its lifetime $t_{\text{life}}$
\begin{equation}
t_{\text{life}} \sim  M^3.
\label{lifetimeBH}
\end{equation}
To be consistent with the result of a more rigorous calculation (see e.g. \cite{Fabbri:2005mw}), we need a factor of about $10^{5}$ in (\ref{lifetimeBH})
\begin{equation}
t_{\text{life}} \simeq 10^{5}  M^3 \sim 10^{75} \left( \frac{M}{M_{\odot}} \right)^3 \text{[sec]},
\end{equation}
which is much much longer than the cosmic age of $\sim 4 \times 10^{17} \ [\text{sec}]$ for astrophysical BHs whose mass are $ \gtrsim M_{\odot}$.
It may be true that BHs evaporate due to the Hawking radiation, at least, until reaching the Planck mass. However, the gravitational curvature near the horizon eventually reaches the Planckian scale and the classical picture of background gravitational field would break down. As such, the possibility of leaving a ``remnant" after the evaporation has been discussed (see e.g. \cite{Aharonov:1987tp,Banks:1992is,Giddings:1994pj,Adler:2001vs}), but the most natural possibility would be that only Hawking radiation is left after the completion of the BH evaporation.

\begin{figure}[t]
\begin{center}
\includegraphics[keepaspectratio=true,height=80mm]{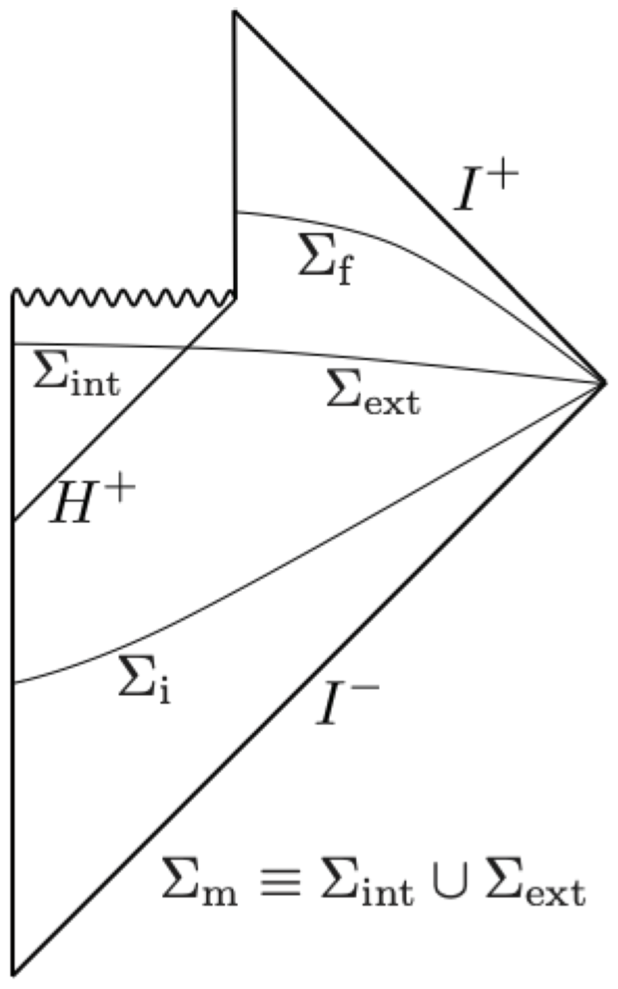}
\end{center}
\caption{The Penrose diagram describing an evaporating BH.
  }%
  \label{haw1}
\end{figure}

If the Hawking evaporation just leaves the ``thermal" radiation afterwards, one can immediately understand why the evaporation process is paradoxical. Let us suppose that a pure quantum state collapses into a BH and it radiates Hawking quanta until the BH evaporates. If the final state is a thermal mixed state, the evaporation is a process which transforms a pure to mixed state. Therefore, if the final state of any BH is a completely thermal state, one can say that the evaporation process is a non-unitary process. The information loss paradox can be also explained from the geometric aspect using the Penrose diagram. In quantum mechanics, the time-evolution of a quantum state is described by a unitary operator, $\hat{U}$, that maps an initial quantum state $\ket{\text{in}}$ on a past Cauchy surface $\Sigma_{\text{i}}$ into a final quantum state $\ket{\text{f}}$ on a future Cauchy surface $\Sigma_{\text{f}}$.
Since the unitary operator gives a reversible process, one can also obtain the initial state from the final state as
\begin{equation}
\ket{\text{in}} = \hat{U}^{\dag} \ket{\text{f}}.
\end{equation}
Although this is true in a flat space, the argument is very controversial in the existence of an evaporating BH. Assuming a gravitational collapse forms a horizon and singularity, then it eventually evaporates, leaving behind a thermal radiation, the Penrose diagram describing the whole process is given by Fig. \ref{haw1}.
Let us consider three quantum states: an initial quantum state $\ket{\text{in}}$ on $\Sigma_{\text{i}}$, an intermediate quantum state $\ket{\text{mid}}$ on $\Sigma_{\text{m}}$, and a final state $\ket{\text{f}}$ on $\Sigma_{\text{f}}$, where $\Sigma_{\text{i}}$, $\Sigma_{\text{m}}$, and $\Sigma_{\text{f}}$ are the Cauchy surfaces and $\Sigma_{\text{m}}$ intersects the future horizon $H^+$ and so one can split it into the exterior and interior regions as $\Sigma_{\text{m}} \equiv \Sigma_{\text{ext}} \cup \Sigma_{\text{int}}$ (see Fig. \ref{haw1}).

The final quantum state $\ket{\text{f}}$ is determined by information on the exterior part of the intermediate Cauchy surface $\Sigma_{\text{ext}}$ rather than that on the whole intermediate Cauchy surface $\Sigma_{\text{m}}$, which leads to the information loss paradox.
To see this in more detail, let us consider an initial pure quantum state
\begin{equation}
\displaystyle \ket{\text{in}} = \sum_{i} c^{\rm in}_i \ket{\psi_i},
\end{equation}
where $\left\{ c^{\rm in}_i \right\}$ is an initial vector in the Hilbert space. The intermediate state is still a pure state due to the unitary evolution of $\ket{\text{in}}$
\begin{equation}
\ket{\text{mid}} = \displaystyle \hat{U} \ket{\text{in}} = \sum_{i,j} c_{i,j} \ket{\psi_i}_{\text{int}} \otimes \ket{\psi_j}_{\text{ext}},
\end{equation}
the time-evolution from $\Sigma_{\text{m}}$ to $\Sigma_{\text{f}}$ is non-unitary, provided that the final state on $\Sigma_{\text{f}}$ is obtained by the unitary evolution of the exterior intermediate state.
The density matrix of the exterior intermediate state, denoted by $\hat{\rho}_{\text{ext}}$, is obtained by tracing over all the internal basis states:
\begin{equation}
\hat{\rho}_{\text{ext}} = \displaystyle \sum_{k} \bra{\psi_k}_{\text{int}} \ket{\text{mid}} \braket{\text{mid} | \psi_k}_{\text{int}} = \displaystyle \sum_{k,j,j'} c_{k,j} c^{\ast}_{k,j'} \ket{\psi_j}_{\text{ext}} \bra{\psi_{j'}}_{\text{ext}}.
\label{densitymatext}
\end{equation}
The resulting density matrix, (\ref{densitymatext}), is independent of the interior orthogonal basis $\left\{ \ket{\psi_j}_{\text{int}} \right\}$ due to the tracing operation. Therefore, the loss of the interior information results in a non-unitary evolution and an initial quantum state evolves to a mixed state after the BH evaporation.

\subsection{BH complementarity}
The BH complementarity has been one of the leading proposals for the retrieval of BH information, which was first put forth by by Susskind, Thorlacius, and Uglum \cite{Susskind:1993if}. According to a distant observer, due to the infinite redshift at a BH horizon, the Hawking radiation involves modes of transplanckian frequency whose energy can be arbitrarily large in the vicinity of the horizon. In the BH complementarity proposal, the energetic modes form the membrane, which can absorb, thermalize, and reemit information, on the BH horizon. They argue that such a picture regarding the retrieval of BH information by the stretched horizon is consistent with the following three plausible postulates:\\ \\
{\textit Postulate 1 (unitarity)}--- According to a distant observer, the formation of a BH and the evaporation process can be described by the standard quantum theory. There exists a unitary S-matrix which describes a process from infalling matter to outgoing non-thermal radiation.\\ \\
{\textit Postulate 2 (semi-classical equations)}--- Outside the stretched horizon of a massive BH, physics can be approximately described by a set of semi-classical field equations.\\ \\
{\textit Postulate 3 (degrees of freedom)}--- For a distant observer, the number of microscopic states of a BH can be estimated by $\exp{S (M)}$, where the exponent $S(M)$ is the Beksntein-Hawking entropy.\\ \\

On the other hand, it has been presumed that a freely infalling observer would not observe anything special when passing through the horizon due to the equivalence principle. In this sense, there are two totally different and seemingly inconsistent scenarios that co-exist in the BH complementarity. However, the contradiction arises only when attempting to compare the experiments performed inside and outside horizon, which might be impossible due to a backreaction of the high-energy modes near the stretched horizon \cite{Susskind:1993mu}.

\subsection{Firewalls}
In 2012, Almheiri, Marolf, Polchinski and Sully (AMPS) argued \cite{Almheiri:2012rt} that the Postulates 1-3 in the BH complementarity and the Equivalence principle of GR are mutually inconsistent for an \textit{old BH} \cite{Page:1993df, Page:1993wv, Page:2013dx}, provided that the monogamy of entanglement is satisfied. Then they argued that the ``most conservative'' resolution is a violation of the equivalence principle  near the BH and its horizon should be replaced by high-energetic quanta, so called ``firewall'', to avoid the inconsistency. Before introducing the original firewall argument in more detail, let us review a theorem in quantum information theory, the monogamy of entanglement. Let us consider three independent quantum systems, A, B, and C. The strong subadditivity relation of entropy is given by
\begin{equation}
S_{AB} + S_{BC} \geq S_{B} + S_{ABC}.
\end{equation}
If A and B is fully entangled, we have
\begin{equation}
S_{AB} = 0 \ \ \text{and} \ \ S_{ABC} = S_C.
\end{equation}
Then the strong subadditivity relation reduces to
\begin{equation}
S_{B} + S_{C} - S_{BC} \leq 0.
\label{mutual1}
\end{equation}
Since the left hand side in (\ref{mutual1}) is the mutual information of $B$ and $C$, denoted by $I_{BC}$, and it is a non-negative quantity, (\ref{mutual1}) reduces to
\begin{equation}
I_{BC} = S_{B} + S_{C} - S_{BC} = 0,
\end{equation}
which means that the quantum system B cannot fully correlate with C when B and A are fully entangled mutually. Therefore, any quantum system cannot fully entangle with other two quantum systems simultaneously. This is the monogamy of entanglement that is an essential theorem in the firewall argument.

Let us consider an old BH, whose origin is a gravitational collapse of a pure state, with early Hawking particles A, late Hawking particle B, and infalling particle inside the horizon C.
In order for the final state of the BH to be pure state, A and B should be fully entangled mutually, that is a necessary condition for the Postulate 1. On the other hand, created pair particles , B and C, are also fully entangled according to the quantum field theory in classical background (Postulate 2). That is, imposing the Postulate 1 and 2 inevitably results in that B is fully and simultaneously entangled with both A and C, which obviously contradicts with the monogamy of entanglement. In order to avoid this contradiction, AMPS argued that there is no interior of BHs and the horizons should be replaced by energetic boundaries that the entanglement of Hawking pairs are broken. They called these boundaries ``firewalls''. According to this proposal, any object falling into a BH would burn up at the firewall, which contradicts the equivalence principle (in vacuum) and replaces the BH complementarity proposal. Although there are some updates of this proposal, based on ER=EPR conjecture \cite{Almheiri:2013hfa,Papadodimas:2012aq,Maldacena:2013xja,Susskind:2013lpa,Bousso:2012as}, backreaction due to gravitational schockwaves \cite{Yoshida:2019qqw}, and quantum decohenrence of Hawking pair due to the interior tidal force \cite{Oshita:2016pbh}), they do remain speculative, and at the level of toy models. However, on general grounds, if quantum effects lead to such an energetic wall at the stretched horizon, it could contribute to the reflectivity of BH which may be observable by merger events leading to the formation of BHs.

\subsection{Gravastars}
The gravitational vacuum condensate star (gravastar) was proposed as a final state of gravitational collapse by Mazur and Mottola \cite{Mazur:2004fk}. According to the proposal, the resulting state of gravitational collapse is a cold compact object whose interior is a de Sitter condensate, which is separated from the outside black hole spacetime by a null surface. In this state, there is no singularity (with the exception of the null boundary) and no event horizon, which avoids the BH information loss paradox. Such gravitational condensation could be caused by quantum backreaction at the Schwarzschild horizon $r=r_g$ even for an arbitrarily large-mass collapsing object. One might wonder why the backreaction can lead to such a drastic effect for any mass since the tidal force which acts on an infalling test body can be arbitrarily weak for an arbitrarily large mass at the Schwarzschild radius. The argument is that considering a photon with asymptotic frequency $\omega$ near the Schwarzschild radius, the (infinite) blue-shift effect by which the local energy is enhanced as $\hbar \omega / \sqrt{1-r_g/r}$, could lead to a drastic effect at the Schwarzschild radius. This is unavoidable since any object is immersed in quantum vacuum fluctuations and virtual particles always exist around them. From this argument, the gravitational condensation has been expected to take place at the final stage of gravitational collapse. The authors in \cite{Mazur:2004fk} also estimate the entropy on the surface of gravastar by starting with a simplified vacuum condenstate model which consists of three different equations of state
\begin{align}
0 \leq r < r_1, ~~~~~~~~~~~ &\rho = -p,\\
r_1 < r < r_1+\delta r, ~~~~~~~~~~~ &\rho=p,\\
r_1+\delta r < r, ~~~~~~~~~~ &\rho=p=0,
\end{align}
where $r_1$ is the radius of interior region and $\delta r$ is the thickness of the thin-shell of the gravastar. Then the obtained entropy of the shell was found out to be $S \sim 10^{57} g k_{\text{B}} \left( M/M_{\odot} \right)^{3/2}$, where $g$ is a dimensionless constant. Recently, the derivation of gravastar-like configuration was performed by Carballo-Rubio \cite{Carballo-Rubio:2017tlh}. He derived the semi-classical Tolman-Oppenheimer-Volkoff (TOV) equation by taking into account the polarization of quantum vacuum and solved it to obtain the exact solution of an equilibrium stellar configuration. It also has its de Sitter interior and thin-shell near the Schwarzschild radius, which is consistent with the original gravastar proposal \cite{Mazur:2004fk}.

From the observational point of view, the shadows of a gravastar was investigated in \cite{Sakai:2014pga} where they argue the shadows of a BH and gravastar could be distinguishable. In addition, tests of gravastar with GW observations have been discussed in e.g. \cite{Pani:2009ss,Cardoso:2016oxy,Conklin:2017lwb}.

\subsection{\label{fuzzball}Fuzzballs}

Samir D. Mathur has proposed fuzzballs \cite{mathur2005fuzzball} as description of true microstates of the quantum BHs from string theory. A fuzzball state has the BH mass inside a horizon-sized region and a smooth (but higher-dimensional) geometry. Here are some crucial features of the conjecture:

\begin{enumerate}
\item  Different fuzzball geometries represent different microstates of the quantum BH --- fuzzball. Application the AdS/CFT duality \cite{Maldacena:1997re} suggests that the counting of the microstates is consistent with the Bekenstein-Hawking entropy. 

\item Fuzzballs do not possess horizons. Instead, they end with smooth "caps" near where the horizons would have been. Every microstate has almost the same geometry outside the would-be horizon matching the classical BH picture for the outside observers. But the microstates differ from each other near the would-be horizons.

\item Fuzzball solves the information paradox by removing the horizon and singularity. The horizon is replaced by fuzzy matter and no longer vacuum. The particles created near the would-be horizon now have access to the information of fuzzball interior. Moreover, the higher-dimensional spacetime ends smoothly around the would-be horizon and is singularity-free. The infalling particles at the low frequencies interact with the ``fuzz'' for a relatively long time scale, while high frequency ones excite the microstates and lose their energy the same as in the classical BHs case. Hence, the traditional horizons only show up effectively from the point of view of an outside observers, over relatively short time scale $\lesssim M \log(M)$.

\end{enumerate}

How do these higher dimensional ``microstates'' with the smooth and horizonless geometries looks like? We, for the first time, show a specific reduced 4D fuzzball solution has an associated 4D effective fluid  near the would-be horizon. The anisotropic pressure of the fluid is crucial to the horizonless geometry. 

Applying Kaluza-Klein reduction of non-supersymmetric microstates of the D1-D5-KK system \cite{giusto2007non}. the metric in 4D is
\begin{eqnarray}
\label{equ1}
ds_4^2=-\frac{f^2}{\sqrt{AD}}(dt+c_1 c_5 \omega)^2+\sqrt{AD}[\frac{dr^2}{\Delta}+d\theta^2+\frac{\Delta}{f^2} \sin^2\theta d\varphi^2]\\
\Delta=r^2-r_0^2, \  f^2=\Delta+r_0^2n^2\sin^2\theta,\\
A=f^2+2p[(r-r_0)+n^2 r_0 (1+\cos\theta) ],\\
B=f^2+2\frac{r_0 (r-r_0) (n^2-1)}{p-r_0  (1+n^2)}[(r-r_0)+n^2 r_0 (1-\cos\theta)], \\
C=2\frac{r_0 \sqrt{r_0(r+r_0)} n (n^2-1)}{p-r_0  (1+n^2)}[(r-r_0)+(p+ r_0) (1-\cos\theta)],\\
G=\frac{Af^2-C^2}{B^2},\  D=B c_1^2 c_5^2- f^2(c_1^2 s_5^2+s_1^2 c_5^2)+ \frac{G f^2}{A} s_1^2 s_5^2,\\
J^2=\frac{r_0^3 p (r+r_0) n^2 (n^2-1)^2}{p-r_0  (1+n^2)}, \ \omega ^2=\frac{2J \sin^2\theta(r-r_0)}{f^2} d\varphi,
\end{eqnarray}
where parameters $c_1$, $c_5$, $s_1$, $s_5$, $r_0$, n and p are related to the mass, angular momentum and charges of the solution.

\subsubsection{\label{sec2.1}Asymptotic behavior}

Here, we study the asymptotic behavior of metric. As shown in Table \ref{table1}, it behaves exactly like Schwarzschild metric when setting K=1, M=$r_{\rm g}$. They have different $g_{t\varphi}$ compared with Kerr metric. As stated, it resembles the Schwarzschild BH far away, but present different geometry close to the horizon.
\begin{table}
\caption{\label{table1}Asymptotical behavior of metric}

\begin{tabular}{ |c||c|c|c|  }
\hline
Metric & Fuzzball & Kerr BH& Sch. BH\\
\hline
$g_{tt}$&$-K+\frac{M}{K^3} \frac{1}{r} +O( \frac{1}{r^2})$&$-1+r_g\frac{1}{r} +O( \frac{1}{r^2})$&$-1+r_g \frac{1}{r} +O( \frac{1}{r^2})$\\
\hline
$g_{rr}$&$K+\frac{M}{K} \frac{1}{r} +O( \frac{1}{r^2})$&$1+r_g\frac{1}{r} +O( \frac{1}{r^2})$&$1+r_g \frac{1}{r} +O( \frac{1}{r^2})$\\
\hline
$g_{\theta\theta}$&$Kr^2+\frac{M}{K}r +O( 1)$&$r^2+O(1)$&$r^2+O(1)$\\
\hline
$g_{\varphi\varphi}$&$Kr^2\sin^2\theta +\frac{M}{K}r+O(1)$&$r^2\sin^2\theta +O(1)$&$r^2\sin^2\theta +O(1)$\\
\hline
$g_{t\varphi}$&$ 0+\frac{4 c_1 c_5 J}{K}\frac{1}{r}+O(\frac{1}{r^2})$&$ N r+ L \frac{1}{r}+O(\frac{1}{r^2})$&0\\
\hline
\end{tabular}

\begin{eqnarray}
K=\sqrt{c_1^2 c_5^2-c_1^2 s_5^2-s_1^2 c_5^2}\\
M=\frac{p}{K}+\frac{c_1^2 c_5^2 J^2}{r_0^2 p n^2 (n^2-1)}\\
N=-2r_g \alpha r \sin^2\theta\\
L=2r_g \alpha^3 \cos^2\theta \sin^3\theta
\end{eqnarray}

\end{table}

\subsubsection{\label{sec2.2}Matter field}

\begin{figure}[htb]
\begin{center}\includegraphics[width=0.8\columnwidth]{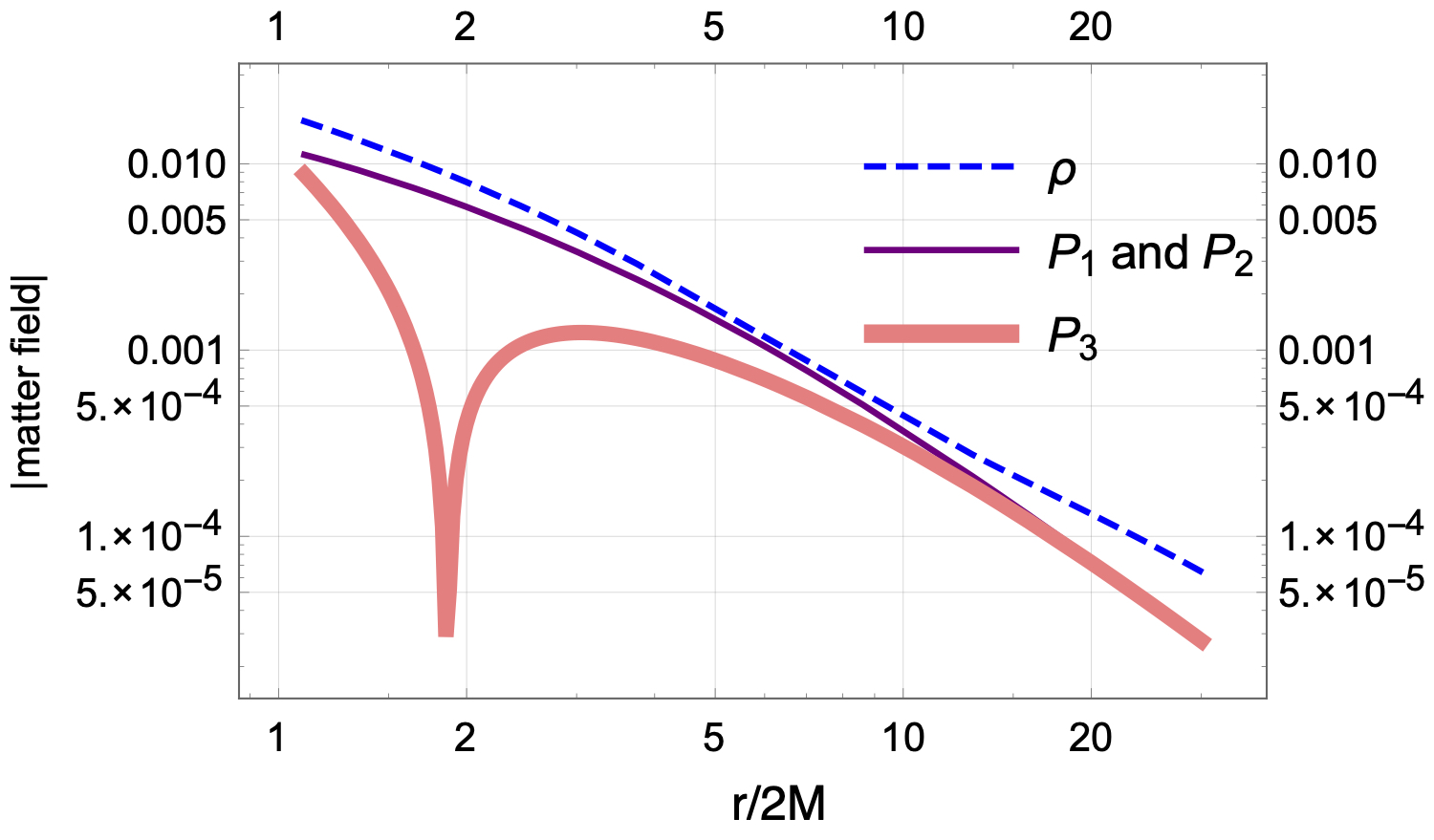}
\caption{\label{fuzzmatter}Near-horizon matter field of fuzzball solution from paper \cite{giusto2007non}. Shown in the figure, $P_1=-P_2$. Asymptotically $P_1=-P_3$.}
\end{center}
\end{figure}

\begin{figure}[htb]
\begin{center}
\includegraphics[width=0.8\columnwidth]{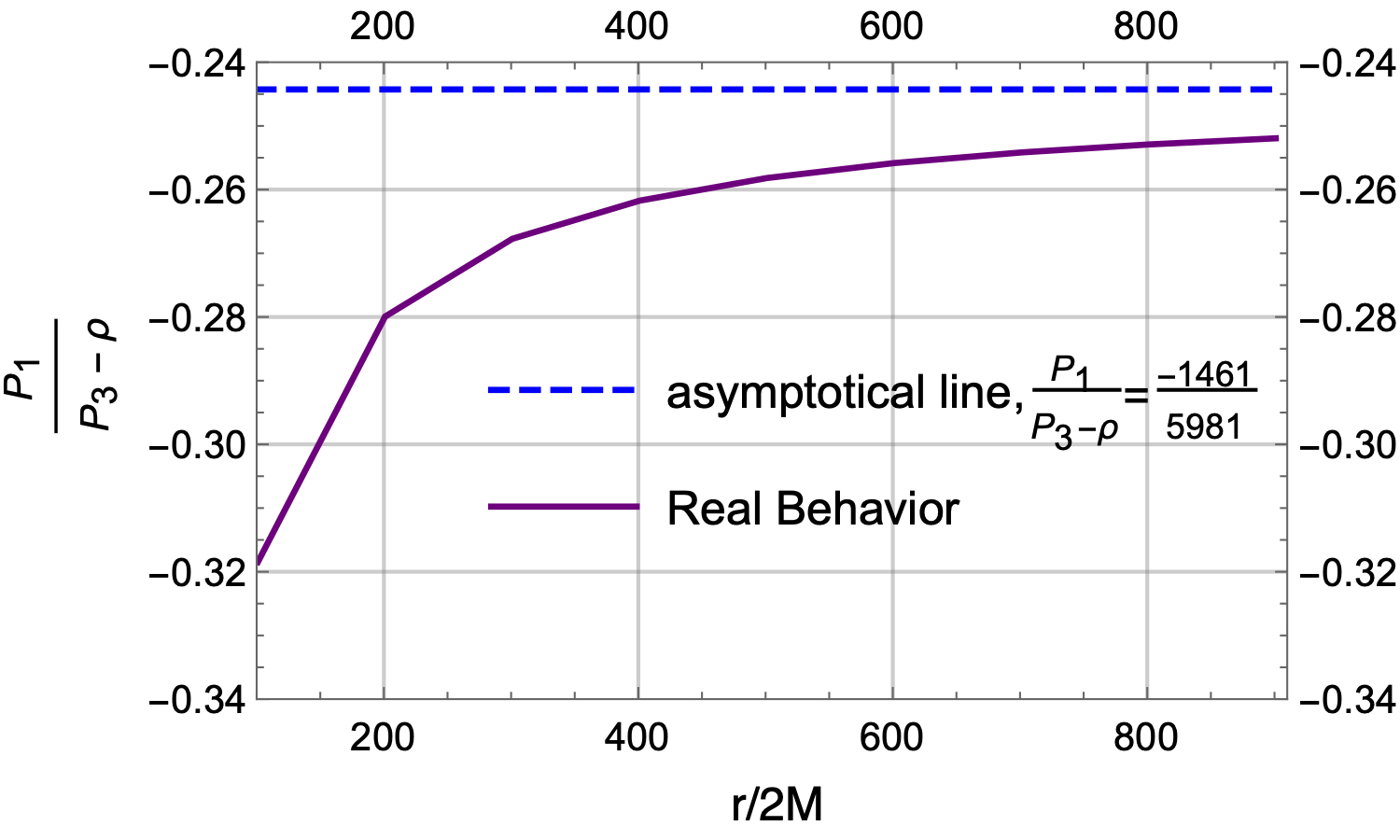}
\caption{\label{fuzzhorizon}Near-horizon relationship between density and pressure of the fuzzball solution from paper \cite{giusto2007non}. The dashed line is the asymptotic behavior, and the solid line is the real behavior of $\frac{P_1}{P_3 - \rho}$}
\end{center}
\end{figure}

We can now study the effective 4d matter stress tensor from the Einstein tensor of the 4d fuzzball geometry (\ref{equ1}). For a sample choice of parameters, outlined in Table \ref{table2}, the (diagonalized) Energy-stress tensor $T^{\mu}_{\ \nu}$ is
\[ \left( \begin{array}{cccc}
-\rho=-\frac{5981}{1461}f(r,\theta)-g(r,\theta) & 0 & 0 & 0 \\
0 & P_1=f(r,\theta) & 0 & 0 \\
0 & 0 & P_2=-f(r,\theta) & 0 \\
0 & 0 & 0 & P_3=g(r,\theta)\\ \end{array} \right)\]
\begin{table}[b]
\caption{\label{table2}Parameters Setting}
\centering
\begin{tabular}{|c|c|c|c|c|c|c|}
\hline
$r_0$&$p$&m&$c_1$&$c_5$&$s_1$&$s_5$ \\
\hline
$1$ & $4$ &2& $2$ & $1$ & $1$ & $\frac{1}{\sqrt{2}}$\\
\hline
\end{tabular}
\end{table}
where $f(r,\theta)$ and $g(r,\theta)$ are functions of coordinate r and $\theta$. The concrete expressions depend on parameter setting. $\rho$, $P_1$, $P_2$ and $P_3$ are the energy density and anisotropic pressure of the matter field. Their behavior near horizon is shown in Fig. \ref{fuzzmatter}. Pressure $P_1=-P_2$ is an analytic result and true for any parameter setting, while relationship between the energy and pressure: $-\rho=-\frac{5981}{1461}f(r,\theta)-g(r,\theta)=-\frac{5981}{1461}P_1-P_3$ is a numerical approximation. The approximation is exact far away from the fuzzball with parameters in Table \ref{table2} and for any given $\theta$ except 0 and $\pi$. The relationship changes near the horizon shown as in Fig \ref{fuzzhorizon} after averaging over $\theta$. At around $r \sim 1000$, we have radial pressure equals tangential pressure $P_1=P_3$. Similar to the metric, matter fields are singular at $r=r_0$, $\theta=0$ and $\pi$. 

The matter field has an anisotropic pressure. It is not traceless so it cannot be a simple electromagnetic field. We also checked that it cannot be a single scalar field.
However, most fuzzball microstates still remain intractable, with no clear dimensional reduction or 4d geometry. To circumvent this obstacle, we have proposed a ``mock fuzzball'' spacetime \cite{wang:2016mock} which captures the horizonless feature of the model with an anisotropic fluid. This conjecture leads to an interesting application to dynamic binary quantum BH merger simulations, as discussed later in Sec. \ref{numerical}.

\subsection{\label{sec3}Mock Fuzzballs}

\begin{figure}[htb]
\begin{center}
\includegraphics[width=\columnwidth]{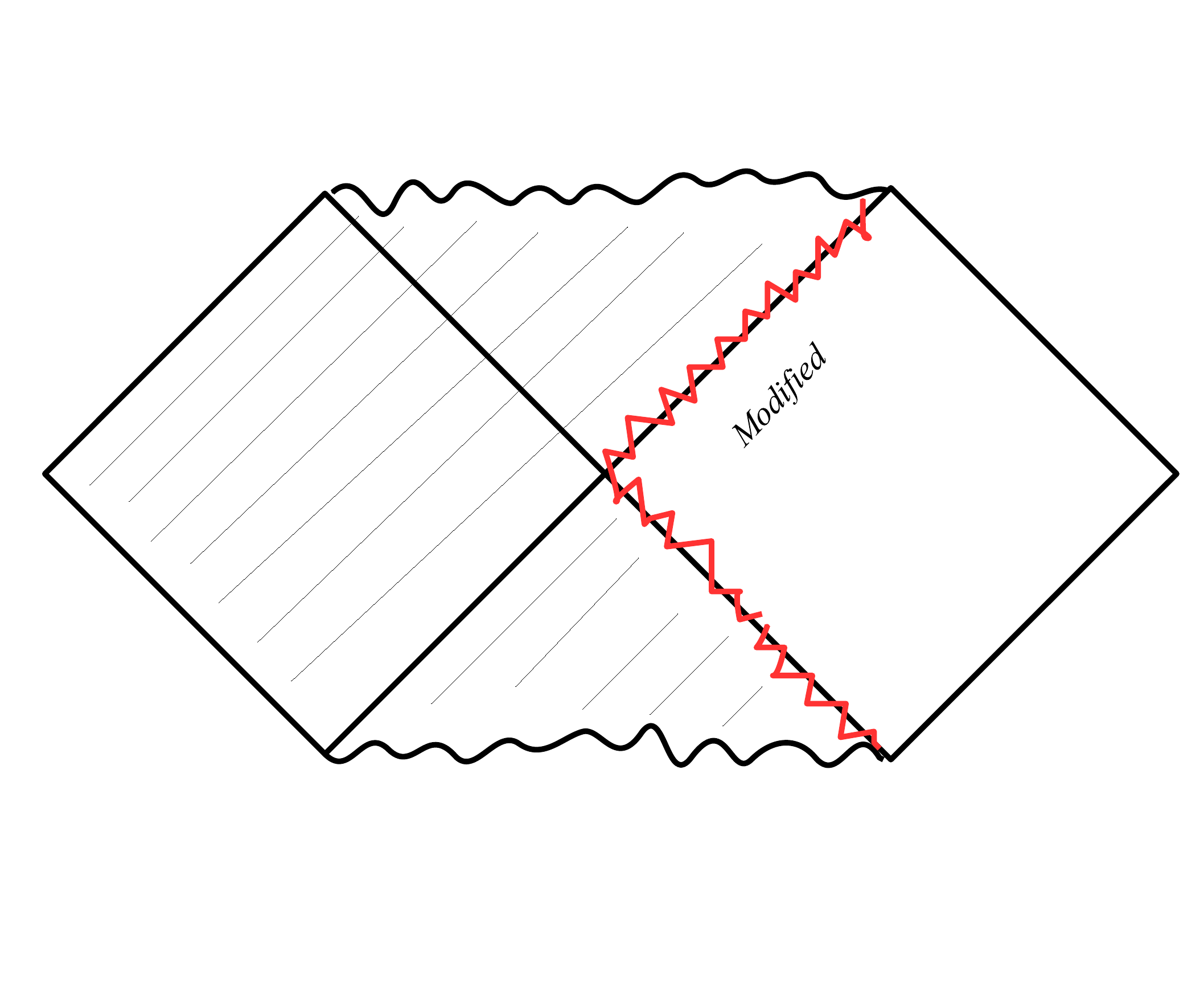}
\caption{\label{figmockf} The geometry of Schwarzschild mock fuzzball. Red curve shows that the metric is modified around the horizon, and the shaded area is removed from the spacetime.}
\end{center}
\end{figure}
\begin{figure}[htb]
\begin{center}
\includegraphics[width=0.45\columnwidth]{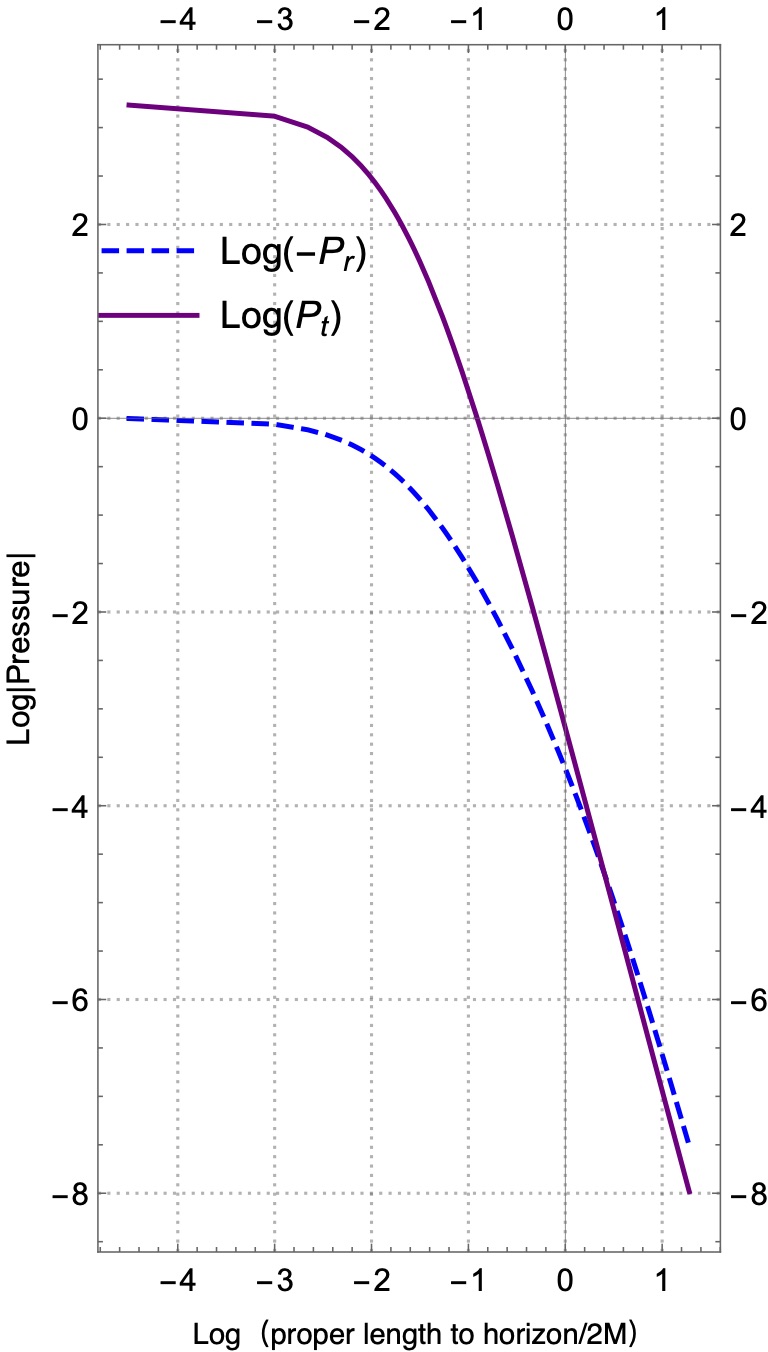}
\caption{\label{schmfuzz} Anisotropic behavior of pressure of Schwarzschild mock fuzzball versus proper distance from horizon. The absolute value of two pressures are equal at $r= \frac{3M}{1+a}$. Radial pressure (dashed line) is much larger than tangential(solid line) pressure near horizon (proper length=0), and opposite far away. The figure shows that when approaching horizon, both pressure reach the extremum, hence have no singularity at r=2M.}
\end{center}
\end{figure}

Here, we introduce a ``mock'' fuzzball geometry, based on the motivation to build a generic and macroscopic metric which captures important, coarse-grained properties of the fuzzball, i.e, the metric has neither horizon nor singularity and spacetime ends around the stretched horizon. Fig. \ref{figmockf} shows how the causal diagram of a Schwarzschild BH is changed to remove the horizon. We study different mock fuzzballs, and check the corresponding matter fields, a swell as potential observable effect. 
The simplest case of the mock fuzzball \footnote{It turns out that this toy model spacetime coincides with the one proposed earlier by \cite{Damour:2007ap}.} is given by:
\begin{eqnarray}
\label{simmock}
ds^2=-(1-\frac{2M}{r}+b) dt^2+\frac{1}{1-\frac{2M}{r}} dr^2+r^2 d\Omega^2, \ r>2M
\end{eqnarray}
where parameter $ 0 < b \ll 1$ is only important around $2M$. It ensures that $g_{tt}$ doesn't vanish and thus remove the horizon. The geometry resembles a traditional BH far away. Besides, this metric is only valid where $r>2M$ to imitate a fuzzball metric which ends around the stretched horizon. The corresponding matter field is
\begin{align}
\label{fakematter}
P_r&=T^{r}_{\ r}= -\frac{bM}{4 \pi r^2(-2M+r+br)}, \\
P_t&=T^{\theta}_{\ \theta} = T^{\varphi}_{\ \varphi}=\frac{bM(-M+r+br)}{8 \pi r^2(-2M+r+br)^2}, \\ &\textnormal{other components vanish}
\end{align}
The anisotropic behavior of the pressure near the stretched horizon is shown as in Fig. \ref{schmfuzz} with $b=0.01$ and $M=\frac{1}{2}$. The absolute value of two pressures are equal at $r= \frac{3M}{1+b}$. Tangential pressure is much larger than radial pressure near horizon, and opposite far away. It doesn't have any singularity at r=2M since all fields reach an extremum there. 

\subsubsection{Other mock fuzzballs}

Besides the simplest case introduced in the last section, we can modify other terms in Schwarzschild metric to recover energy density which fuzzballs in Sec. \ref{fuzzball} actually have. We also study charged and rotating BHs.

\begin{itemize}
    \item Schwarzschild metric
    
The simplest mock fuzzball studied above has no energy density. However, we can recover energy density by changing $g_{\theta\theta}$ and $g_{\varphi\varphi}$ within spherical symmetry:
\begin{align}
ds^2=-(1+b-\frac{2M}{r}) dt^2+\frac{1}{1-\frac{2M}{r}} dr^2+(2 d M + (1 - d) r)^2 d\Omega^2, \ r>2M\\
\rho=-T^{t}_{\ t}=\frac{d}{4M^2}+O(r-2M)\\ 
P_r= T^{r}_{\ r}=-\frac{1}{4M^2}+O(r-2M)\\
P_t=T^{\theta}_{\ \theta}=T^{\varphi}_{\ \varphi}=\frac{1 + 2 b - 2 b d}{16 b M^2}+O(r-2M)\\
\textnormal{other components vanish}
\end{align}
Small b and d ensures that $\rho \ll P_r \ll P_t$. In addition, $b>0,\ d<1$ ensure finite Ricci scalar without curvature singularity at $r>2M$.

Another possible modification to recover energy density is to assume that mass has a time dependence:
\begin{align}
ds^2=-(1+b-\frac{2M(t)}{r}) dt^2+\frac{1}{1-\frac{2M(t)}{r}} dr^2+r^2 d\Omega^2, \ r>2M\\
\rho=-G^{t}_{\ t}=-\frac{\sqrt{-b M(t)^3 M'(t)^2}}{\sqrt{2 b M(t)^3 \sqrt{r-2M(t)}} }+\frac{1}{8M(t)^2}+)(r-2M(t))^\frac{1}{2}\\ 
P_r= G^{r}_{\ r}=-\frac{\sqrt{-b M(t)^3 M'(t)^2}}{\sqrt{2 b M(t)^3 \sqrt{r-2M(t)}} }-\frac{1}{8M(t)^2}+O(r-2M(t))^\frac{1}{2}\\
P_t=G^{\theta}_{\ \theta}=G^{\varphi}_{\ \varphi}=-\frac{3M'(t)^2}{b (r-2M(t))^2}+\frac{M'(t)^2-b M(t)m''(t)}{a^2M(t)(r-M(t))}\\
+\frac{b^2+2b^3-4M'(t)^2-8bM'(t)^2+8bM(t)m''(t)}{16b^3M(t)^2}+O(r-2M(t))\\
\textnormal{other components vanish}
\end{align}

\item Extremal BH metric

For an extremal BH, the fuzzball has another interesting property: Proper length from somewhere near stretched horizon to ``horizon'' is finite, in contrast to the infinite throat in the traditional picture. Parameter $c$ here captures the finite throat.  
\begin{align}
ds^2=-(b+(1-\frac{r_q}{r})^2) dt^2+\frac{1}{c+(1-\frac{r_q}{r})^2} dr^2+r^2 d\Omega^2, \ r>2M\\
\rho=-T^{t}_{\ t}=\frac{1-c}{{r_q}^2}+O(r-r_q)\\ 
P_r= T^{r}_{\ r}=-\frac{-1+c}{{r_q}^2}+O(r-r_q)\\
P_t=T^{\theta}_{\ \theta}=T^{\varphi}_{\ \varphi}=\frac{c}{b {r_q}^2}+O(r-r_q)\\
\textnormal{other components vanish}
\end{align}

\item Non-Extremal BH metric
\begin{align}
ds^2=-(a+1-\frac{r_s}{r}+(\frac{r_q}{r})^2) dt^2+\frac{1}{c+1-\frac{r_s}{r}+(\frac{r_q}{r})^2} dr^2+r^2 d\Omega^2, \ r>2M\\
\rho=-T^{t}_{\ t}=\frac{c r^2 - rq^2}{{r}^4}\\ 
P_r= T^{r}_{\ r}=-\frac{(1 + a) c r^4 + (-1 + a) r^2 {r_q}^2 - c r^2 {r_q}^2 -{r_q}^4 - a r^3 rs + 
 r {r_q}^2 rs}{r^4 ({r_q}^2 + r (r + a r - rs))}\\
Pt=T^{\theta}_{\ \theta}=T^{\varphi}_{\ \varphi} \textnormal{ dropped for simplicity}\\
\textnormal{other components vanish}
\end{align}
\end{itemize}

\subsubsection{What does an infalling observer see?}

Assuming Einstein field equations, mock fuzzball geometries can only be sourced by matter fields with exotic (and anisotropic) equations of state. Considering simplest Schwarzschild mock fuzzball:
\begin{equation}
ds^2=-(1-\frac{2M}{r}+b) dt^2+\frac{1}{1-\frac{2M}{r}} dr^2+r^2 d\Omega^2,
\end{equation}
There is no longer vacuum outside the horizon, which may potentially lead to observable effects, depending on how strongly the ``fuzz'' matter can interact with the detectors. To visualize the signal, we assume a geodesic observer radially falling towards the stretched horizon with zero velocity at infinity. We calculate (see Appendix \ref{fuzzballflux} for details) two observable scalars: energy density $\mathcal{U}$ and energy flux $\mathcal{F}$, as seen by the observer:
\begin{align}
 \mathcal{U}&= T_{\mu \nu} u^{\mu}  u^{\nu}=-\frac{4 b M^2}{r^2 (b r-2 M+r)^2},\\
 \mathcal{F}&= -T_{\mu \nu} u^{\mu}  a^{\nu}=-\frac{ b \left(\frac{2M}{r}\right)^{3/2}}{(b r-2 M+r)^2},
\end{align}
where $T_{\mu \nu}$ is from Einstein field equation of the mock fuzzball, $a^{\nu}$ is the unit detector area vector and $u^{\mu}$ is the four-velocity of the observer with $a_\mu u^\mu=0$. Both energy density and flux are finite and vanish when parameter $b$ vanishes.

\begin{figure}[htb]
\begin{center}
\includegraphics[width=\columnwidth]{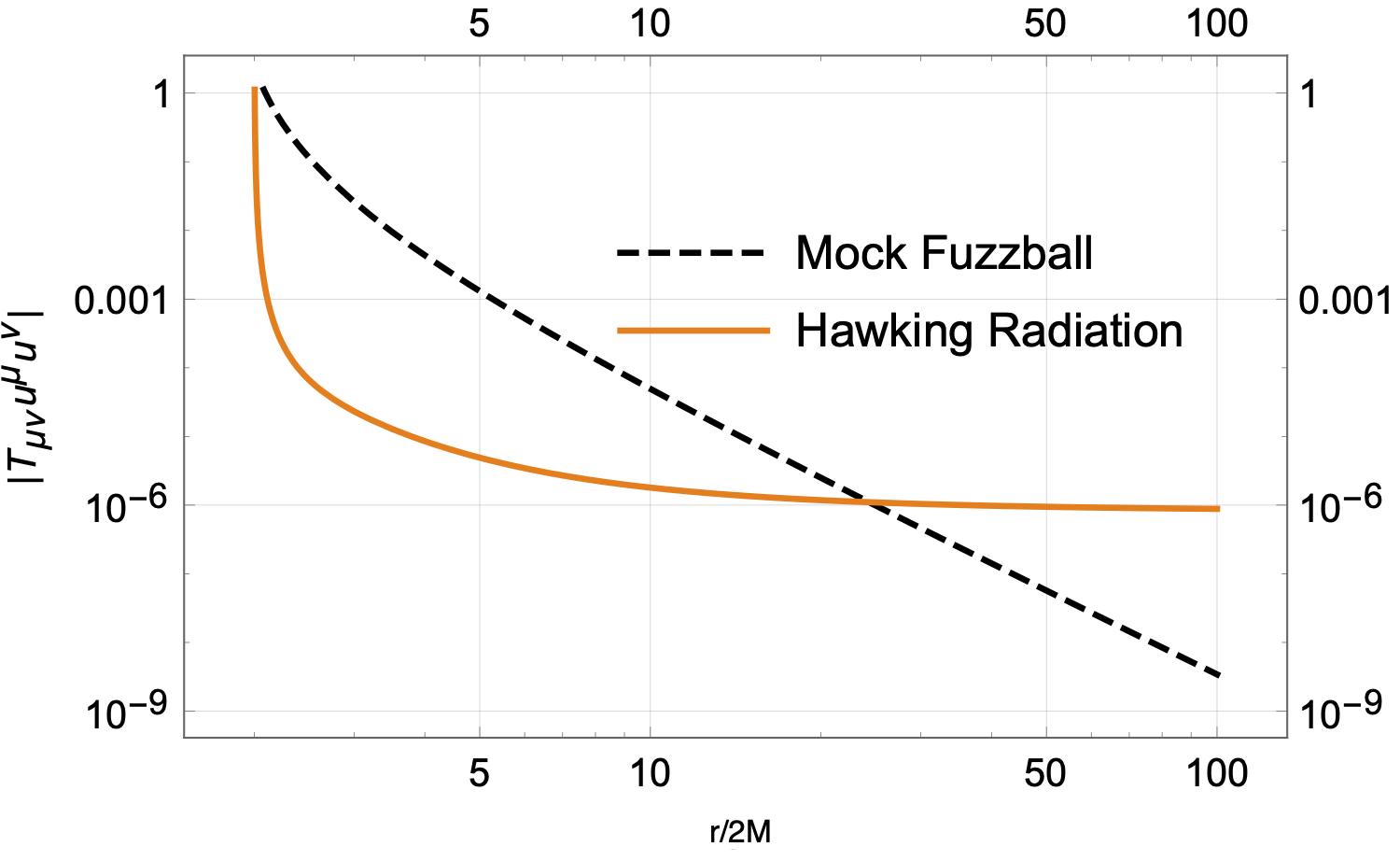}
\caption{\label{energy} The energy density of mock fuzzball, compared with that of Hawking radiation as seen by a radially infalling observer along a geodesic. Here the orange curve is positive and the black curve is negative. The energy density of mock fuzzball is larger than the Hawking radiation near horizon and drops faster than the Hawking radiation away from the horizon.}
\end{center}
\end{figure}

\begin{figure}[htb]
\begin{center}
\includegraphics[width=\columnwidth]{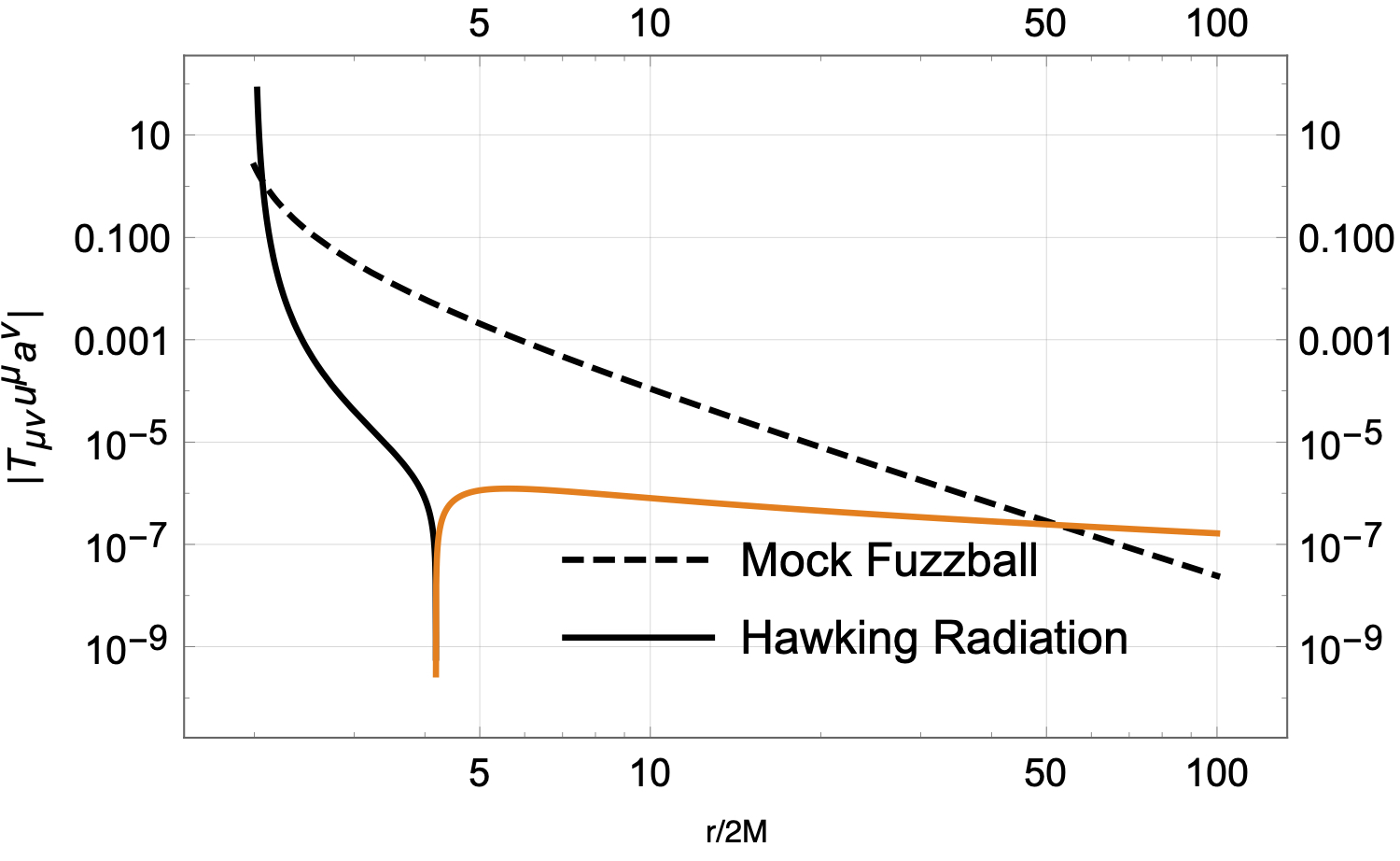}
\caption{\label{flux} The energy flux seen by an observer falling through fuzzball geometry, compared to that of Hawking radiation. Here the orange curve is positive, while the black curves are negative. The flux of mock fuzzball is larger than the Hawking radiation near horizon and drops faster than the Hawking radiation away from the horizon.}
\end{center}
\end{figure}

To visualize observable energy density and flux, we choose parameters $M=1$ and $b=0.1$ and compare it with the signal of Hawking radiation as Fig. \ref{energy} and Fig. \ref{flux}. Both energy density and flux of the mock fuzzball are larger than those of Hawking radiation near the horizon, but drops faster away from the horizon. Energy density of the mock fuzzball is negative while Hawking radiation is positive. Flux of the mock fuzzball flows into BH while Hawking radiation flows into BH near horizon and changes direction away from the horizon. 

Specially, energy density and flux of fuzzball are finite near horizon while Hawking radiation diverges. This is because we use Page's approximation of the Hawking radiation \cite{page1982thermal}, which assumes the Hartle-Hawking state for expectation value of stress-energy tensor:
\begin{align}
T_{tr}&=\frac{-7.44 10^5}{4 \pi r(r-2M)\frac{1}{M^2}},\\
T_{rr}&=\frac{1}{1-2M/r} \alpha^{-1} (\beta+960\frac{M^6}{r^6}),\\
T_{tt}&=(1-2M/r) 3 \alpha^{-1} (\beta-2112\frac{M^6}{r^6}),\\
\alpha&=368640 {\pi}^2 M^2,\\
\beta&=192 \left(\frac{M}{r}\right)^5+80 \left(\frac{M}{r}\right)^4+32 \left(\frac{M}{r}\right)^3+12 \left(\frac{M}{r}\right)^2+\frac{4M}{r}+1,
\end{align}

The Hartle-Hawking state is a thermal equilibrium states of particles while the Hawking radiation is not in equilibrium. The inconsistency leads to infinite flux and energy density of Hawking radiation. More precise correction can be found in \cite{parker2009quantum}.

\subsection{Aether Holes and Dark Energy}

In 2009, Prescod-Weinstein, Afshordi, and Balogh \cite{prescod2009stellar} studied the spherically symmetric solutions of the Gravitational Aether proposal for solving the old cosmological constant problem \cite{Afshordi:2008xu,Aslanbeigi:2011si}. Surprisingly, they showed that if one sets Planck-scale boundary conditions for aether near the horizons of stellar mass BHs, its pressure will match the observed pressure of dark energy at infinity. 

In the Gravitational Aether proposal \cite{Afshordi:2008xu,Aslanbeigi:2011si}, the modified Einstein field equation is given by
\begin{align}
\frac{1}{8 \pi G'}G_{\mu \nu}=T_{\mu \nu}-\frac{1}{4}T ^{\alpha}_{\ \alpha}g_{\mu \nu}+T'_{\mu \nu},\\
T'_{\mu \nu}=p'(u'_{\mu}u'_{\nu}+g_{\mu \nu}),
\end{align}
where $G'=\frac{4}{3}G_N$, and then energy-momentum tensor of aether is assumed to be a perfect fluid with stress-energy tensor $T'_{\mu \nu}$ without energy density. Here, quantum vacuum energy decouples from the gravity, as only the traceless part of the matter energy-momentum tensor appears on the right-hand side of the field equations. It can be shown that the Bianchi identity and energy-momentum conservation completely fix the dynamics, and thus the theory has no additional free parameters, or dynamical degrees of freedom, compared to General Relativity. 

The modified Schwarzschild metric is the vacuum solution with spherical symmetry in modified equations, and identical to a traditional equations sourced by the aether perfect fluid. Far away from the would-be horizon but close enough to the origin ($2M \ll r \ll |p_0|^{-1/2}$), the solution has the form
\begin{equation}
ds^2=-(1+4\pi p_0 r^2)dt^2 +dr^2 +r^2 d\Omega^2
\end{equation}
which can be compared to the de Sitter metric
\begin{equation}
ds^2=-(1-\frac{8}{3}\pi \rho_\Lambda r^2)dt^2 + (1-\frac{8}{3}\pi \rho_\Lambda r^2)^{-1} dr^2 +r^2 d\Omega^2
\end{equation}

We see that assuming $p_0=-\frac{2}{3}\rho_\Lambda$, the $g_{\rm tt}$'s agree with each other. Therefore, the Newtonian observers (for $2M \ll r \ll |p_0|^{-1/2}$) will experience the same acceleration as in the de-Sitter metric with the cosmological constant. However, on larger scales, one has to take into account the effects of multiple black holes and other matter in the Universe. The Planckian boundary conditions at the (would-be) horizon relates the pressure of the aether to the mass of the astrophysical BHs, $-p_0 \sim M^{-3}$  \cite{prescod2009stellar}. In particular, the BH masses within the range $10~ M_{\odot}-100~ M_{\odot}$, which correspond to the most astrophysical BHs in galaxies, yield aether pressures comparable to the pressure of Dark Energy, inferred from cosmic acceleration.   Moreover, Ricci scalar is inversely proportional to $g_{tt}$, so the event horizon where $g_{tt}=0$ has a curvature singularity, which is reminiscent of the firewall and fuzzball proposals discussed above.  

In particular, the fuzzball paradigm is a good approach to remove the singularity. On the one hand, fuzzball gives an extra anisotropic matter field similar to the aether theory, which stands as a good evidence that quantum effects can modify the Einstein field equation with extra sources of 4d energy-momentum like aether. Furthermore, fuzzball is a regular and horizonless geometry, which might indicate the singularity is removable in the full quantum picture of BHs.

\subsection{2-2 holes}
In general relativity, gravitational collapse of ordinary matter will always leads to singularities behind trapping horizons \cite{1965PhRvL..14...57P}. In \cite{Holdom:2016nek}, Holdom and Ren revisited this problem with the asymptotically free quadratic gravity, which could be regarded as a UV completion of general relativity \cite{Holdom:2016nek}. The quantum quadratic gravity (QQG), whose action is given by
\begin{equation}
S_{\text{QQG}} = \int d^4 x \sqrt{-g} \left( \frac{1}{2} {\cal M}^2 R - \frac{1}{2 f_2^2} C_{\mu \nu \alpha \beta} C^{\mu \nu \alpha \beta} + \frac{1}{3 f_0^2} R^2 \right),
\label{QQG_action}
\end{equation}
is famously known to be not only asymptotically free, but also perturbatively renormalizable \cite{Stelle:1976gc,Voronov:1984,Fradkin:1981iu,Avramidi:1985ki}. However, it suffers from a spin-2 ghost due to the higher derivative terms, which is commonly regarded as a pathology of the theory. In \cite{Holdom:2016nek}, it is proposed that the ghost may not be problematic when ${\cal M}$ is sufficiently small, so that the poles in the perturbative propagators fall into the non-perturbative regime, and the perturbative analysis of ghosts is not reliable. Then it is conjectured \cite{Holdom:2015kbf} that the full graviton propagator in the IR, when ${\cal M} \lesssim \Lambda_{QQG}$, the spin-2 ghost pole is absent in an analogy with the quantum chromodynamics (QCD) where the gluon propagator, describing off-shell gluons, also does not have a pole. Here $\Lambda_{QQG}$ is a certain critical value in QQG, analogous to confinment scale $\Lambda_{QCD}$ in QCD. Based on this conjecture, the asymptotically free quadratic action in (\ref{QQG_action}) may involve small quadratic corrections at super-Planckian scale, and so the super-Planckian gravity might be governed by the classical action
\begin{equation}
S_{CQG} = \frac{1}{16 \pi} \int d^4x \sqrt{-g} \left( M_{\text{Pl}}^2 R -\alpha C_{\mu \nu \alpha \beta} C^{\mu \nu \alpha \beta} + \beta R^2 \right).
\label{CQG_action}
\end{equation}
Since gravitational collapse would involve the super-Planckian energy scale, applying the classical action (\ref{CQG_action}) to such a situation is interesting from a point of view of the quantum gravitational phenomenology. Then the authors in \cite{Holdom:2016nek} found a solution of horizonless compact object, so-called 2-2 hole, in the classical quadratic gravity. 2-2 holes have an interior with a shrinking volume and a timelike curvature singularity at the origin. It also has a thin-shell configuration, leading to non-zero reflectivity at the would-be horizon, which may cause the emission of GW echoes \cite{Conklin:2017lwb}. Recently, 2-2 holes sourced by thermal gases were also investigated in \cite{Holdom:2019ouz,Ren:2019afg}.

\subsection{Non-violent Unitarization}\label{sec:non-violent}

A separate class of possible approaches to the BH information paradox involves a violation Postulate 2 in BH complementarity, i.e. non-locality of field equations well outside the stretched horizon, which is dubbed as ``nonviolent unitarization'' by Steve Giddings \cite{Giddings:2017mym}. Such a possibility would allow for transfer of information outside horizon around the Page time (e.g., \cite{Bardeen:2018omt,Bardeen:2018frm}), but could also lead to large scale observable deviations from general relativistic predictions in GW and electromagnetic signals \cite{Giddings:2019vvj}. However, it is not clear whether this non-locality is only limited to BH neighborhoods, and if not, how it could affect precision experimental/observational tests in other contexts. Moreover, in contrast to GW echoes that we shall discuss next, it is hard to provide concrete predictions for astrophysical observations in the nonviolent unitarization scenarios.  

\section{Gravitational Wave Echoes: Predictions}
\label{sec:echo_predictions}
GW echoes may be one of the observable astrophysical signals, a smoking gun, so to speak, for the quantum gravitational processes near BH horizons. A number of models of Exotic Compact Objects (ECOs) that we discussed above are expected to emit GW echoes. Some examples are wormholes \cite{Cardoso:2016rao}, gravastars \cite{Cardoso:2016oxy}, and 2-2 holes \cite{Holdom:2016nek}. Moreover, even Planckian correction in the dispersion relation of gravitational field \cite{Oshita:2018fqu,Oshita:2019sat,Wang:2019rcf} and the BH area quantization \cite{Cardoso:2019apo} may also lead to echo signals. Not only the specific models to reproduce GW echoes but also comprehensive modeling of echo spectra in non-spinning case \cite{Mark:2017dnq}, in spinning case \cite{Wang:2019rcf,Conklin:2019fcs}, and in a semi-analytical way \cite{Testa:2018bzd,Maggio:2019zyv} have been investigated, which enable us to easily obtain echo spectra. In this section, we review the details of GW echoes by starting with the Chandrasekhar-Detweiler (CD) equation \cite{Chandrasekhar:1976zz,Detweiler:1977gy} that is a wave equation with a purely real angular momentum barrier in the Kerr spacetime. We also provide a short review of the GW ringdown signal, that is followed by the GW echo, and the superradiance of spinning BHs. The superradiance with a high reflectivity at the would-be horizon may cause the ergoregion instability, which we shall also discuss separately.
\subsection{On the equations governing the gravitational perturbation of spinning BHs}
\begin{figure}[t]
\begin{center}
\includegraphics[keepaspectratio=true,height=45mm]{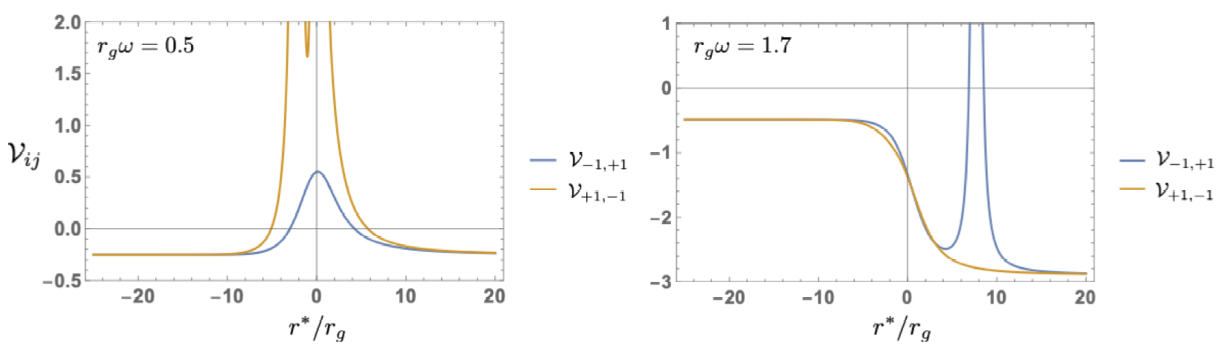}
\end{center}
\caption{The CD potentials with $(i,j) = (-1,+1)$ and $(+1,-1)$ for $\bar{a} = 0.8$ and $\ell = m = 2$. The potential ${\cal V}_{+1,-1}$ (${\cal V}_{-1,+1}$) is singular for $r_g \omega = 0.5$ $(1.7)$ in this case.
}%
  \label{CD_potential}
\end{figure}

The GW ringdown is one of the most important signals to probe the structure of BH since it mainly consists of discrete QNMs of BH characterized by mass and spin. In this subsection, we review that the QNMs can be obtained by looking for specific complex frequencies such that the mode functions of GWs satisfy the outgoing boundary condition. Let us start with the CD equation \cite{Chandrasekhar:1976zz,Detweiler:1977gy} that is the wave equation for a spin-$s$ field and has a purely real angular momentum barrier:
\begin{equation}
\left[ \frac{d^2}{dr^{\ast} {}^2} - {\cal V}_{ij} \right] {}_s X_{lm} (r^{\ast}, \omega) = - T,
\label{SN_equation}
\end{equation}
where $T$ is the source term and the potential ${\cal V}_{ij}$ with $i,j = \pm 1$ is given by
\begin{align}
\begin{split}
{\cal V}_{ij} &= \frac{-K^2}{(r^2+a^2)^2} + \frac{\rho^4 \Delta}{(r^2+a^2)^2} \left[ \frac{\lambda (\lambda+2)}{g+ b_i \Delta} -b_i \frac{\Delta}{\rho^8} + \frac{(\kappa_{ij} \rho^2 \Delta - h) (\kappa_{ij} \rho^2 g -b_i h)}{\rho^4 (g+b_i \Delta) (g-b_i \Delta)^2} \right]\\
&+ \left[ \frac{r \Delta a m/\omega}{(r^2+a^2)^2 \rho^2} \right]^2 - \frac{\Delta}{(r^2+a^2)} \frac{d}{dr} \left[ \frac{r \Delta a m/\omega}{(r^2+a^2)^2 \rho^2} \right].
\end{split}
\label{CD_potential_1}
\end{align}
The functions in (\ref{CD_potential_1}) are defined by
\begin{align}
b_{\pm1} &\equiv \pm 3 (a^2 - am/\omega),\\
\kappa_{ij} &\equiv j \left\{ 36 M^2 -2\lambda \left[ (a^2-am/\omega) (5\lambda+6) -12 a^2 \right] + 2 b_i \lambda (\lambda +2) \right\}^{1/2},\\
\rho^2 &\equiv r^2+a^2-am/\omega,\\
\Xi_{i} & \equiv \frac{\Delta^2}{\rho^8} (F+b_i),\\
\Theta_{ij} &\equiv i \omega +\frac{1}{F-b_i} \left( \frac{\Delta}{\rho^2} \frac{dF}{dr} -\kappa_{ij} \right),\\
\begin{split}
\kappa &\equiv (\lambda^2 (\lambda+2)^2 +144 a^2\omega^2 (m-a\omega)^2 -a^2 \omega^2 (40 \lambda^2-48 \lambda)+a\omega m (40 \lambda^2 +48 \lambda))^{1/2}\\
&~~~~+12 i \omega M,
\end{split}\\
F & \equiv \frac{\lambda \rho^4 +3 \rho^2 (r^2-a^2) -3r^2 \Delta}{\Delta},\\
g & \equiv \lambda \rho^4 +3 \rho^2 (r^2-a^2) -3r^2 \Delta,\\
h & \equiv g' \Delta - g \Delta'.
\end{align}
Equation (\ref{CD_potential_1}) gives four potentials, $(i,j) = (-1,+1), \ (+1, -1), \ (+1,+1)$, and $(-1,-1)$. One has to use the different potentials in order to cover the whole frequency space with the CD potentials because $1/(g+ b_{+1} \Delta)$ and $1/(g+ b_{-1} \Delta)$ in (\ref{CD_potential_1}) are singular in different frequency regions (see FIG. \ref{CD_potential}).

The CD equation is obtained as the generalized Darboux transformation of the Teukolsky equation \cite{Chandrasekhar:1975zza}\footnote{Recently, it was found out that the CD equation with $(i,j) = (+1, \pm 1)$ is related to $(-1,\mp1)$ by the Darboux transformation.}.
In the asymptotic regions, $r^{\ast} \to \pm \infty$, the CD equation reduces to the following wave equation
\begin{align}
\begin{cases}\displaystyle
\left( \frac{d^2}{dr^{\ast} {}^2} + \tilde{\omega}^2 \right) {}_s X_{lm} = - T, \ &\text{for} \ r^{\ast} \to - \infty,\\
\displaystyle
\left( \frac{d^2}{dr^{\ast} {}^2} + \omega^2 \right) {}_s X_{lm} = -T, \ &\text{for} \ r^{\ast} \to + \infty,
\end{cases}
\label{SN_equation_asmp}
\end{align}
where $\tilde{\omega} \equiv \omega - m \Omega_{\rm H}$, in terms horizon angular frequency $\Omega_{\rm H} \equiv a / (2 M r_+)$, and horizon outre radius $r_+ \equiv M + \sqrt{M^2 - a^2}$ of the Kerr BH. In the following, we will omit the subscripts of $l$, $m$, and $s$ for brevity. One can read that the homogeneous solutions of (\ref{SN_equation_asmp}) are given by the superposition of ingoing and outgoing modes
\begin{equation}
X=
\begin{cases}
A e^{i \tilde{\omega} r^{\ast}} + B e^{-i \tilde{\omega} r^{\ast}} & \text{for} \ r^{\ast} \to - \infty,\\
C e^{i \omega r^{\ast}} + D e^{-i \omega r^{\ast}} & \text{for} \ r^{\ast} \to  +\infty,
\end{cases}
\end{equation}
where $A$, $B$, $C$, and $D$ are arbitrary constants. The QNMs can be found by looking for the complex frequencies at which the homogeneous solution satisfies the outgoing boundary condition of $A=D = 0$. This is equivalent to looking for the zero-points of the Wronskian between the two homogeneous solutions $X_{+}$ and $X_{-}$
\begin{equation}
W_{\text{BH}} \equiv X_{-} \frac{d X_{+}}{dr^{\ast}} - X_+ \frac{d X_-}{d r^{\ast}} = 2i \omega A_{\text{in}} (\omega) = 2i\tilde{\omega} B_{\text{out}} (\omega),
\end{equation}
where the two homogeneous solutions satisfy the following boundary conditions
\begin{align}
&X_- \sim
\begin{cases}
e^{-i \tilde{\omega} r^{\ast}} \ &\text{for} \ r^{\ast} \to - \infty,\\
A_{\text{out}} e^{i \omega r^{\ast}} + A_{\text{in}} e^{-i \omega r^{\ast}} \ &\text{for} \ r^{\ast} \to + \infty,
\label{modein}
\end{cases}\\
&X_+ \sim
\begin{cases}
B_{\text{in}} e^{-i \tilde{\omega} r^{\ast}} + B_{\text{out}} e^{i \tilde{\omega} r^{\ast}} \ &\text{for} \ r^{\ast} \to - \infty,\\
e^{i \omega r^{\ast}} \ &\text{for} \ r^{\ast} \to + \infty.
\label{modeout}
\end{cases}
\end{align}
Recently, the QNMs of Kerr spacetime were precicely investigated in \cite{Casals:2019vdb} by using the method developed by Mano, Suzuki, and Takasugi \cite{10.1143/PTP.95.1079,Mano:1996mf,Sasaki:2003xr,Casals:2018eev} that enables us to obtain the solution of the Teukolsky equation in an analytic way.

\subsection{\label{super}Transmission and reflection coefficients of the angular momentum barrier}
\begin{figure}
\begin{center}
\includegraphics[width=0.7\textwidth]{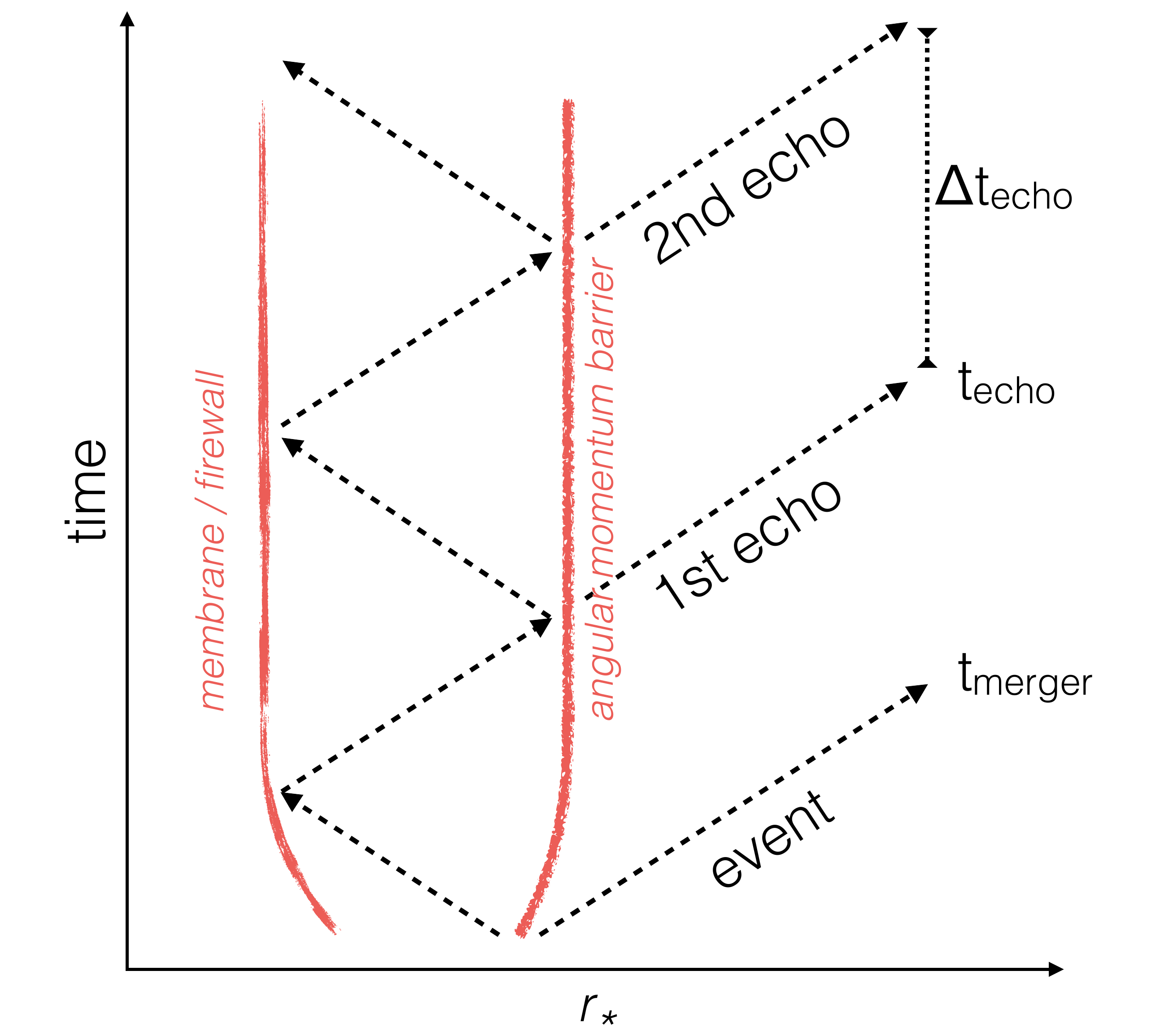}
\caption{GW echoes following a BBH merger from a cavity of membrane/firewall-angular momentum barrier \cite{Abedi:2016hgu}. }
\label{echo_pic_1}
\end{center}
\end{figure}

The GW echoes are results of multiple reflections in the {\textit cavity} between the would-be horizon (e.g., fuzzball/firewall) and angular momentum barrier (see Fig. \ref{echo_pic_1}) and so the amplitude of echoes is mainly determined by the reflectivities of the would-be horizon and angular momentum barrier. In this subsection, we review the calculation of the reflectivity of angular momentum barrier.

\begin{figure}[t]
\begin{center}
\includegraphics[keepaspectratio=true,height=45mm]{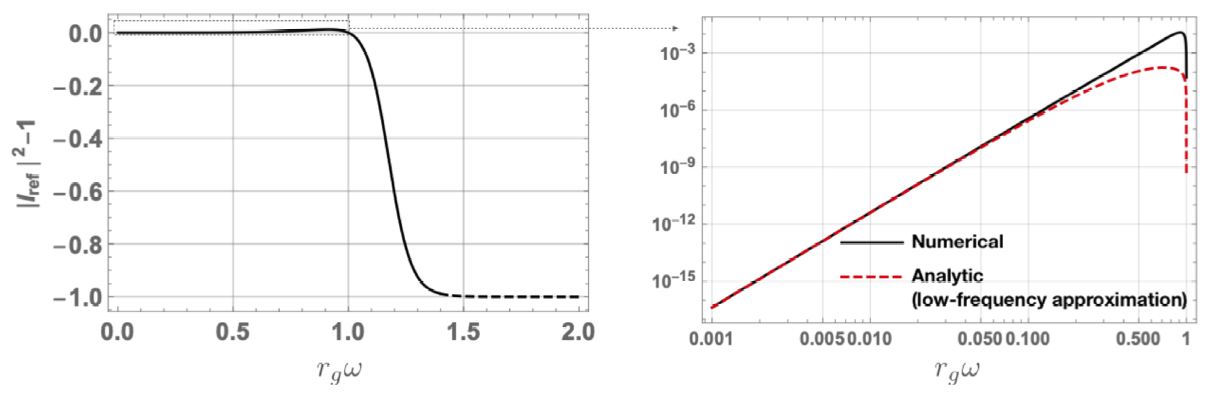}
\end{center}
\caption{The amplification factor for $\bar{a} = 0.8$ and $\ell = m = 2$. In the right panel, low frequency region (solid) is calculated with the potential ${\cal V}_{-1,+1}$ and the higher frequency region (dashed) is calculated from ${\cal V}_{+1,-1}$ in the CD equation.}
  \label{ampl1}
\end{figure}

From the mode functions (\ref{modein}, \ref{modeout}), one can obtain the energy conservation law for the incident, reflected, and transmitted waves by using another Wronskian relation
\begin{equation}
\tilde{W}_{\text{BH}} \equiv X \frac{d X^{\ast}}{dr^{\ast}} - X^{\ast} \frac{d X}{dr^{\ast}},
\end{equation}
which is constant for real frequency due to the reality of the angular momentum barrier in the CD equation. Then we obtain the following relations by using $\tilde{W}_{\text{BH}} (- \infty) = \tilde{W}_{\text{BH}} (+ \infty)$ for $X_-$ and $X_+$
\begin{align}
&1-\left| \frac{A_{\text{out}}}{A_{\text{in}}} \right|^2 = \frac{\tilde{\omega}}{\omega} \frac{1}{|A_{\text{in}}|^2},\\
&1-\left| \frac{B_{\text{in}}}{B_{\text{out}}} \right|^2 = \frac{\omega}{\tilde{\omega}} \frac{1}{|B_{\text{out}}|^2}.
\end{align}
From the above relations, one can read that the energy reflectivity and transmissivity for inward incident waves
\begin{equation}
I_{\text{ref}}^{\leftarrow} \equiv |A_{\text{out}}/A_{\text{in}}|^2, \ \  I_{\text{trans}}^{\leftarrow} \equiv (\tilde{\omega}/\omega) |{1}/A_{\text{in}}|^2,
\label{ref/tra_in}
\end{equation}
respectively, and those for outward incident waves are given by 
\begin{equation}
I_{\text{ref}}^{\rightarrow} \equiv |B_{\text{in}}/B_{\text{out}}|^2, \ \  I_{\text{trans}}^{\rightarrow} \equiv (\omega/\tilde{\omega}) |{1}/B_{\text{out}}|^2.
\label{ref/tra_out}
\end{equation}
From (\ref{ref/tra_in}, \ref{ref/tra_out}), one can calculate the reflectivity/transmissivity of the angular momentum barrier by numerically solving the homogeneous CD equation (\ref{SN_equation}). In the spinning case $\bar{a} > 0$, the energy reflectivity is greater than $1$ for $-m \Omega_H< \tilde{\omega} < 0$, a phenomenon that is often referred to as BH {\it superradiance}. The superradiance can be characterized by the amplification factor $Z \equiv (I_{\rm ref}^{\leftarrow/\rightarrow})^2-1$, and when only interested in the low frequency region, one can use the analytic expression \cite{1973JETP...37...28S}
\begin{equation}\displaystyle
Z \simeq 4 Q \beta_{sl} \prod_{k=1}^{l} \left( 1+ \frac{4 Q^2}{k^2} \right) [\omega (r_+ -r_-)]^{2l +1},
\label{sta-san}
\end{equation}
where $r_-$ is the radius of inner horizon, $\sqrt{\beta_{sl}} \equiv \frac{(l-s)! (l+s)!}{(2l)! (2l+1)!!}$ and $Q \equiv- \frac{r_+^2 +a^2}{r_+-r_-} \tilde{\omega}$.
To give a few examples of $I_{\text{ref}}^{\leftarrow}/I_{\text{ref}}^{\rightarrow}$, we numerically calculate it in the frequency range $0.001 \leq 2M \omega \leq 2$, which is shown in the FIG. \ref{ampl1}.
As can be seen from FIG. \ref{ampl1}, the energy flux reflectivity exceeds $1$, which means that the energy of a spinning BH is extracted by reflected radiation, within the superradiance regime. We will discuss the ergoregion instability caused by the superradiance and the reflectivity of would-be horizon in Sec. \ref{subsec_ergoinst}.

\subsection{Transfer function of echo spectra and geometric optics approximation}
When the GW echo is caused by an incident wave packet repeatedly reflected between the cavity, one can use the geometric optics approximation to predict the GW echo signal, which was first pioneered in \cite{Mark:2017dnq}. Let us start with the calculation of the Green's function of GW ringdwon signal $G_{\text{BH}}$ by using the CD equation. It satisfies
\begin{equation}
\left( \frac{d^2}{dr^{\ast} {}^2} - {\cal V} \right) G_{\text{BH}}(r^{\ast}, r^{\ast} {}') = \delta (r^{\ast}-r^{\ast} {}'),
\end{equation}
where we omit the subscripts of ${\cal V}_{ij}$. Once imposing the outgoing boundary condition, the Green's function $G_{\text{BH}}$ is uniquely determined as
\begin{equation}
G_{\text{BH}} (r^{\ast}, r^{\ast} {}') = \frac{X_- (r^{\ast}_{<}) X_+ (r^{\ast}_>)}{W_{\text{BH}}},
\end{equation}
where $r^{\ast}_{<} \equiv \text{min} (r^{\ast}, r^{\ast} {}')$ and $r^{\ast}_{>} \equiv \text{max} (r^{\ast}, r^{\ast} {}')$. Therefore, when there is no reflectivity at the horizon, the Fourier mode of GWs at infinity and at the horizon can be obtained as
\begin{align}
\lim_{r^{\ast} \to \infty} X(r^{\ast}, \omega) &= -X_+ (r^{\ast}) \int^{+ \infty}_{-\infty} d r^{\ast} {}' \frac{X_- (r^{\ast} {}') T (r^{\ast} {}')}{W_{\text{BH}}} \equiv X_+ Z_{\infty} (\omega),
\label{Z_inf}\\
\lim_{r^{\ast} \to -\infty} X(r^{\ast}, \omega) &= -X_- (r^{\ast}) \int^{+ \infty}_{-\infty} d r^{\ast} {}' \frac{X_+ (r^{\ast} {}') T (r^{\ast} {}')}{W_{\text{BH}}} \equiv X_-
Z_{\text{BH}} (\omega),
\label{Z_BH}
\end{align}
where $T(r^*)$ is the source for the inhomogeneous CD equation. 
If there is no reflection near the horizon, the relevant observable spectrum is only $Z_{\infty}$, and $Z_{\text{BH}}$ is irrelevant for observation. On the other hand, if reflection at the would-be horizon is caused by a certain mechanism, $Z_{BH}$ is also observable in addition to $Z_{\infty}$.

One can obtain echo spectra by using the geometric optics approximation, which should be reliable as long as the would-be horizon and angular momentum barrier are well separated in tortoise coordinates, $r^*$. The amplitude of the first echo, $Z_{\text{echo}}^{(1)}$, can be estimated by
\begin{equation}
Z_{\text{echo}}^{(1)} \simeq {\cal T}_{\text{BH}}^{\rightarrow} {\cal R} e^{-2i \tilde{\omega} r^{\ast}_0} Z_{\text{BH}} (\omega),
\end{equation}
and the second echo may have the amplitude of
\begin{equation}
Z_{\text{echo}}^{(2)} \simeq {\cal T}_{\text{BH}}^{\rightarrow} {\cal R}^2 {\cal R}_{\text{BH}}^{\rightarrow} e^{-2\times 2i \tilde{\omega} r^{\ast}_0} Z_{\text{BH}} (\omega).
\end{equation}
As such, one can obtain the amplitude of $n$-th echo as
\begin{equation}
Z_{\text{echo}}^{(n)} = {\cal T}_{\text{BH}}^{\rightarrow} {\cal R}^n ({\cal R}_{\text{BH}}^{\rightarrow})^{n-1} e^{-2 n i \tilde{\omega} r^{\ast}_0} Z_{\text{BH}} (\omega),
\end{equation}
where ${\cal R}_{\rm BH}^{\rightarrow} \equiv B_{\rm in}/B_{\rm out}$ and ${\cal T}_{\rm BH}^{\rightarrow} \equiv \sqrt{\omega/ |\tilde{\omega}|} B_{\rm out}^{-1}$. Since only the reflectivity and transmissivity of outgoing waves are involved in the echoes, we will not use ${\cal R}_{\rm BH}^{\leftarrow}$ and ${\cal T}_{\rm BH}^{\leftarrow}$ in the following, and so omit the symbol $\rightarrow$. Summing up all contributions from $n=1$ to $n=\infty$, one obtains 
\begin{equation}\displaystyle
\sum_{n=1}^{\infty} Z_{\text{BH}}^{(n)} = \frac{{\cal T}_{\text{BH}} {\cal R} e^{-2i \tilde{\omega} r^{\ast}_0}}{1-{\cal R} {\cal R}_{\text{BH}} e^{-2i\tilde{\omega} r^{\ast}_0}} Z_{\text{BH}} \equiv {\cal K} (\omega) Z_{\text{BH}}.
\end{equation}
Note that here we assume that $|{\cal R} {\cal R}_{\text{BH}}| < 1$, as otherwise the infinite sum of the geometric series does not converge.
Finally, the spectrum taking into account the reflection at the would-be horizon is obtained
\begin{equation}
X (r^{\ast}, \omega) = (Z_{\infty} + {\cal K} Z_{\text{BH}}) e^{i \omega r^{\ast}} = (1+ {\cal K} Z_{\text{BH}} / Z_{\infty}) Z_{\infty} e^{i \omega r^{\ast}}.
\end{equation}
When ${\cal R} = 0$ we have ${\cal K} = 0$ and so it reduces to (\ref{Z_inf}). Once we specify a specific form of the source term $T$, one can obtain $Z_{\text{BH}}/ Z_{\infty}$. For example, let us assume the source term located at $r^{\ast} = r^{\ast}_s$
\begin{equation}
T(r^{\ast}) = S (\omega) \delta (r^{\ast}-r^{\ast}_s),
\end{equation}
where $S (\omega)$ is a non-singular function in terms of frequency. Substituting this source term in (\ref{Z_inf}) and (\ref{Z_BH}), one obtains
\begin{equation}
\frac{Z_{\text{BH}}}{Z_{\infty}} = \frac{{\cal R}_{\text{BH}} + e^{-2i\tilde{\omega} r^{\ast}_s}}{{\cal T}_{\text{BH}}}.
\end{equation}
Note that this is independent of the function $S(\omega)$. Therefore, we finally obtain the following transfer function
\begin{align}
X &= e^{i \omega r^{\ast}} Z_{\infty} \left( 1+ {\mathcal K} \frac{Z_{\text{BH}}}{Z_{\infty}} \right) =e^{i \omega r^{\ast}}
Z_{\infty} \left( 1+ {\mathcal K}^{+}_{\text{echo}} + {\mathcal K}^{-}_{\text{echo}} \right),
\label{Xtilde}\\
{\mathcal K}^{+}_{\text{echo}} &\equiv \frac{{\mathcal R}_{\text{BH}} {\mathcal R} e^{-2i \tilde{\omega} r_0^{\ast}}}{1-{\mathcal R} {\mathcal R}_{\text{BH}} e^{-2i \tilde{\omega} r_0^{\ast}}},
\label{K+}\\
{\mathcal K}^{-}_{\text{echo}} &\equiv \frac{e^{-2i\tilde{\omega} r^{\ast}_s} {\mathcal R} e^{-2i \tilde{\omega} r_0^{\ast}}}{1-{\mathcal R} {\mathcal R}_{\text{BH}} e^{-2i \tilde{\omega} r_0^{\ast}}}.
\label{K-}
\end{align}
As discussed in \cite{Oshita:2019seis1}, actually ${\cal K}_{\text{echo}}^+$ and ${\cal K}_{\text{echo}}^-$ represent two different trajectories of GWs in the cavity. Here we are interested in outgoing incident waves that is related to ${\cal K}_{\text{echo}}^+$ and so in the following we discard ${\cal K}_{\text{echo}}^-$ from the transfer function, which does not change the qualitative feature of resulting echo signals.

Once we determine the spectrum of injected GWs, $Z_{\infty}$, one can obtain a template of GW echoes. Using the spectrum of GW ringdown may be a good approximation to obtain a realistic template. In this case, $Z_{\infty}$ is given by \cite{Berti:2005ys}
\begin{align}
Z_{\infty} &= \frac{2 G M}{D_o} \tilde{A}_{lm0} \left[ e^{i \phi_{lm0}} S_{lm0} (\theta) \alpha_+ + e^{-i \phi_{lm0}} S^{\ast}_{lm0} (\theta) \alpha_- \right],
\label{Z_inf_QNM}\\
\alpha_{\pm} &\equiv \frac{-\text{Im}[\omega_{lm0}]}{\text{Im}[\omega_{lm0}]^2 + (\omega \pm \omega_{lm0})^2},
\end{align}
where $D_o$ is the distance between the GW source and observer, $\omega_{lm0}$ is the most long-lived QNM, $\tilde{A}_{lm0}$ is the initial ringdown amplitude, $\phi_{lm0}$ is the phase of ringdown GWs, and $\theta$ is the observation angle. The amplitude $\tilde{A}_{lm0}$ is proportional to $\sqrt{\epsilon_{\text{rd}}}$, where $\epsilon_{\text{rd}} \equiv E_{\text{GW}}/M$ and $E_{\text{GW}}$ is the total energy of GW ringdown \cite{Berti:2005ys}. To give a few examples, the ringdown $+$ echo spectra are shown in FIG. \ref{spe_echo_ring}.

\subsection{Ergoregion Instability and the QNMs of quantum BH}
\label{subsec_ergoinst}
\begin{figure}[t]
\begin{center}
\includegraphics[keepaspectratio=true,height=55mm]{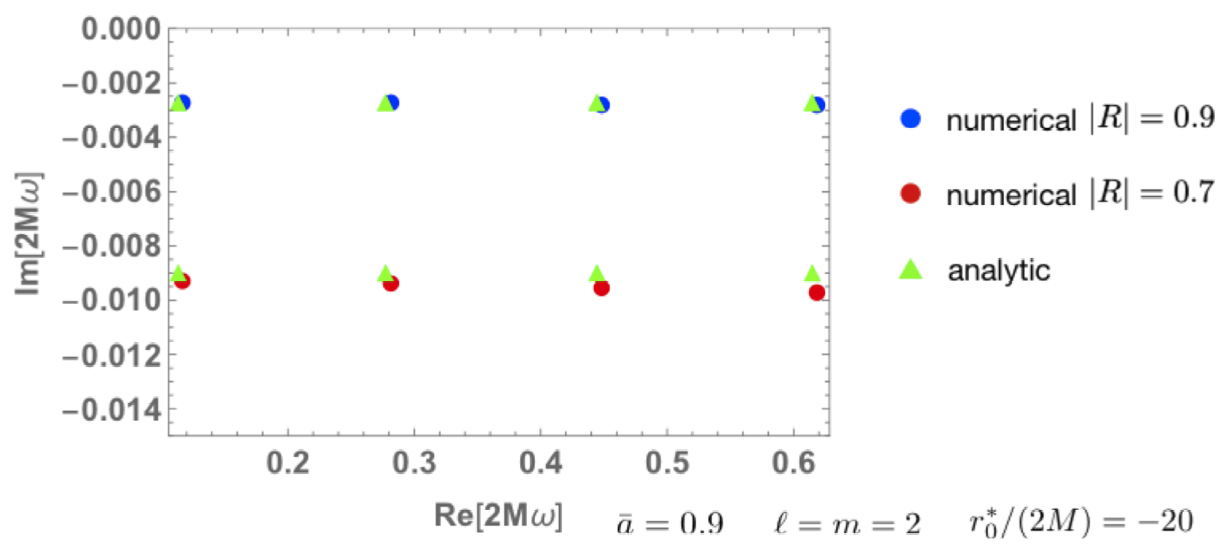}
\end{center}
\caption{QNMs with $|R| = 0.7$ (red filled circles) and $0.9$ (blue filled circles), $r^{\ast}_0 = -40 M$, $\bar{a}=0.9$, and $\ell = m= 2$.}
  \label{qnm1}
\end{figure}

As pointed out in the previous subsection, one should check if $|{\cal R} {\cal R}_{\text{BH}}| < 1$ is satisfied when calculating the transfer function in the geometric optics picture. This is physically important to understand the ergoregion instability caused by the reflection at the would-be horizon. Since the common ratio of the geometric series is ${\cal R} {\cal R}_{\text{BH}}$, the echo amplitude may be amplified and diverges when $|{\cal R} {\cal R}_{\text{BH}}| > 1$. This is nothing but the ergoregion instability that prevents BHs from having high spins. One can also derive the criterion from the QNMs of ECOs. The echo QNMs $\omega_n$ can be obtain by looking for the poles of the Green's function of echo GWs. That is, one can look for the poles from the zero points of the denominator of the transfer function
\begin{align}
1- {\cal R} {\cal R}_{\text{BH}} e^{-2i (\omega_n - m \Omega_H) r^{\ast}_0}=0,
\end{align}
and we obtain
\begin{equation}
\omega_n = \frac{2 \pi n +(\delta+ \delta')}{\Delta t_{\text{echo}}} + m \Omega_H + i \frac{\ln{|{\cal R} {\cal R}_{\text{BH}}|}}{\Delta t_{\text{echo}}},
\end{equation}
where $\Delta t_{\text{echo}} \equiv 2 |r^{\ast}_0|$, $\delta \equiv \text{arg} [{\cal R}]$ and $\delta' \equiv \text{arg} [{\cal R}_{\text{BH}}]$.
Then we obtain the real and imaginary parts of the echo QNMs
\begin{align}
\text{Re} [\omega_n] &\simeq \frac{2 \pi n +(\delta+ \delta')}{\Delta t_{\text{echo}}} + m \Omega_H,\\
\text{Im} [\omega_n] &\simeq \left. \frac{\ln{|{\cal R} {\cal R}_{\text{BH}}|}}{\Delta t_{\text{echo}}} \right|_{\omega = \text{Re}[\omega_n]}.
\end{align}
The positivity of the imaginary part of QNMs, which leads to the instability, is equivalent to having $|{\cal R} {\cal R}_{\text{BH}}| > 1$. Furthermore, we can see that the real parts of the QNM frequencies depend on the phases of ${\cal R}$ and ${\cal R}_{\rm BH}$, while their imaginary part depends on their absolute values. We can also rewrite the imaginary part in terms of the amplification factor
\begin{equation}
\text{Im} [\omega_n] \simeq \frac{\ln|{\cal R}|}{\Delta t_{\text{echo}}} + \frac{\ln{(1+Z)}}{2 \Delta t_{\text{echo}}} \simeq \frac{\ln|{\cal R}|}{\Delta t_{\text{echo}}} + \frac{Z}{2 \Delta t_{\text{echo}}}.
\end{equation}
Then we obtain the analytic form of the imaginary part of QNMs in the low-frequency regime $(M \omega \ll 1)$
\begin{equation}
\text{Im} [\omega_n] \simeq \frac{\ln|{\cal R}|}{\Delta t_{\text{echo}}} + \frac{2Q}{\Delta t_{\text{echo}}} \beta_{sl} \prod_{k=1}^{l} \left( 1+ \frac{4 Q^2}{k^2} \right) \left[\left(  \frac{2 \pi n +(\delta+ \delta')}{\Delta t_{\text{echo}}} + \Omega_H \right) (r_+ -r_-) \right]^{2l +1},
\end{equation}
where we used (\ref{sta-san}). This is the generalization of the analytic form of QNMs \cite{Oshita:2019seis1}. This analytic form is well consistent with numerically obtained QNMs in the low frequency region as is shown in FIG. \ref{qnm1}.

\subsection{Echoes from Planckian correction to dispersion relation and Boltzmann Reflectivity}
\label{subsec:boltzmann}
The Boltzmann reflection of a BH horizon has been discussed in the context of (stimulated) Hawking radiation from the path integral approach \cite{PhysRevD.13.2188}, quantum tunneling approach \cite{Srinivasan:1998ty,Vanzo:2011wq}, and Feynman propagator approach \cite{Padmanabhan:2019yyg}. Recently,  two of us studied the reflectivity of a BH for incident GWs from a Lorentz violating dispersion relation and argued that it can be approximated by a  Boltzmann-like reflectivity \cite{Oshita:2018fqu}.  More recently, three of us used general arguments from thermodynamic detailed balance, fluctuation-dissipation theorem, and CP-symmetry to show that the reflectivity of quantum BH horizons should be universally given by a Boltzmann factor \cite{Oshita:2019sat,Wang:2019rcf}: 
\begin{equation}\left. 
\frac{{\cal F}_{\rm out}}{{\cal F}_{\rm in}}\right|_{\rm horizon} = \exp\left(-\frac{\hbar \tilde{\omega}}{k_B T_{\rm H}}\right)   \label{Bolt_flux}
\end{equation}
The reflection of quantum BH might be understood as Hawking radiation stimulated by enormous number of incoming gravitons, and if that is so, having the dependence of the reflectivity on the Hawking temperature $T_{\rm H}$ is natural. Furthermore, one can also avoid the ergoregion instability in this model \cite{Oshita:2019sat,Wang:2019rcf}. In this subsection, we briefly review the Boltzmann reflectivity model from both theoretical and phenomenological aspects.

\subsubsection{Boltzmann reflectivity from dissipation}
The dissipative effects at the apparent horizon have been discussed from the point of view of the membrane paradigm \cite{Thorne:1986iy,Jacobson:2011dz}, the fluctuating geometry around a BH \cite{Parentani:2000ts,Barrabes:2000fr}, and the minimal length uncertainty principle \cite{Brout:1998ei}. Our approach to derive the Boltzmann reflectivity starts with a heuristic assumption to model the dissipative effects, which are expected in any thermodynamic system from fluctuation-dissipation theorem. Let us assume that the wave equation governing the perturbation of BH is given by \cite{Oshita:2019sat}:
\begin{equation}
\left[ -i \frac{\tilde{\gamma} \Omega (r^{\ast})}{E_{\text{Pl}}} \frac{d^2}{dr^{\ast} {}^2} + \frac{d^2}{dr^{\ast} {}^2} + \tilde{\omega}^2 - V(r^{\ast}) \right]
\psi_{\tilde{\omega}} (r^{\ast}) = 0,
\label{we1}
\end{equation}
where $\tilde{\gamma}$ is a dimensionless dissipation parameter, $\Omega (x) \equiv |\tilde{\omega}|/ \sqrt{|g_{00} (x)|}$ is the blueshifted (or proper) frequency, and $V$ is the angular momentum barrier. The form of the dissipation term is expected from the fluctuation-dissipation theorem near the horizon, where the Hawking radiation (quantum fluctuation/dissipation) and the incoming GWs (stimulation) are blue shifted. This dissipative modification to the dispersion relation becomes dominant only when the blueshift effect is so intense that the proper frequency is comparable to the Planck energy, $\Omega \sim E_{\text{Pl}}$. Furthermore, from a phenomenological point of view, the dissipative term in (\ref{we1}) is similar to the viscous correction to sound wave propagation in terms of  shear viscosity, $\nu$, in Navier-Stokes equation, $-i (4/3) \nu \Omega \nabla^2$ (e.g., \cite{Liberati:2013usa}).

\begin{figure}[t]
\begin{center}
\includegraphics[keepaspectratio=true,height=75mm]{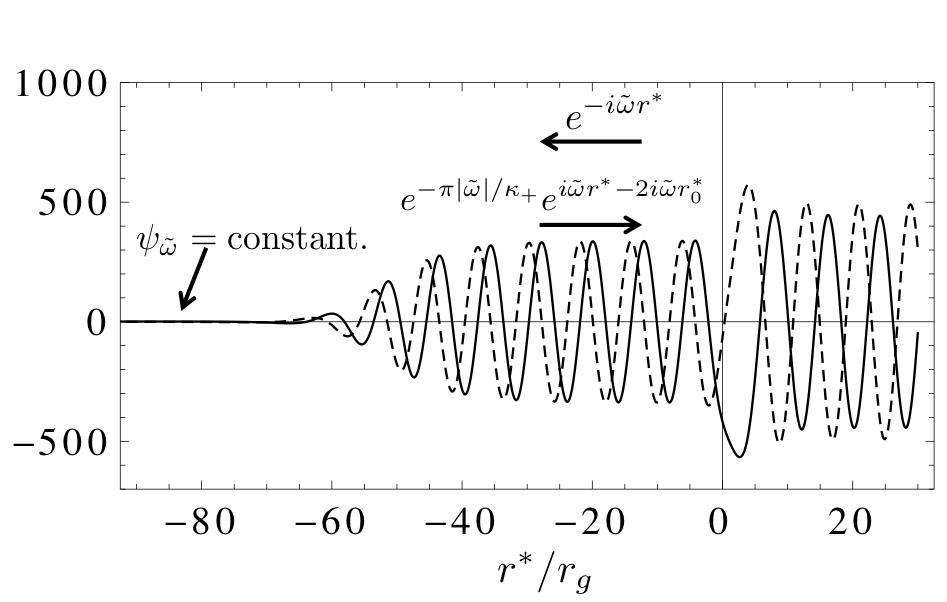}
\end{center}
\caption{Mode function obtained by solving (\ref{we1}) with the boundary condition of $\psi_{\tilde{\omega}} = \text{constant.}$ for $r^{\ast} \to - \infty$.}
  \label{mode_func}
\end{figure}

Let us solve the modified wave equation by imposing a physically reasonable boundary condition (see FIG. \ref{mode_func}):
\begin{align}
\psi_{\tilde{\omega}} \sim \text{constant.} \ \ \text{for} \ \ r^{\ast} \to - \infty,
\end{align}
The constant boundary condition in the limit of $r^{\ast} \to - \infty$ means that the energy flux carried by the ingoing GWs cannot go through the horizon, and is either absorbed or reflected. That is consistent with the BH complementarity \cite{Susskind:1993if} or the membrane paradigm \cite{Thorne:1986iy,PhysRevD.18.3598}.
Although there is no unique choice of wave equation around a Kerr BH, we here choose the CD equation that has a purely real angular momentum barrier. The modified CD equation is assumed to have the form of
\begin{equation}
\left( \frac{- i \tilde{\gamma} |\tilde{\omega}|}{\sqrt{\delta (r)} E_{\text{Pl}}} \frac{d^2}{dr^{\ast} {}^2} + \frac{d^2}{dr^{\ast} {}^2} - {\cal V} \right) \psi_{\tilde{\omega}} =0,\label{SNeq}
\end{equation}
where $\sqrt{\delta(r)} \equiv \sqrt{1-r_g/r +(a/r)^2}$ is the blue shift factor in terms of the co-rotating frame \cite{Frolov:2014dta,Poisson:2009pwt}. In the near horizon limit ($r^{\ast} \to - \infty$, see below for details), the CD equation reduces to the following form in the limit of $r^{\ast} \to -\infty$:
\begin{equation}
\left(-i \frac{\tilde{\gamma} |\tilde{\omega}|}{Q E_{\text{Pl}}} e^{-\kappa_+ r^{\ast}} \frac{d^2}{dr^{\ast} {}^2} + \frac{d^2}{dr^{\ast} {}^2} - \tilde{\omega}^2 \right) \psi_{\tilde{\omega}} = 0,
\label{SN_near_horizon}
\end{equation}
where $\kappa_+$ is the surface acceleration at the outer horizon, $Q$ is defined as
\begin{equation}
Q \equiv  \exp \left[ \frac{1}{2} \frac{\sqrt{1-\bar{a}^2}}{r_+^2/r_g^2 +\bar{a}^2/4} \left( - \frac{r_+}{r_g} + \frac{r_-^2/r_g^2 +\bar{a}^2/4}{2\sqrt{1-\bar{a}^2}} \log{(1-\bar{a}^2)} \right) + \frac{1}{2} \log \sqrt{1-\bar{a}^2} -\log \left(\frac{r_+}{r_g}\right) \right],
\end{equation}
and $r_{\pm} \equiv M(1 \pm \sqrt{1 - \bar{a}^2})$. The solution of (\ref{SN_near_horizon}) which satisfies the aforementioned boundary condition is
\begin{equation}\displaystyle
\lim_{r^{\ast} \to -\infty} \psi_{\tilde{\omega}} = {}_2 F_1 \left[ -i \frac{\tilde{\omega}}{\kappa_+}, i \frac{\tilde{\omega}}{\kappa_+}, 1, -i \frac{Q E_{\text{Pl}} e^{\kappa_+ r^{\ast}}}{\tilde{\gamma} |\tilde{\omega}|} \right],
\end{equation}
and one can read that in the intermediate region, $- \kappa_+^{-1} \log \left[Q E_{\text{Pl}} / (\tilde{\gamma} |\tilde{\omega}|) \right] \ll r^{\ast} \ll \pm \kappa_{+}^{-1}$, $\psi_{\tilde{\omega}}$ can be expressed as the superposition of outgoing and ingoing modes
\begin{align}
\psi_{\tilde{\omega}}=
\begin{cases}
e^{\pi \tilde{\omega}/(2 \kappa_+)}A_+ e^{-i \tilde{\omega} r^{\ast}} + e^{-\pi \tilde{\omega}/(2 \kappa_+)} A_+^{\ast} e^{i \tilde{\omega} r^{\ast}} \ &\text{for} \ \tilde{\omega} > 0,\\
e^{-\pi \tilde{\omega}/(2 \kappa_+)}A_- e^{-i \tilde{\omega} r^{\ast}} + e^{\pi \tilde{\omega}/(2 \kappa_+)} A_-^{\ast} e^{i \tilde{\omega} r^{\ast}} \ &\text{for} \ \tilde{\omega} < 0,
\end{cases}
\end{align}
where $A_{\pm}$ has the form of
\begin{equation}
A_{\pm} \equiv \left(\frac{\tilde{\gamma} |\tilde{\omega}|}{Q E_{\text{Pl}}} \right)^{i \tilde{\omega}/ \kappa_+} \times \frac{\Gamma (-2i \tilde{\omega}/ \kappa_+)}{\Gamma (-i \tilde{\omega}/\kappa_+) \Gamma (1-i \tilde{\omega}/\kappa_+)} = e^{i \tilde{\omega} r^{\ast}_0} \times \frac{\Gamma (-2i \tilde{\omega}/ \kappa_+)}{\Gamma (-i \tilde{\omega}/\kappa_+) \Gamma (1-i \tilde{\omega}/\kappa_+)},
\end{equation}
and
\begin{align}
r^{\ast}_0 \equiv \frac{1}{\kappa_+} \ln{\left( \tilde{\gamma} | \tilde{\omega}|/(Q E_{\text{Pl}}) \right)}.
\label{xf}
\end{align}
Therefore, the energy reflectivity is given by
\begin{align}
|{\cal R}|^2 =
\begin{cases}
e^{-2\pi \tilde{\omega}/\kappa_+} \ &\text{for} \ \tilde{\omega} > 0,\\
e^{2\pi \tilde{\omega}/\kappa_+} \ &\text{for} \ \tilde{\omega} < 0,
\end{cases}
\end{align}
and finally we obtain
\begin{equation}
{\cal R} = \exp{\left[- \frac{|\tilde{\omega}|}{2 T_H} +i\delta_{\text{wall}} \right]},\label{Bolt_amp}
\end{equation}
where $\delta_{\text{wall}}$ is the phase shift at the would-be horizon and it is determined by $A_+$ or $A_-$. Equation (\ref{Bolt_amp}) then reproduces the Boltzmann energy flux reflectivity in (\ref{Bolt_flux}). As we noted earlier, the same result can be independently derived using thermodynamic detailed balance or CP symmetry near BH horizons. 

When we further modify the dispersion relation by adding a quartic correction term
\begin{equation}
\tilde{\Omega}^2 = \tilde{K}^2 +i \tilde{\gamma} \tilde{\Omega} \tilde{K}^2 -C_d^2 \tilde{K}^4,
\label{dispersion}
\end{equation}
where $C_d$ is a constant parameter and $\tilde{\Omega}$ and $\tilde{K}$ are the proper frequency and proper wavenumber, respectively, the exponent of the Boltzmann factor is modified, and the analytic form can be obtain for $C_d \gg \tilde{\gamma}$ by using the WKB approximation \cite{Oshita:2018fqu}
\begin{equation}\label{Ref_dis}
|{\cal R}| \simeq \exp{\left[ - \frac{\sqrt{2 + 4 C_d^2 /\tilde{\gamma}^2}}{\pi (1+4 C_d^2/\tilde{\gamma}^2)} \left( \frac{|\tilde{\omega}|}{2 T_H} \right) \right]}.
\end{equation}
One of the essential differences between the modified dispersion relation model, that could give the Boltzmann reflectivity at the would-be horizon, and the Exotic Compact Object (ECO) model is the reflection radius $r^{\ast}_0$. In the former case, $r^{\ast}_0$ depends on the frequency of incoming GWs, $\tilde{\omega}$, and so the reflection surface is not uniquely determined. This is because the reflection takes place when the frequency $|\tilde{\omega}|$ reaches the Planckian frequency at which the modification in the dispersion relation becomes dominant. Therefore, the reflection radius depends on the initial (asymptotic) frequency of incoming GWs (see Equation \ref{xf}). On the other hand, in the ECO scenario, the reflection radius would be fixed and it would stand at $\sim$ a Planck proper length outside the horizon. In this case, the reflection radius is given by
\begin{equation}
r^{\ast}_0 \simeq \frac{1}{\kappa_+} \ln{\left( \frac{M}{E_{\text{Pl}}} \right)},
\end{equation}
which depends only on the mass of BH. For a detailed discussion of how one can observationally distinguish these scenarios, we refer the reader to \cite{Oshita:2019seis1}. 

\subsubsection{Phenomenology of Boltzmann reflectivity}
Here we summarize some interesting phenomenological aspects of the Boltzmann reflectivity model, and refer the reader to \cite{Wang:2019rcf} for more details. 

\begin{figure}[t]
\begin{center}
\includegraphics[keepaspectratio=true,height=55mm]{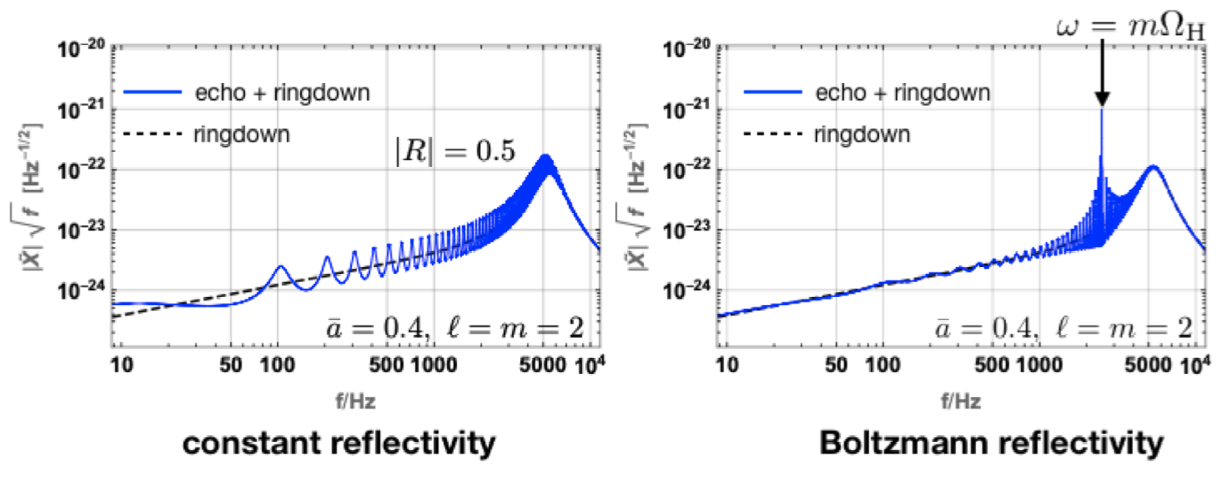}
\end{center}
\caption{Spectra of echo $+$ ringdown with $\bar{a} = 0.4$, $\ell=m=2$, and $M= 2.7 M_{\odot}$, $\epsilon_{\text{rd}} = 0.01$, $D_o = 40$ Mpc, and $\theta = 90^\circ$. The left panel shows the spectrum in the constant reflectivity model with $|{\cal R}|=0.5$ and the right panel shows the spectrum in the Boltzmann reflectivity model with $\tilde{\gamma} =1$.}
  \label{spe_echo_ring}
\end{figure}

\begin{figure}[t]
\begin{center}
\includegraphics[keepaspectratio=true,height=65mm]{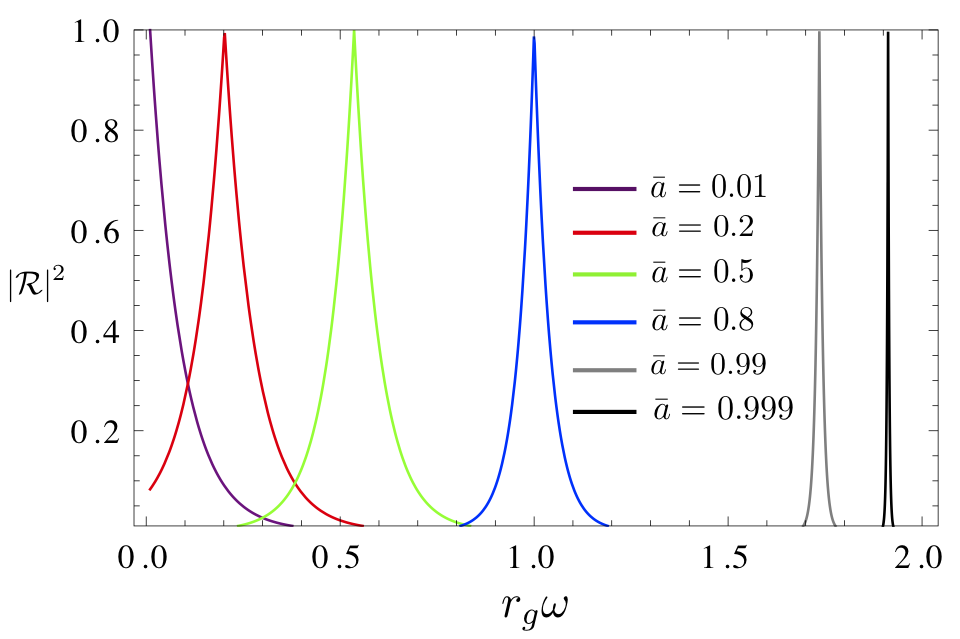}
\end{center}
\caption{The energy reflection rate $|{\cal R}|^2$ in the Boltzmann reflectivity model.}
  \label{ref_bol}
\end{figure}

Nearly all the previous studies of GW echoes assume a constant reflectivity model, in which the ratio of outgoing to ingoing flux at the horizon is assumed to be independent of frequency. In contrast, the echo spectrum in the Boltzmann reflectivity model can be significantly different. This difference can be seen in the sample echo spectra for both models, shown in FIG. \ref{spe_echo_ring}. As can be seen from the spectra, the echo amplitude is highly excited near $m \times$ the horizon frequency,  since the Boltzmann reflectivity is sharply peaked around $\omega \simeq m\Omega_{\rm H} \pm T_{\rm H}$ and is exponentially suppressed outside this range. In the extremal limit $\bar{a}\to 1$, the Hawking temperature becomes zero and so the frequency range in which $|{\cal R}| \sim 1$ vanishes. Therefore, the peaks in echo spectrum is highly suppressed for a highly spinning BH (see FIG. \ref{ref_bol}) \footnote{Note that in order to set the initial conditions of the QNMs of the quantum BH cavity, we choose a superposition that reproduces the time-evolution of the dominant QNM of the classical BH for $t \lesssim \Delta t_{\rm echo}$. }.

\begin{figure}[t]
\begin{center}
\includegraphics[keepaspectratio=true,height=85mm]{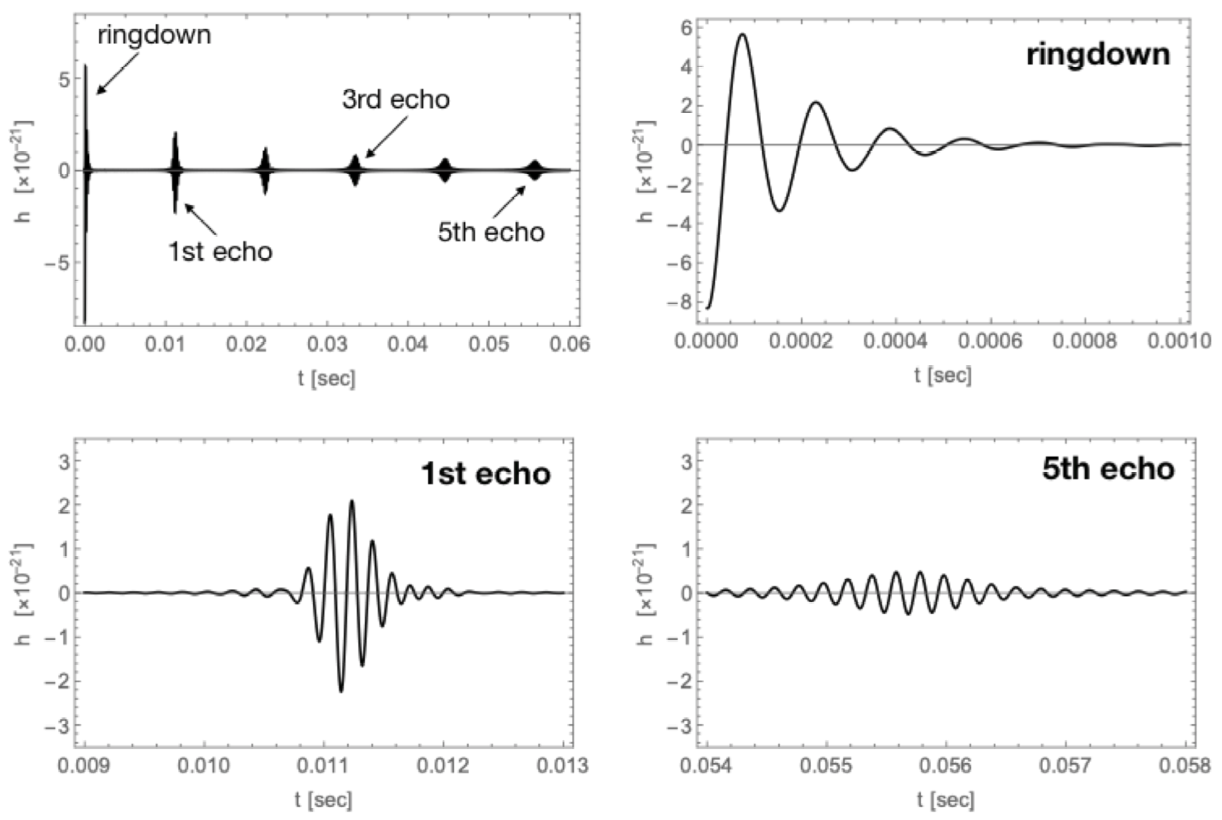}
\end{center}
\caption{The time domain function with $M=2.7 M_{\odot}$, $\bar{a} = 0.7$, $D_o=40$ Mpc, $\epsilon_{\text{rd}} =0.01$, $\theta=90^\circ$, and $\ell =m = 2$ in the Boltzmann reflectivity model with $\tilde{\gamma}=1$ and $T_{\rm H} / T_{\rm QH} = 0.6$.}
  \label{time_domain}
\end{figure}

This nature of the Boltzmann reflectivity suppresses the ergoregion instability at least up to the Thorne limit $\bar{a} \leq 0.998$. In \cite{Oshita:2019seis1}, a more general case is investigated, where the Hawking temperature in the Boltzmann factor is replaced by the quantum horizon temperature, $T_{\rm H} \to T_{\rm QH}$ (e.g., as in Equation \ref{Ref_dis} above)
\begin{equation}
{\cal R} = \exp{\left(- \frac{|\tilde{\omega}|}{2T_{\rm QH}} \right)},
\end{equation}
and the ratio $T_{\rm H} / T_{\rm QH}$ is constrained from the ergoregion instability by using $|{\cal R} {\cal R}_{\text{BH}}| < 1$. The constraint is $T_{\rm H} / T_{\rm QH} \gtrsim 0.5$ up to the Thorne limit \cite{Oshita:2019seis1} and so the Boltzmann reflectivity $(T_{\rm H} / T_{\rm QH} =1)$ is safe up to $\bar{a} \lesssim 0.998$. As an example, we show a time domain function of ringdown and echo phases with $T_{\rm H} / T_{\rm QH} = 0.6$ in FIG. \ref{time_domain} by implementing the inverse Fourier transform of $X = Z_{\infty} (1+ {\cal K}_{\text{echo}}^{+})$, where we choose $Z_{\infty}$ so that it reproduces the ringdown phase \cite{Oshita:2019seis1}.

Other notable phenomenological properties of quantum BHs with Boltzmann echoes are \cite{Wang:2019rcf}:
\begin{itemize}
    \item The QNMs of the quantum BH are approximately those of a cavity with a {complex} length $|r_0^*|+i(4 T_{\rm QH})^{-1}$.
    \item For $\tilde{\gamma} \sim 1$ (i.e. Planck-scale modifications), the first $\sim 20$ echo amplitudes decay as inverse time $1/t$, and then exponentially.
    \item Each QNM of the classical BH can be written as a superposition QNMs of the quantum BH for $t < \Delta t_{\rm echo}$. The superposition can be approximated as a geometric series, leading to a closed-form expression for echo waveforms. In particular, the first 20 echoes have approximate temporal  Lorentzian envelopes around their peaks, whose width grows linearly width echo number. 
\end{itemize}

The parameter dependence of echo spectrum in the Boltzmann reflectivity model and its consistency with the tentative detection of echo in GW170817 are also investigated in \cite{Oshita:2019seis1} in more detail.

\section{Gravitational Wave Echoes: Observations}
\label{sec:echo_searches}
\textit{For the first time in modern science history we are able to probe the smallest possible theoretical scales or highest possible theoretical energies through GW echoes.}

The direct observation of GWs \cite{Abbott:2016blz} was a scientific breakthrough that has opened a vast new frontier in astronomy, providing us with  possible tests of General relativity in the extreme physical conditions near the BH horizons. Motivated by the resolutions of BH information paradox that propose alternatives to BH horizons (see Section \ref{sec:QBHs} above), several groups have searched the LIGO/Virgo public data for GW echoes \cite{Cardoso:2016rao,Cardoso:2016oxy} (see Section \ref{sec:echo_predictions} and \cite{Cardoso:2019rvt} for a review), which has led to claims (and counter-claims) of tentative evidence and/or detection  \cite{Abedi:2016hgu, Conklin:2017lwb, Westerweck:2017hus,Nielsen:2018lkf, Abedi:2018npz,Salemi:2019uea,Uchikata:2019frs, Holdom:2019bdv}. While the origins of these tentative signals remain controversial \cite{Westerweck:2017hus,Nielsen:2018lkf, Ashton:2016xff, Abedi:2017isz, Abedi:2018pst,Salemi:2019uea} they motivate further investigation using improved statistical and theoretical tools, and well as new observations.

\begin{figure}
\centering
\includegraphics[width=0.7\textwidth]{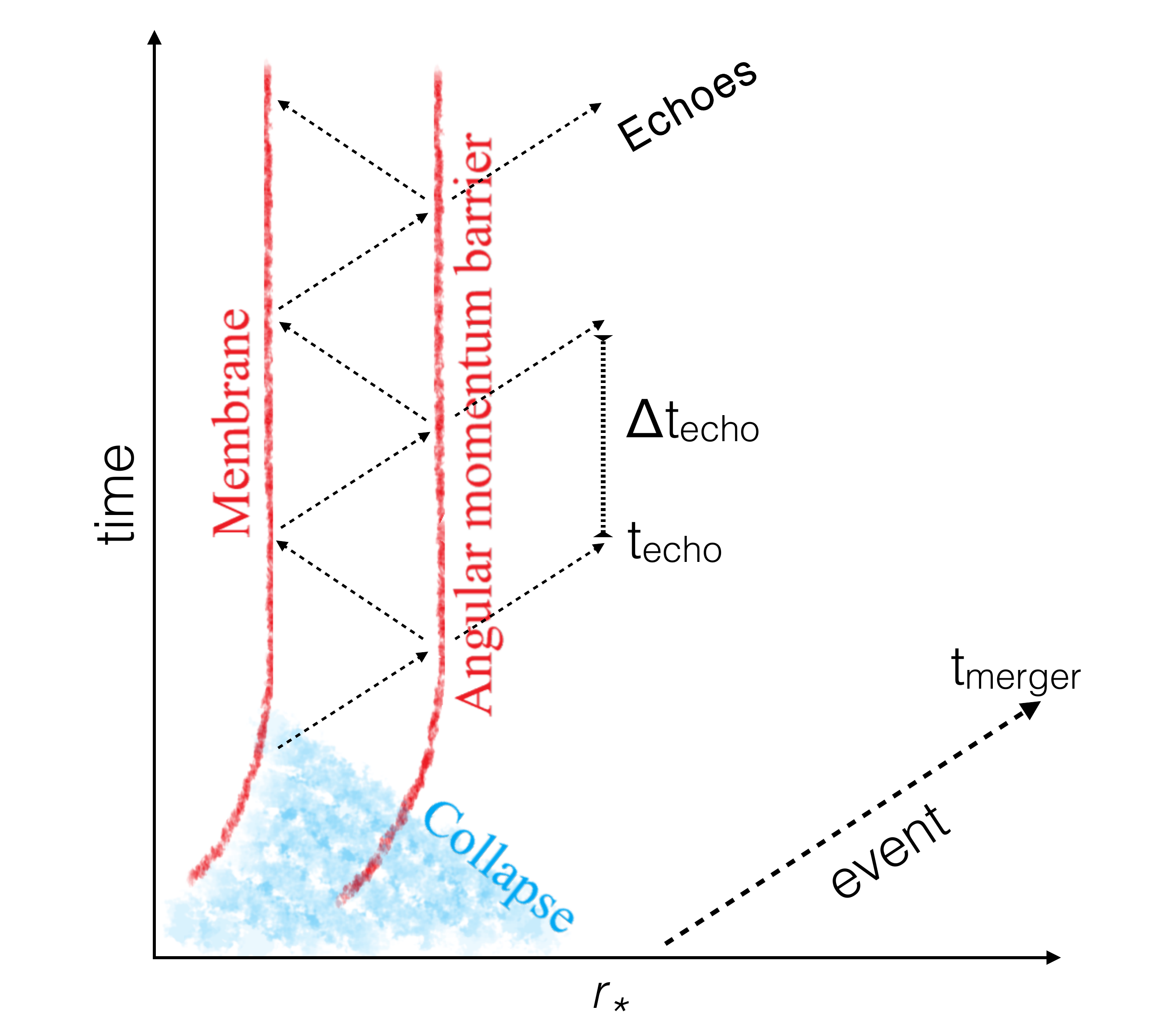}
\caption{GW echoes following a collapse of binary neutron star merger event from a cavity of membrane-angular momentum barrier \cite{Abedi:2018npz}. }
\label{echo_pic_2}
\end{figure}

In astrophysics, GW echoes from quantum BHs can be seen as a transient signal, coming from the post-coalescence phase of the binary BH merger (Fig. \ref{echo_pic_1}) or formation of a BH (e.g., via collapse of a hypermassive neutron star; Fig. \ref{echo_pic_2}). This section will summarize the current status of observational searches for echoes \cite{Abedi:2020sgg}.

In order to properly model echoes, we need a full knowledge of quantum BH nonlinear dynamics, which is so far nonexistent. Therefore, any strategy to search for echoes requires parametrizing one's ignorance, which has so far taken many shapes and form.  Indeed, we need to keep a balance between having a simple tractable model (which may simply miss the real signal), or an exhaustive complex model (which may dilute a weak signal with look-elsewhere effects).  
Current search methods can be generally split into two: Parametrized template-based  methods
\cite{Abedi:2016hgu,Uchikata:2019frs,Westerweck:2017hus,Nielsen:2018lkf,Lo:2018sep,Tsang:2018uie}, and ``model-agnostic'' coherent methods \cite{Conklin:2017lwb, Abedi:2018npz, Salemi:2019uea, Holdom:2019bdv}. 

Out of these, 8 studies find some observational evidence for echoes \cite{Abedi:2016hgu,Conklin:2017lwb,Westerweck:2017hus,Abedi:2018npz,Nielsen:2018lkf,Salemi:2019uea, Uchikata:2019frs, Holdom:2019bdv}, 3 are comment notes \cite{Ashton:2016xff,Abedi:2017isz,Abedi:2018pst}, and 3 more \cite{Uchikata:2019frs,Tsang:2019zra,Lo:2018sep} found no significant echo signals in the binary BH merger events. We can sort them into eight independent groups with 1. positive \cite{abedi2016echoes,Abedi:2018npz,Uchikata:2019frs,Conklin:2017lwb,Holdom:2019bdv}, 2. mixed \cite{Westerweck:2017hus,Nielsen:2018lkf,Salemi:2019uea}, and 3. negative \cite{Uchikata:2019frs,Lo:2018sep,Tsang:2019zra} results.

\subsection{\label{Positive}Positive Results}
\subsubsection{\label{Echoes from Abyss:O1}Echoes from the Abyss: Echoes from binary BH mergers O1 by Abedi, Dykaar, and Afshordi (ADA) \cite{Abedi:2016hgu}}

The first search for echoes from Planck-scale modifications of general relativity near BH event horizons using the public data release by the Advanced LIGO GW observatory was developed by Abedi, Dykaar, and Afshordi (ADA)  \cite{Abedi:2016hgu}. In this search, a naive phenomenological template for echoes was introduced, leading to tentative evidence at false detection probability of 1\% (or $\simeq 2.5\sigma$ significance level\footnote{In \cite{Abedi:2016hgu}, 2-tailed gaussian probability assigned to significance, e.g., p-value $= 68\%$ and $95\%$ correspond to $1\sigma$ and $2\sigma$ respectively.} shown in Figs. \ref{SNR}, \ref{SNR_fig} and \ref{Histogramloglog}) for the presence of echoes \cite{Abedi:2016hgu}. This work was followed by comments, discussion, and controversy about the origin of this signal \cite{Ashton:2016xff,Westerweck:2017hus, Abedi:2017isz, Abedi:2018pst,Nielsen:2018lkf,Salemi:2019uea}. The ADA model was also later tested for LIGO/Virgo O2 independent events\cite{Uchikata:2019frs}, which interestingly, yielded a similar percent-level p-value as O1 (see Section \ref{Uchikata} below). The ADA search was the first phenomenological time-domain echo template search applied to real GW observations \cite{Abedi:2016hgu}. Using a standard GR inspiral-merger-ringdown template $M(t)$, a naive model including five free parameters was proposed:
\begin{figure}[!tbp]
\begin{center}
    \includegraphics[width=0.7\textwidth]{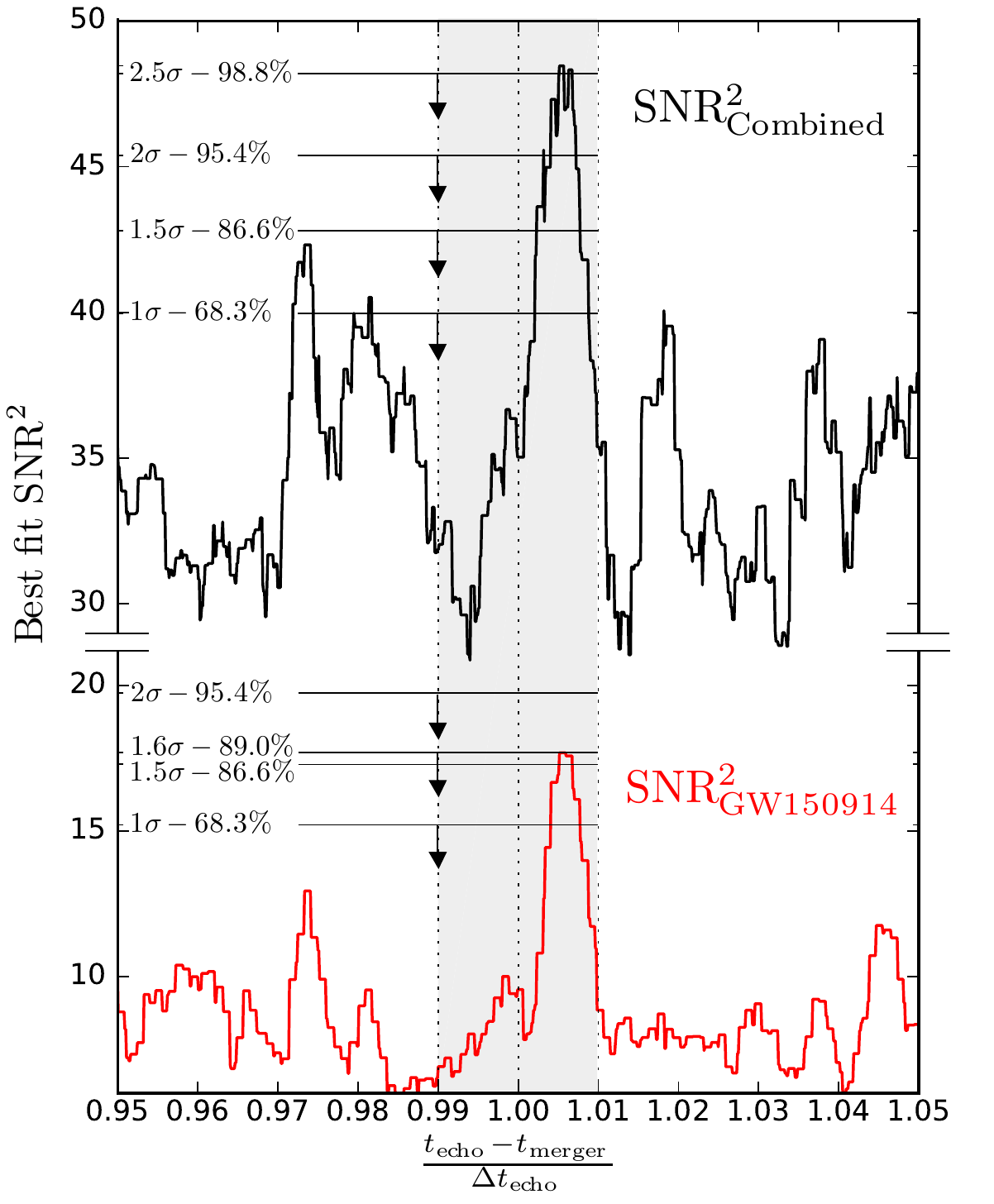}
 \caption{Maximized SNR$^2$ around the expected time of merger echoes Eq. (\ref{t_echo_meas}), for the combined (top) and GW150914 (bottom) events. The significance and p-values of the peaks within the gray rectangle are specified in this plot \cite{Abedi:2016hgu}.}\label{SNR}
\end{center}
\end{figure}

\begin{figure*}
  \centering
    \includegraphics[width=\textwidth]{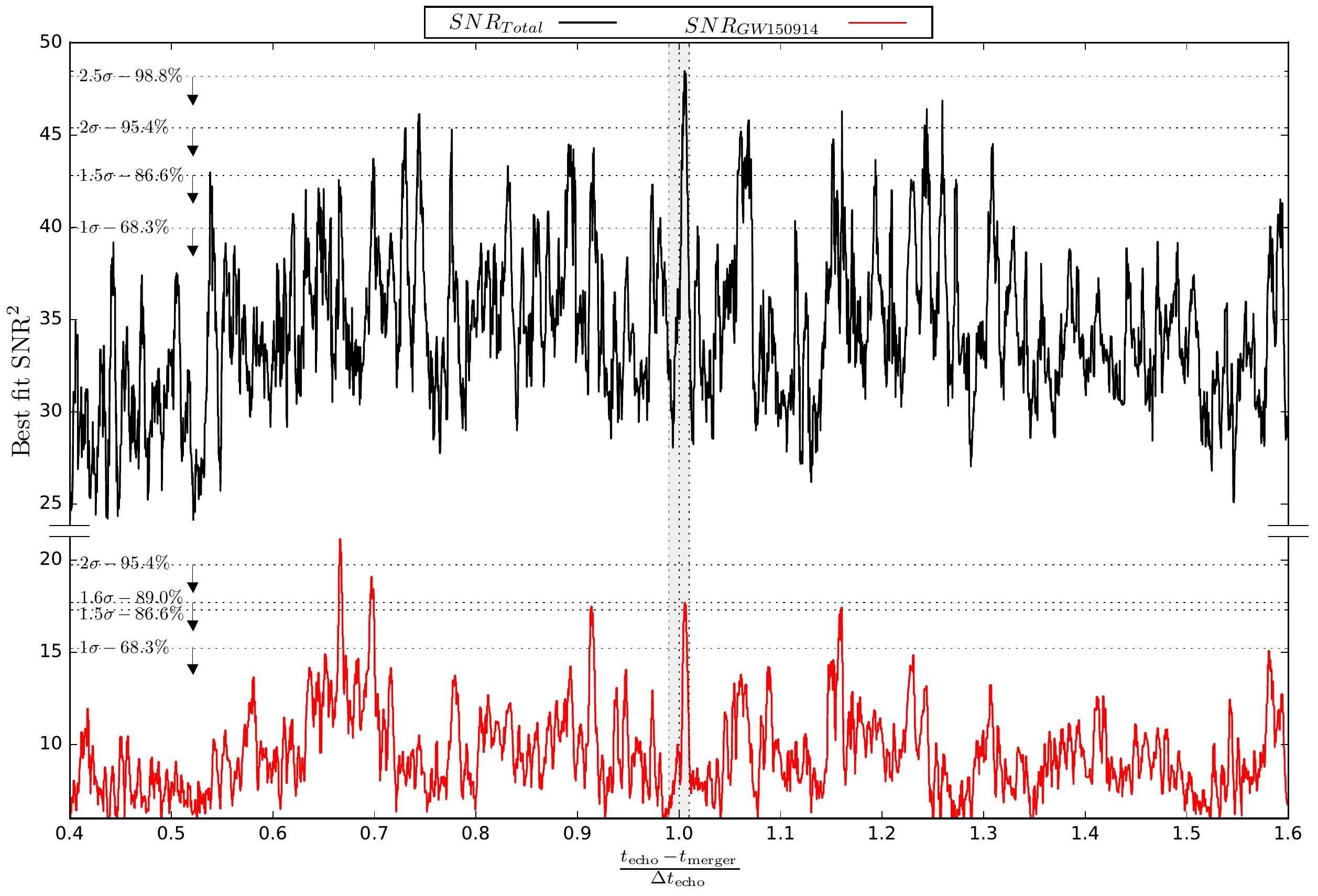}
    \caption{Same as Fig. \ref{SNR}, but over an extended range of $x=\frac{t_{\rm echo} - t_{\rm merger}}{\Delta t_{\rm echo}}$. The SNR peaks at the predicted value of $1-0.01<x<1+0.01$ within gray rectangle have false detection probability of 0.11 (0.011) and significance of 1.6$\sigma$ (2.5$\sigma$), for GW150914 (combined events) \cite{Abedi:2016hgu} (See also \cite{Abedi:2017isz}). } \label{SNR_fig}
\end{figure*}

\begin{figure}[!tbp]
\begin{center}
    \includegraphics[width=0.7\textwidth]{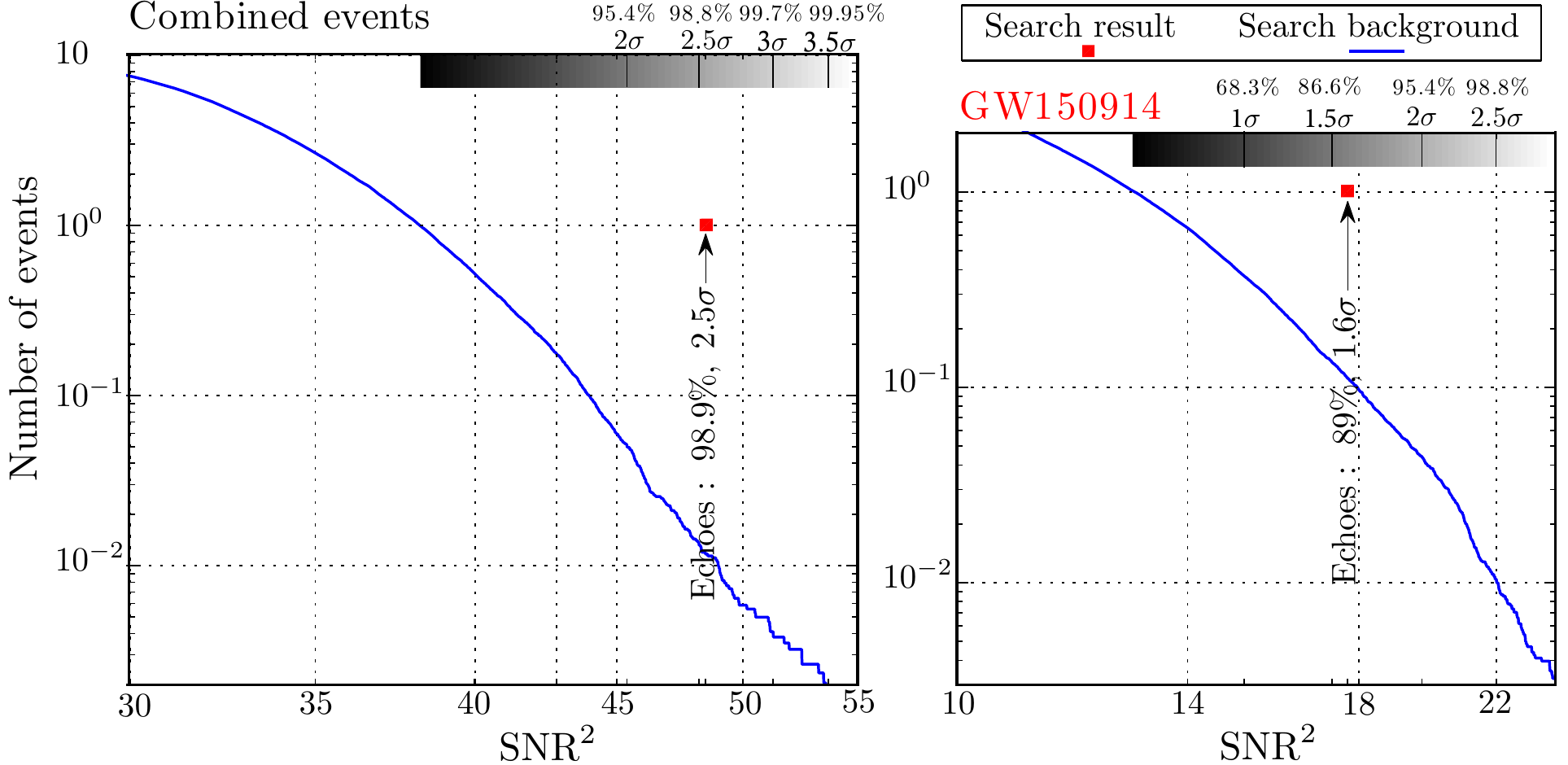}
    \caption{Average number of background peaks higher than a particular SNR-value within a time-interval $2\% \times   \overline{\Delta t}_{\rm echo}$ (gray rectangle in Figs. \ref{SNR} and \ref{SNR_fig}) for combined (left) and GW150914 (right) events \cite{Abedi:2016hgu}.  The red dots show the observed SNR peak at $t_{\rm echo} = 1.0054 \Delta t_{\rm echo}$ (Figs. \ref{SNR} and \ref{SNR_fig}). The correspondence between SNR values and their significance is indicated in horizontal bar.}
 \label{Histogramloglog}
\end{center}
\end{figure}

For a Kerr BH with final mass $M_{\rm BH}$ and dimensionless spin parameter $a$, the time delay of echoes from Planck-scale  modifications of general relativity is: \cite{Abedi:2016hgu,Abedi:2018npz}:
\begin{eqnarray}
\Delta t_{\rm echo} \simeq \frac{4 G M_{\rm  BH}}{c^3} \left(1+\frac{1}{\sqrt{1-\bar{a}^2}}\right) \times \ln\left(M_{\rm  BH} \over M_{\rm planck}\right) \nonumber \\
\simeq 0.126~ {\rm sec} \left(\frac{M_{\rm BH}}{67~ M_{\odot}}\right) \left(1+\frac{1}{\sqrt{1-\bar{a}^2}}\right). \label{eq.0.1}
\end{eqnarray}
For the final BH (redshifted) masses and spins reported by the LIGO collaboration for each merger event, echo time delays $\Delta t_{\rm echo}$ and their errors constrained inside $1\sigma$ error are as follows \cite{Abedi:2016hgu}:
\begin{eqnarray}
\Delta t_{{\rm echo}, I }({\rm sec})
=\left\{
 \begin{matrix}
  0.2925 \pm 0.00916 & I= {\rm GW150914} \\
  0.1778 \pm 0.02789 & I={\rm GW151012} \\
  0.1013 \pm 0.01152 & I={\rm GW151226} 
 \end{matrix}
 \right. \ \ \ \ \ \label{t_echo_meas}
 \end{eqnarray}

\subsubsection{Search} \label{sec_template}

In this analysis, ADA devised an echo waveform using theoretical best-fit waveform of Hanford $M_{H,I}(t)$ and Livingston $M_{L,I}(t)$ detectors (in real time series) for the BBH events, provided by the  LIGO and Virgo collaborations. The search used the observed data release for the two detectors, $h_{H,I}(t)$ and $h_{L,I}(t)$ respectively, at 4096 Hz and for 32 sec duration. The devised phenomenological echo waveform which was then constructed using 
five free parameters:\begin{enumerate}
\item $\Delta t_{\rm echo}$: Time-interval between successive echoes, within their $1\sigma$ range (Eq. \ref{t_echo_meas}).

\item $t_{\rm echo}$: Time of arrival of the first echo, which is related to $\Delta t_{\rm echo}$ with corrections $\sim\pm \mathcal{O}(1\%) \times \Delta t_{\rm echo}$ due to the non-linear dynamics of the merger.

\item $t_0$: Truncation time for GR template with a smooth cut-off function,
\begin{equation}
\Theta_I(t, t_{0})\equiv\frac{1}{2}\left\{1+ \tanh\left[\frac{1}{2} \omega_I(t)(t-t_{\rm merger}-t_{0})\right]  \right\},
\end{equation}
where $\omega_I(t)$ is frequency of GR template as a function of time \cite{TheLIGOScientific:2016src} and $t_{\rm{merger}}$ is the time of maximum amplitude of the template. It is assumed that $t_{0}$ vary within the range $t_{0} \in (-0.1,0) \overline{\Delta t}_{\rm echo}$. Having this definition, a truncated template is introduced:
\begin{eqnarray}
{\cal M}_{T,I}^{H/L} (t, t_{0}) \equiv\Theta_I(t, t_{0}) {\cal M}_{I}^{H/L} (t).
\end{eqnarray}

\item $\gamma$: Damping factor of successive echoes, varying between $0.1$ and $0.9$.

\item $A$: Over-all amplitude of the echo waveform with respect to the main event. This free parameter is fitted assuming a flat prior.

\end{enumerate}

\begin{figure}[!tbp]
  \centering
  \begin{minipage}[b]{0.7\textwidth}
    \includegraphics[width=1\textwidth]{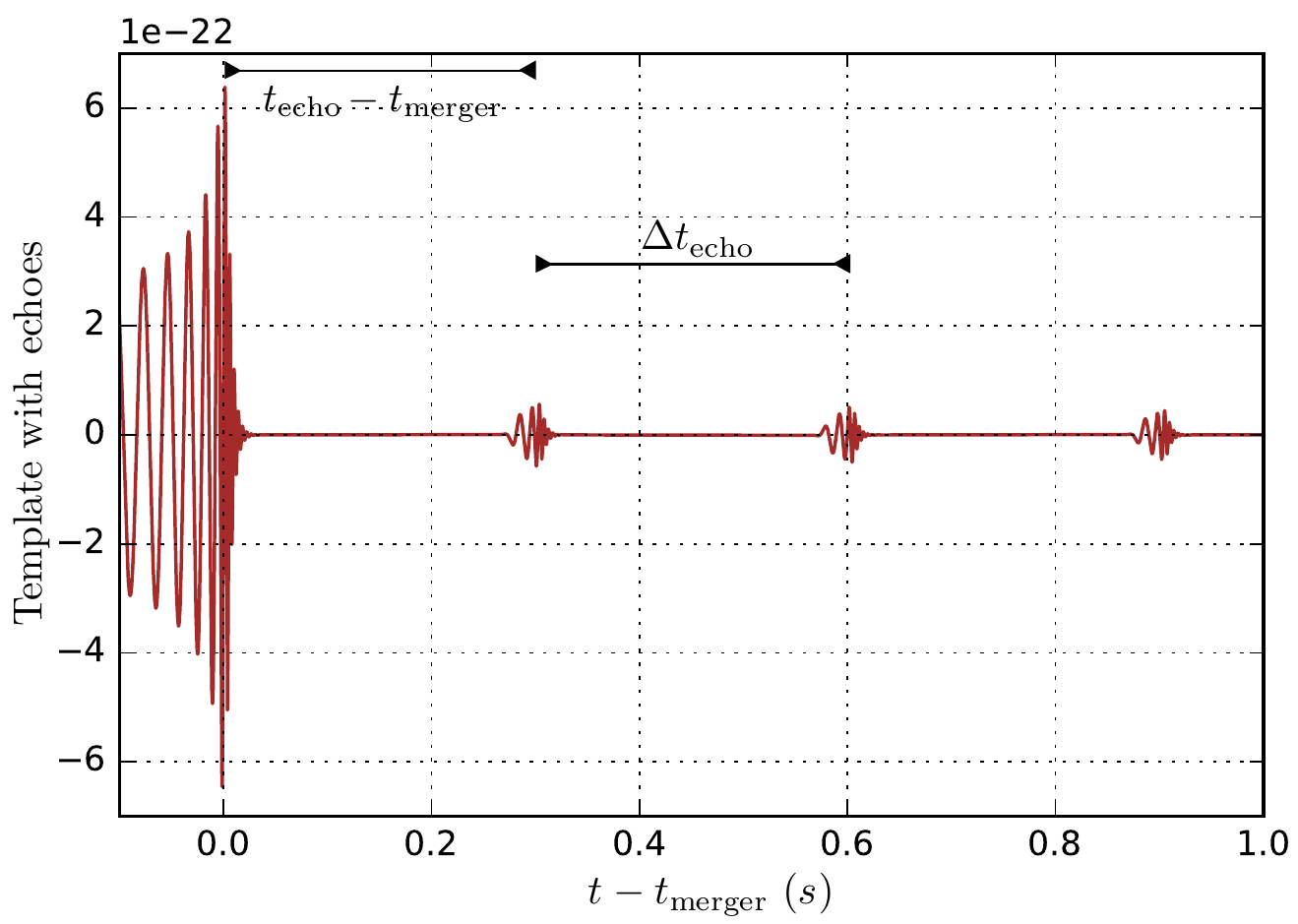}
  \end{minipage}
\caption{Original GW template of GW150914 \cite{Abedi:2016hgu}, along with best fit echoes template \ref{template}.\label{template_echoes}}
\end{figure}

The search model of echoes having all the free parameters, assuming a $(-1)^{n+1}$ factor due to the phase flip at each reflection, is given by:
\begin{equation}
M_{TE,I}^{H/L}(t) \equiv 
~ A\displaystyle\sum_{n=0}^{\infty}(-1)^{n+1}\gamma^{n} {\cal M}_{T,I}^{H/L}(t+t_{\rm merger}-t_{\rm echo}-n\Delta t_{\rm echo},t_{0}). \label{template}
\end{equation}
Fig. (\ref{template_echoes}) shows this template using the best fit parameters within the range given above along with the main merger event GW150914.

Once this analysis has been completed for GW150914 (loudest event of O1), it has been repeated for rest of the events combined via SNR maximization:
\begin{eqnarray}
 {\rm SNR}^2_{total} \equiv \sum_I {\rm SNR}^2_I. \label{snr_total}
 \end{eqnarray}
The proposed combination takes same $\gamma$ and $t_0/\overline{\Delta t}_{\rm echo}$ for all events, keeping $\Delta t_{\rm echo}$ and $A$'s as free. The results are shown in Fig's (\ref{SNR}-\ref{Histogramloglog}) and Tables \ref{table_1}-\ref{table_2}.

\begin{table}
\begin{center}
\begin{tabular}{ |c|c|c|c|c| }
\hline
 & Range& GW150914& Combined  \\
\hline
$(t_{\rm echo} -t_{\rm merger})/ \Delta t_{\rm echo}$& (0.99,1.01)& 1.0054& 1.0054  \\
$\gamma$& (0.1,0.9)& 0.89& 0.9 \\
$t_{0}/\overline{\Delta t}_{\rm echo}$& (-0.1,0)& -0.084& -0.1 \\
Amplitude&   &0.0992 &0.124 \\
SNR$_{\rm max}$  &   & 4.21 & 6.96 \\
\hline
p-value &  &    $0.11$ &$0.011$ \\
significance &  & 1.6$\sigma$ & 2.5$\sigma$\\
\hline
\end{tabular}
\caption{Values of best fit echo parameters of the model Eq. \ref{template} of the highest SNR peak near the predicted $\Delta t_{\rm echo}$ (gray rectangle in Fig. \ref{SNR}), and their significance \cite{Abedi:2016hgu}.}\label{table_1}
\end{center}
\begin{center}
\begin{tabular}{ |c|c|c|c| }
\hline
\ \  & GW150914 & GW151012 & GW151226 \\
\hline
$\Delta t_{\rm echo, pred}$(sec) & 0.2925 & 0.1778 &  0.1013 \\
 & $\pm$ 0.00916 & $\pm$ 0.02789 & $\pm$ 0.01152 \\
\hline
$\Delta t_{\rm echo, best}$(sec) & 0.30068 & 0.19043 & 0.09758 \\ 
$|A_{\rm best, I}|$ & 0.091 & 0.34 & 0.33 \\ 
SNR$_{\rm best, I}$ & 4.13 & 4.52 & 3.83 \\
\hline
\end{tabular}
\caption{Comparing the expected theoretical values of echo time delays $\Delta t_{\rm echo}$'s of each merger event (Eq. \ref{t_echo_meas}), to their best combined fit within the 1$\sigma$ credible region, and the contribution of each event to the combined SNR for the echoes (Eq. \ref{snr_total}) \cite{Abedi:2016hgu}.
}\label{table_2}
\end{center}
\end{table}

\begin{itemize}
\item \textit{Energy estimation}\cite{Abedi:2016hgu}:
Given a best-fit template for the echoes, one can provide an estimate for their total GW energy:

\begin{eqnarray}
E^{I}_{\rm{echoes}}/(M_{\odot}c^{2})=\left\{
 \begin{matrix}
  0.029 & I=\rm{GW150914}, \\
  0.16 & I=\rm{GW151012}, \\
  0.047 & I=\rm{GW151226}, \\
 \end{matrix}
 \right.\nonumber \\ \label{angle}
\end{eqnarray}
\end{itemize}

\subsubsection{\label{Uchikata}Uchikata et al. \cite{Uchikata:2019frs} analysis based on the template of Abedi, Dykaar, and Afshordi (ADA) \cite{Abedi:2016hgu} for O1 and O2}

Uchikata et al. \cite{Uchikata:2019frs} have examined GW echo signals for nine binary BH merger events observed by Advanced LIGO and Virgo during the first and second observation runs (O1 and O2 respectively). They have used several models for a number of searches leading to positive and negative results. In this part we bring their positive results and discuss the rest in part \ref{Uchikatanegative}. In this search the critical p-value as 0.05 corrsponding to $2\sigma$ significance, (p-value below/above this value) indicates echo signals (are likely/unlikely) to be present in the data.

\begin{figure}[!tbp]
  \centering
  \begin{minipage}[b]{0.5\textwidth}
    \includegraphics[width=1\textwidth]{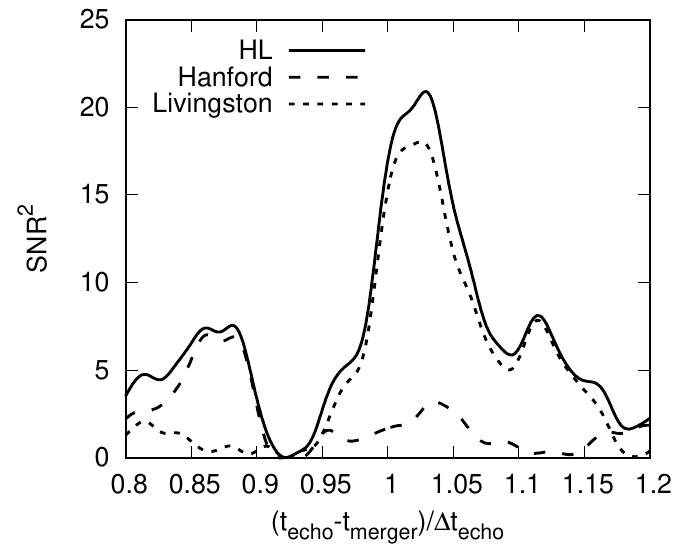}
  \end{minipage}
\caption{$\rm SNR^{2}$ with respect to $T=(t_{\rm{echo}}-t_{\rm{merger}})/\Delta t_{\rm{echo}}$. Solid, dashed, and dotted lines correspond to $\rho^{2}$ for combined (Hanford and Livingston), Hanford, and Livingston, respectively, for the best fit parameters of the event GW150914 \cite{Uchikata:2019frs}.\label{Uchikata1}}
\end{figure}

SNR is evaluated using a matched filter analysis defined using
\begin{eqnarray}
\rho \equiv (x|h)=4\rm{Re} \left( \int_{f_{\rm{min}}}^{f_{\rm{max}}} df \frac{\tilde{x}(f)\tilde{h}^{*}(f)}{S_{n}(f)} \right)
\end{eqnarray}
where $\tilde{\rm{x}}(\rm{f})$ and $\tilde{\rm{h}}(\rm{f})$ are observed data and template in frequency domain respectively, and $\rm{S}_{\rm{n}}(\rm{f})$ is the noise power spectrum of detector. In this analysis they assume frequency band of $\rm f_{max}=2048$ Hz, $\rm f_{min}=40$ Hz and normalization condition of $\rm (h|h)=1$ \cite{Uchikata:2019frs}. They then perform an echo search by maximizing SNR, following \cite{Abedi:2016hgu}. Fig. \ref{Uchikata1} presents $\rm SNR^{2}$ for the best fit parameters of the event GW150914 (C01)\footnote{There are two versions of noise subtraction in LIGO open data, called C01 and C02 \cite{gw-openscience}} with additional consideration of the best fit initial phase\footnote{Additional consideration of the best fit initial phase has given negative results and is discussed in negative result part below \ref{Uchikatanegative}.  Therefore, this plot only justifies Uchikata et al. \cite{Uchikata:2019frs} reevaluation of ADA search \cite{Abedi:2016hgu} in Fig. \ref{SNR}, while  belongs to the discussion in part \ref{Uchikatanegative}.} to the ADA \cite{Abedi:2016hgu} template.

In this part, we show the results using the same template given by ADA \cite{Abedi:2016hgu} except the cut-off parameter $t_{0}$ as described in \ref{sec_template} has been fixed to its best fit value, and set the search region of $\Delta t_{techo}$ to its 90\% (rather than 68\%) credible regions in $(a, M)$ space. Similar to ADA, the initial phase of the template is also fixed to zero. Since variation of $t_{0}$ weakly affects SNR and has an advantage in saving computational costs, they fixed $t_{0} = -0.1\Delta t_{echo}$.

Here are the results:
\begin{enumerate}
\item \textit{O1 events (reanalysis of Westerweck et al.} \cite{Westerweck:2017hus}):
Since Uchikata et al. \cite{Uchikata:2019frs} have followed the same background estimation as Westerweck et al. \cite{Westerweck:2017hus} the results are compared to their O1 results for p-values in Table \ref{table_3}.
It is seen that they are almost consistent within the Poisson errors for all events, confirming a marginal p-value of 3\%-5\% for ADA echoes \cite{Abedi:2016hgu}. 
\begin{table}
\begin{center}
\begin{tabular}{ |c|c|c| }
\hline
Event & Westerweck et al. \cite{Westerweck:2017hus} & Uchikata et al. \cite{Uchikata:2019frs}  \\
\hline
GW150914 & $0.238 \pm 0.043$ & $0.157 \pm 0.035$ \\
\hline
GW151012 & $0.063 \pm 0.022$ & $0.047 \pm 0.019$ \\
\hline
GW151226 & $0.476 \pm 0.061$ & $0.598 \pm 0.069$ \\
\hline
Total & $0.032 \pm 0.016$ & $0.055 \pm 0.021$ \\
\hline
\end{tabular}
\caption{P-values along with Poisson errors for O1 events \cite{Uchikata:2019frs}.}\label{table_3}
\end{center}
\begin{center}
\end{center}
\end{table}

\item \textit{O2 events}:

Analysis of Uchikata et al. \cite{Uchikata:2019frs} show that the six independent BBH O2 events in Table \ref{table_4} have similarly small p-values for ADA echoes as O1. As shown in this table, the total p-value for the six O2 events is 0.039. Combining O2 with O1 events shown in Table \ref{table_3}, leads to the total p-value of 0.047.

\begin{table}
\begin{center}
\begin{tabular}{ |c|c|c| }
\hline
Event & Uchikata et al. \cite{Uchikata:2019frs}  \\
\hline
GW170104 & 0.071 \\
\hline
GW170608 & 0.079 \\
\hline
GW170729 & 0.567 \\
\hline 
GW170814 & 0.024 \\
\hline
GW170818 & 0.929 \\
\hline
GW170823 & 0.055 \\
\hline
Total & 0.039 \\
\hline
\end{tabular}
\caption{P-values for O2 events \cite{Uchikata:2019frs}.}\label{table_4}
\end{center}
\begin{center}
\end{center}
\end{table}

\end{enumerate}

\subsubsection{Echoes from the Abyss: Binary neutron star merger GW170817 \cite{Abedi:2018npz}}

A binary neutron star merger event collapsing into a black hole (Fig. \ref{echo_pic_2}) can also enable us to test general relativity through GW echoes. Although, the current LIGO/Virgo/KAGRA detector sensitivity is blind to post-merger ringdown frequency of GWs, they can be sensitive to low frequency echoes harmonics where the ringdown frequency is suppressed by $ln(M_{\rm BH}/M_{\rm planck})$ in Eq. \ref{eq.0.1} \cite{Cardoso:2016oxy}. In other words, since the final mass of BNS merger (2-3 $M_{\odot}$) is much smaller than that of the binary BH mergers \cite{Abedi:2016hgu}, the lowest harmonics $ n/\Delta t_{\rm echo}$ ($\simeq n \times 80$ Hz) of echo chamber are shifted to the regime of LIGO sensitivity, for small $n$. Therefore, as first suggested by \cite{Conklin:2017lwb}, an optimal model-agnostic search strategy could consist of looking for periodically spaced-harmonics in the frequency space.

\begin{figure*}[!tbp]
    \includegraphics[width=1\textwidth]{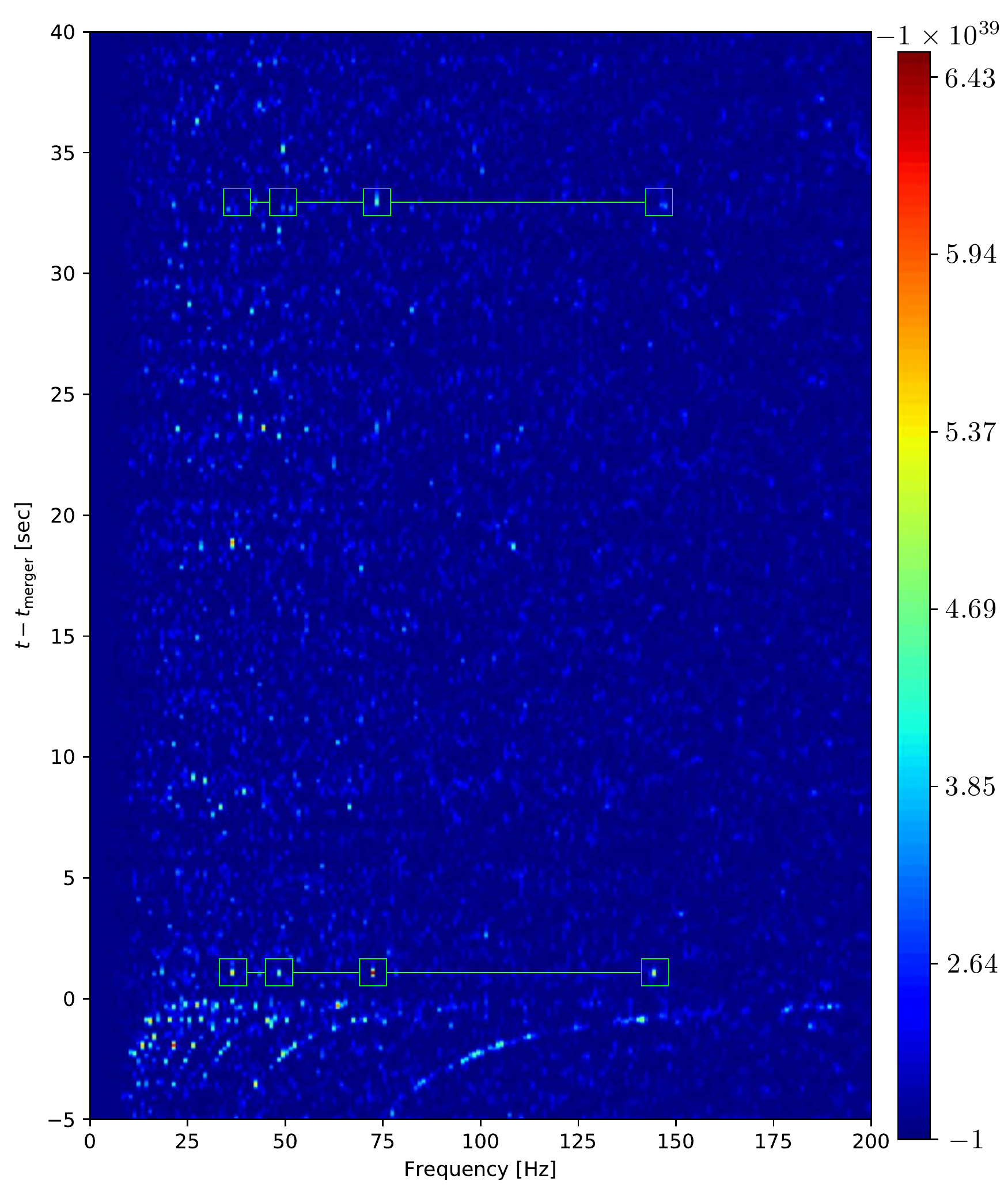}
 \caption{Time-frequency representations of $X(t,f)$ (Equation \ref{x_def}; combining all harmonics with frequencies $n\times f$, with $n\in \mathbb{N}$) around the merger for the BNS gravitational-wave event GW170817, observed through cross-correlating the two LIGO detectors \cite{Abedi:2018npz}. The possible resonance peaks of echoes found in this plot are marked with a green squares. The color scale shows the peak at $f_{\rm{peak}}=72\ (\pm 0.5)$ Hz and $t-t_{\rm{merger}} \simeq $ 1.0 sec, is the highest peak in this diagram, from before and after the BNS merger (see Figs. \ref{NS-NS_3d}, \ref{NS-NS_4} and \ref{NS-NS_9}).  A secondary peak at the same frequency but $t-t_{\rm{merger}} \simeq $ 32.9 sec is also highlighted in this plot.}
 \label{NS-NS_11}
\end{figure*}

\begin{figure*}[!tbp]
    \includegraphics[width=\textwidth]{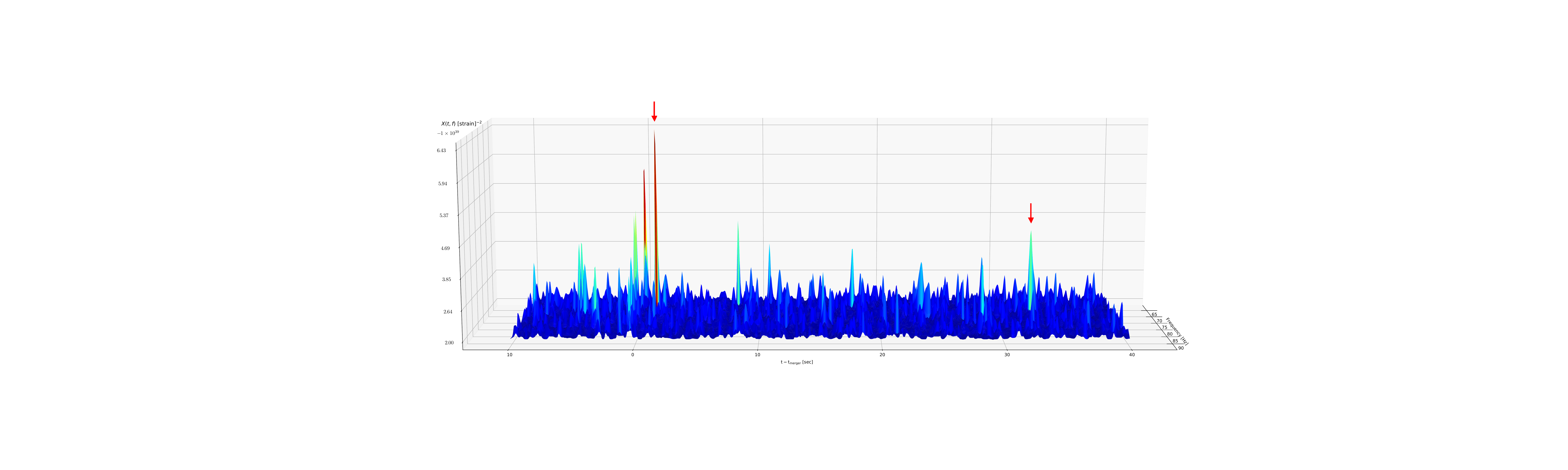}
 \caption{A 3d representation of Fig. (\ref{NS-NS_11}) within echo search frequency window $f=63-92$ Hz \cite{Abedi:2018npz}. This plot shows the tentative detection of echoes at  $f_{\rm{peak}}=72\ (\pm 0.5)$ Hz and $t-t_{\rm{merger}} \simeq $ 1.0 sec clearly stands above noise.}
 \label{NS-NS_3d}
\end{figure*}

\begin{figure*}[!tbp]
    \includegraphics[width=1\textwidth]{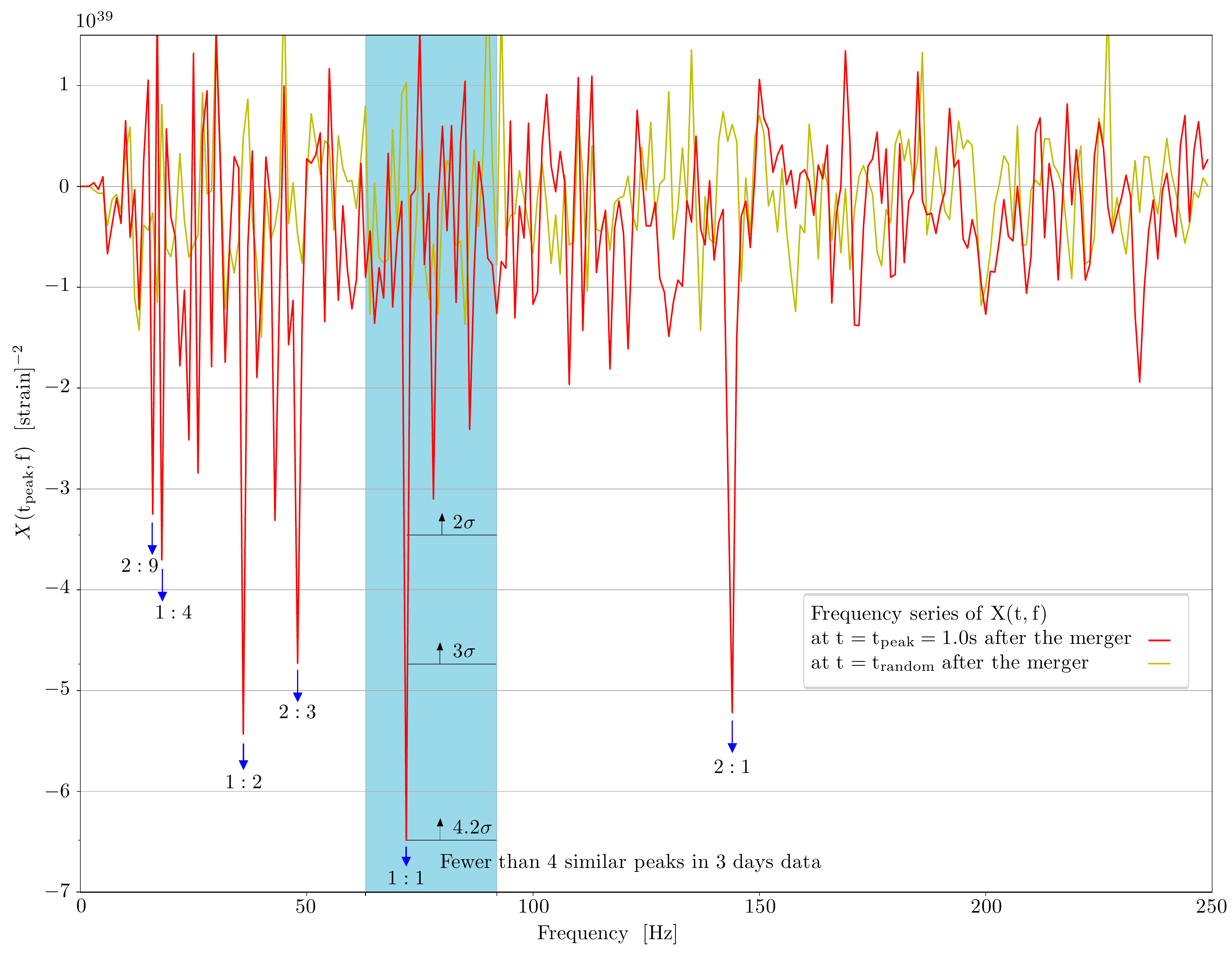}
 \caption{Amplitude-frequency plot of $X(t,f)$ (Equation \ref{x_def}; combining all harmonics with frequencies $n\times f$, with $n\in \mathbb{N}$) for the first peak (red) at 1.0 sec after the merger for the BNS merger gravitational-wave event GW170817, observed by x-correlating the LIGO detectors \cite{Abedi:2018npz}. The same amplitude-frequency plot for a random time in data (yellow), is also shown for comparison. Solid area between 63 Hz and 92 Hz was the search frequency prior range for Planckian echoes.}
 \label{NS-NS_4}
\end{figure*}

\begin{figure}[!tbp]
\centering
    \includegraphics[width=0.5\textwidth]{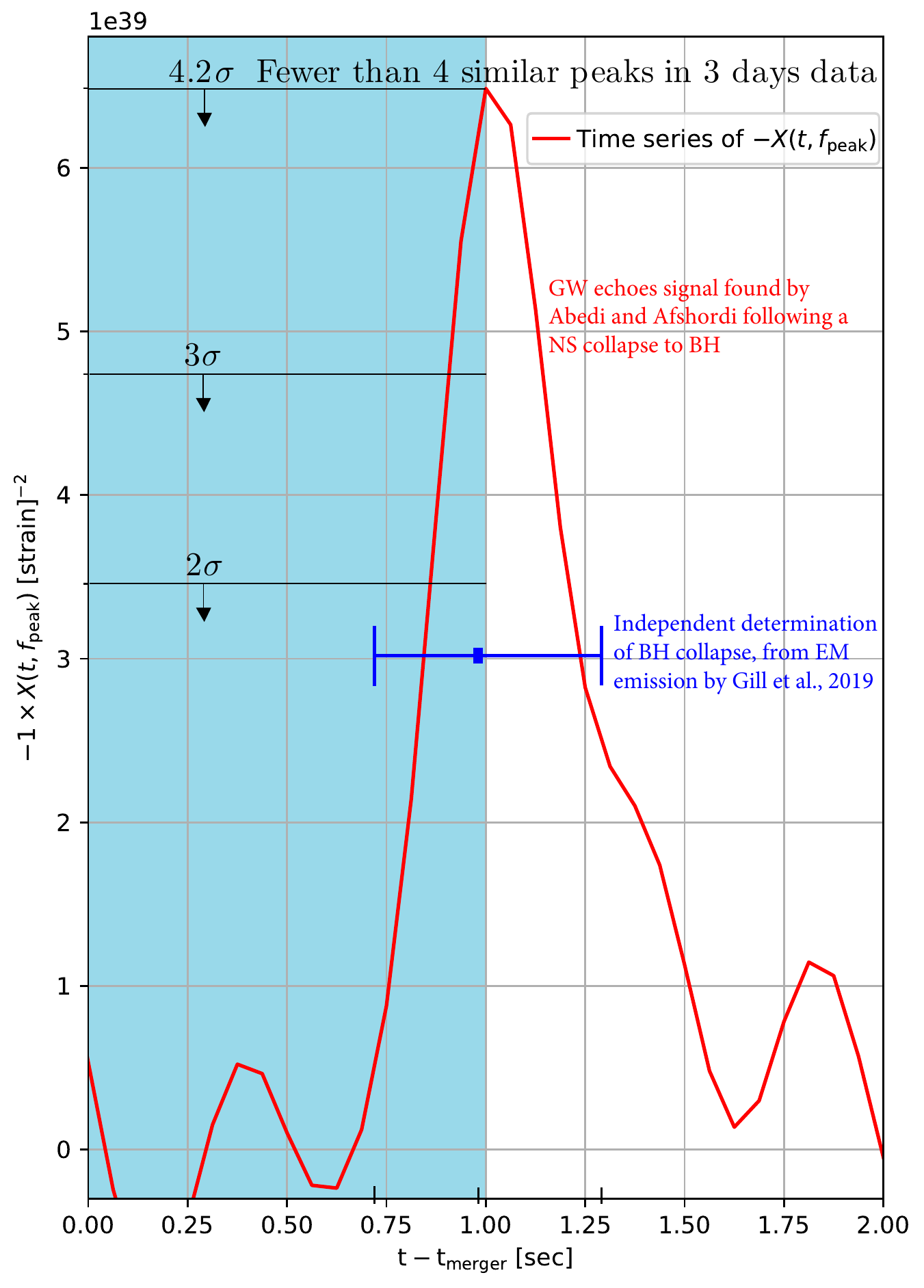}
 \caption{Amplitude-time plot of first echo peak at 1.0 sec after the merger at frequency of 72 Hz \cite{Abedi:2018npz}. After this detection Gill et al. \cite{Gill:2019bvq} with independent Astrophysical considerations have also determined that the remnant of GW170817 must have collapsed to a BH after $t_{\rm coll}=0.98^{+0.31}_{-0.26}$ sec. Error-bar (in blue) is the time of collapse considering this independent observation in \cite{Gill:2019bvq} compared to the detected signal of echoes which is also as a consequence of BH collapse. The shaded region is 0-1 sec prior range after the merger.}
 \label{NS-NS_9}
\end{figure}

\begin{figure}[!tbp]
\centering
    \includegraphics[width=0.5\textwidth]{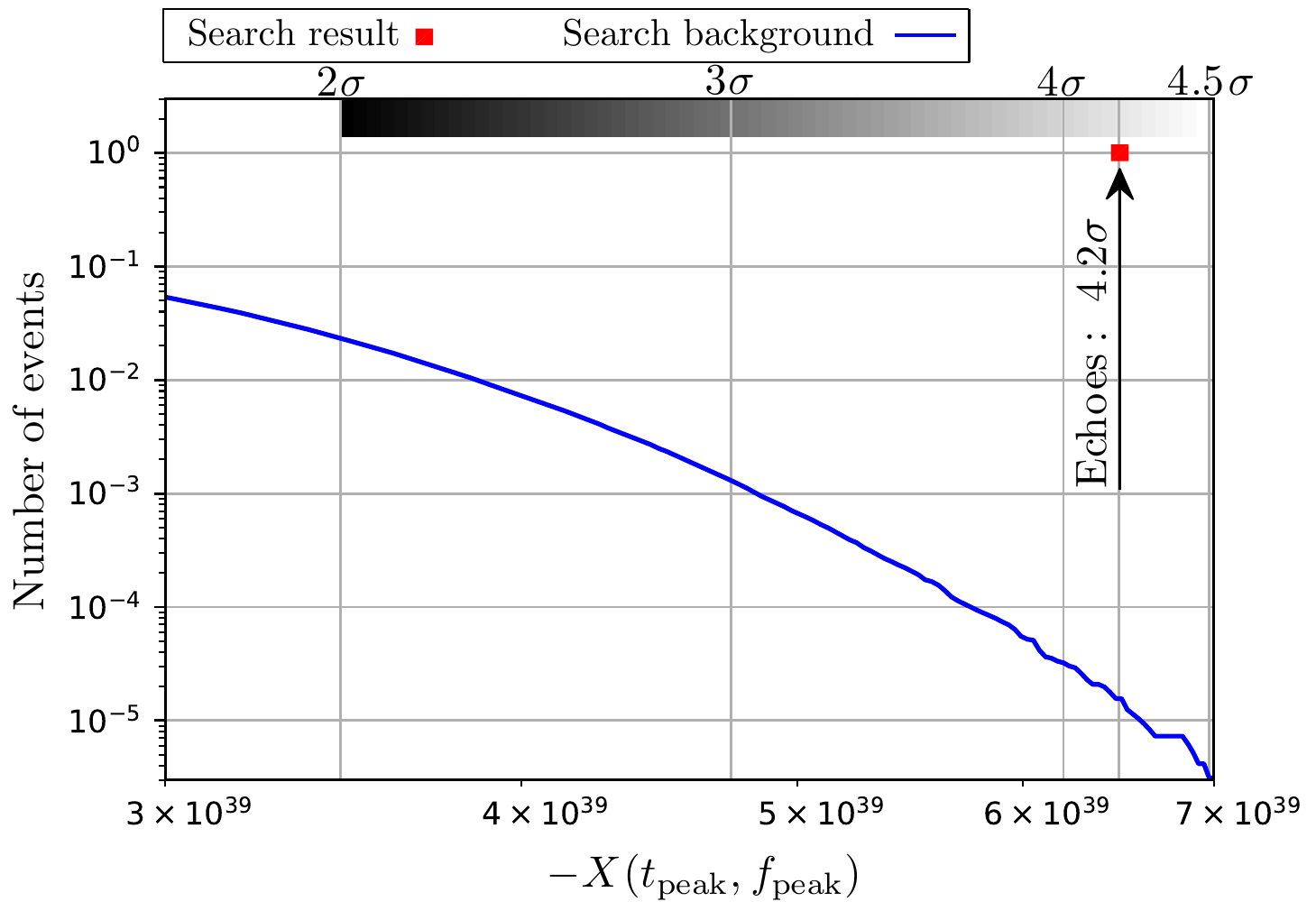}
 \caption{Average number of background peaks higher than a particular -X(t,f) within a frequency-intervals of 63-92 Hz and time-intervals of 1 sec. the observed $-X(t_{\rm{peak}},f_{\rm{peak}})$ peak at 1.0 sec after the merger is marked by red square. The horizontal bar shows the relation between $X(t,f)$ values and their significance \cite{Abedi:2018npz}.}
 \label{NS-NS_10}
\end{figure}

Using this model-agnostic search applied to the cross-power spectrum of the two LIGO detectors, Abedi and Afshordi \cite{Abedi:2018npz} found a tentative detection of echoes around 1.0 sec after the BNS merger, at $f_{echo}\simeq 72$ Hz (see Figs. \ref{NS-NS_11}, \ref{NS-NS_3d}, \ref{NS-NS_4}, and \ref{NS-NS_9}) using GW event data GW170817 provided by LIGO/Virgo collaboration \cite{TheLIGOScientific:2017qsa,Abbott:2017dke}. As it is shown in Figs. \ref{NS-NS_11} and \ref{NS-NS_3d}, the main signal is also accompanied by secondary lower significance resonances at 73 Hz and $t-t_{\rm{merger}}$ = 32.9 sec. It is worth noting that after this detection, Gill et al. \cite{Gill:2019bvq} used independent Astrophysical considerations, based GW170817 electromagnetic follow-ups, to determine that the remnant of GW170817 must have collapsed into a BH after $t_{\rm coll}=0.98^{+0.31}_{-0.26}$ sec, which coincides with the detected GW echo signal at 1.0 second (see Fig. \ref{NS-NS_9}). This fining of Abedi and Afshordi is consistent with a $2.6-2.7$ $M_{\odot}$ "BH" remnant with dimensionless spin $0.84-0.87$. For this signal, considering all the "look-elsewhere" effects, a significance of $4.2\sigma$\footnote{In this paper, a 1-tailed gaussian probability to assign a significance to a p-value is used, e.g., p-value= 84\% and 98\% correspond to $1\sigma$ and $2\sigma$ respectively.} (see Fig. \ref{NS-NS_10}), or a false alarm probability of $1.6\times10^{-5}$ has been reported, i.e. a similar cross-correlation within the expected frequency/time window after the merger cannot be found more than 4 times in 3 days of GW data. Total energy of detected GW echoes signal using simple assumptions is around $\sim 10^{-2}\ M_{\odot} c^2$.

\subsubsection{MODEL-AGNOSTIC SEARCH FOR ECHOES}

In this part, we describe the method that leads to the detection of \cite{Abedi:2018npz}, in some detail.
The final mass of GW170817 is within $\sim 2-3 M_{\odot}$ which could form either a BH or a neutron star (NS). A BNS merger can end up in four possible ways: \cite{Abbott:2017dke}:
\begin{enumerate}
\item A prompt collapse to a BH.

\item A formation of a BH within $\lesssim$ 1 sec from a hypermassive NS.

\item Collapse to a BH on timescales of $10$ - $10^{4}$ sec from a supramassive NS.

\item A stable neutron star. \cite{TheLIGOScientific:2017qsa}.

\end{enumerate}

Abedi and Afshordi \cite{Abedi:2018npz} have considered the possibility of first and second scenario.

As it was pointed out in part \ref{Echoes from Abyss:O1} the time delay for Planck scale echoes is given by, \cite{Abedi:2018npz},
\begin{eqnarray}
&\Delta t_{\rm echo} \simeq \frac{4 G M_{\rm BH}}{c^3}\left(1+\frac{1}{\sqrt{1-\bar{a}^2}}\right) \times \ln\left(M_{\rm BH} \over M_{\rm planck}\right)& \nonumber\\ &\simeq 4.7~ {\rm msec} \left(M_{\rm BH} \over 2.7~ M_\odot \right) \left(1+\frac{1}{\sqrt{1-\bar{a}^2}}\right).& \label{delay}
\end{eqnarray}

There are two natural frequencies for the waveform of the echoes: The resonance frequencies (natural harmonics) of the echo chamber (formed by the angular momentum barrier and the near-horizon quantum structure), and the BH ringdown (or classical QNM) frequencies. The high frequency harmonics that are initially excited by the merger event decay quickly, while the low frequency harmonics live for longer time \cite{Maggio:2017ivp,Bueno:2017hyj,Wang:2018gin}. While the former captures the repeat period of the echoes, the latter describes the echo internal structure. Given that the ringdown frequencies are not resolved by LIGO detector, we can roughly approximate the observable signal as a sum of Dirac delta functions, repeating with the period $\Delta t_{\rm echo}$:
\begin{eqnarray}
h(t) \propto \sum_n \delta_D(t-n\Delta t_{\rm echo} -t_0) \Rightarrow
h_{f} \propto \sum_n \delta_D(f- n f_{\rm echo}). \label{h_f_res}
\end{eqnarray}
Therefore, the method searches for coherent periodic peaks of equal amplitude in cross-power spectrum of the two detectors at integer multiples of $f_{\rm echo} \equiv \Delta t_{\rm echo}^{-1} $, in following steps:
\begin{enumerate}
\item Fundamental frequency of echoes  $f_{\rm echo}= \Delta t_{\rm echo}^{-1}$ within the 90\% credible region range for final BH mass and spin is given by:
$$
63 \leq f_{{\rm echo} }({\rm Hz}) \leq 92.
$$

\item The prior range for echoes search is $0 <t-t_{\rm{merger}} \leq 1~ \rm{sec}$.

\item Using amplitude spectral density (ASD) Wiener filter (rather than whiten) the data by dividing by noise variance PSD=ASD$^2$ (rather than ASD):
\begin{eqnarray}
H(t,f)={\rm Spectrogram}\left[ {\rm IFFT}\left( \frac{{\rm FFT}(h_{H}(t- \delta t))}{PSD_{H}} \right) \right], \nonumber\\ 
L(t,f)={\rm Spectrogram}\left[ {\rm IFFT}\left( \frac{{\rm FFT}(h_{L}(t))}{PSD_{L}} \right) \right].
\end{eqnarray}
where $\delta t$ is the time shift between detectors.

\item Cross-correlating the obtained spectrograms and sum over all the resonance frequencies of $n\times f$,
\begin{eqnarray}
X(t,f)=\sum_{n=1}^{10} \Re\left[H(t,nf) \times L^*(t,nf)\right]. \label{x_def}
\end{eqnarray}
Since the polarizations of the LIGO detectors are opposite for GW170817, the real GW signals appears as peaks in $-X(t,f)$ (see Figs. \ref{NS-NS_11} and \ref{NS-NS_4}).

\end{enumerate}

The simplicity of the method (not having any arbitrary or {\it ad-hoc} cuts or parameters) along with its high significance are reasons for making this finding more interesting and reliable.

\subsubsection{``GW echoes through new windows'' by Conklin et al. \cite{Conklin:2017lwb, Holdom:2019bdv}}

\begin{figure}
\begin{center}
\includegraphics[width=0.9\textwidth]{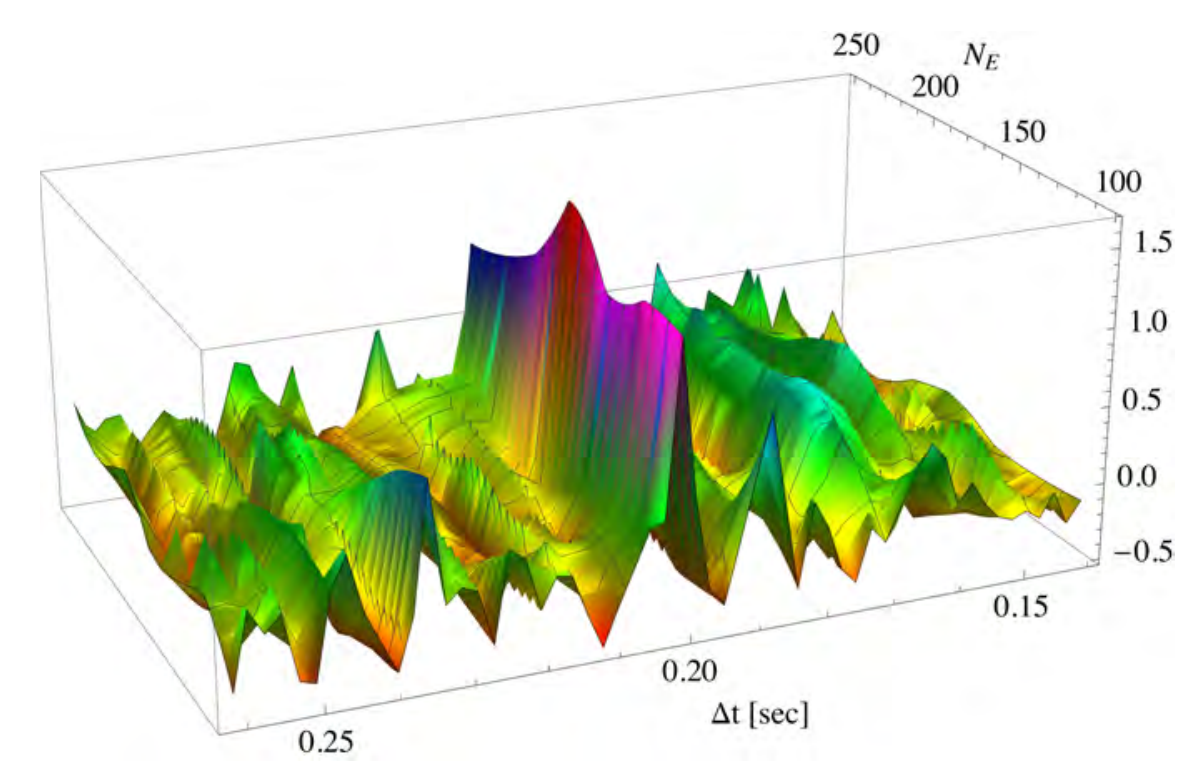}
\caption{Correlation vs. echo time delay $\Delta t_{\rm echo}$ vs. $\rm N_{E}$ for GW170104 using method II \cite{Conklin:2017lwb}. Here $N_{E}$ is the number of frequency steps between spikes.}
\label{Holdom2}
\end{center}
\end{figure}

In this search, three methods (named as I, II, III) that are based on general properties of echoes has been suggested.
Conklin et al. \cite{Conklin:2017lwb} have mostly focused on Method II, which is based on frequency windows, while the other two methods use time windows. Window functions in these methods allow us to find quasiperiodic structures in time and/or frequency domains. Method II turns out to be the most successful one. Accordingly, in this paper we just review this method.  The search methods become more optimal using correlations of data in multiple detectors. Using the suggested method and search \cite{Conklin:2017lwb} find significant evidence for GW echoes, which is shown in Table \ref{table_5} (and see Fig. \ref{Holdom2} for GW170104) for both O1 and O2 LIGO/Virgo observations.

\begin{table*}
\begin{center}
\begin{tabular}{ |c|c|c|c|c| }
\hline
Event (method) & Best-fit $\Delta t_{\rm echo}$ (sec) & p-value & Bandpass $(f_{min},f_{max})\Delta t_{\rm echo}$ & Window parameters for average \\
\hline
GW151226 (I) & 0.0786 & $<0.0013$\footnote{upper bounds are just limited by the number of trials} & (34,62)\footnote{the bandpass ranges in units of Hz are: (433, 789), (152, 733), (80, 308), (185, 794), (52, 251), (132, 351)} & $N_{E}$=(1-29), (5-29), (9-29)\footnote{$(i-j)$ means echoes $i$ through $j$ were used}  \\
\hline 
GW151226 (II) & 0.0791 & 0.0076 & (12,58) & $N_{E}$=(260,270) \\
\hline 
GW170104 (II) & 0.201 & $<0.0018$ & (16,62) & $N_{E}$=(100,125,150,175,200) \\
\hline 
GW170608 (II) & 0.0756 & $<0.004$ & (14,60) & $N_{E}$=(140,200,260) \\
\hline 
GW170814 (II) & 0.231 & 0.04 & (12,58) & $N_{E}$=(170,190)\footnote{the whole time range used was shifted 10 seconds later} \\
\hline 
GW170814 (III) & 0.228 & 0.0077 & (30,80) & $N_{E}=10-17$, $t_{w}=40,80$\footnote{the explicit sets used: ($N_{E}$, $t_{w}/M$) = (15,40),(10,80),(15,80),(3-15,40),(5-17,40),(3-15,80)} \\
\hline 
\end{tabular}
\caption{The best-fit $\Delta t_{\rm echo}$, p-value, bandpass and window parameters for the six signals \cite{Conklin:2017lwb}. Here $N_{E}$ is the number of frequency steps between spikes which has been chosen using injection and echoes model properties in \cite{Conklin:2017lwb}.}\label{table_5}
\end{center}
\begin{center}
\end{center}
\end{table*}

The time delay of echoes, $\Delta t_{\rm echo}$ which is similar to the Planckian echoes in Eq. \ref{eq.0.1}, is paramaterized in \cite{Conklin:2017lwb} using
\begin{eqnarray}
\left. \Delta t_{\rm echo}/M \right|_{CHR}=-\eta \left(1+\frac{1}{\sqrt{1-\bar{a}^{2}}}\right)\ln\left( r_{0}-r_{+} \over M\right)
\end{eqnarray}
where $r_{0}$ is the location of the quantum structure outside $r_{+}$. Here $\eta=2$ corresponds to proper Planck length (Eq. \ref{eq.0.1}).

\begin{figure}
\begin{center}
\includegraphics[width=0.9\textwidth]{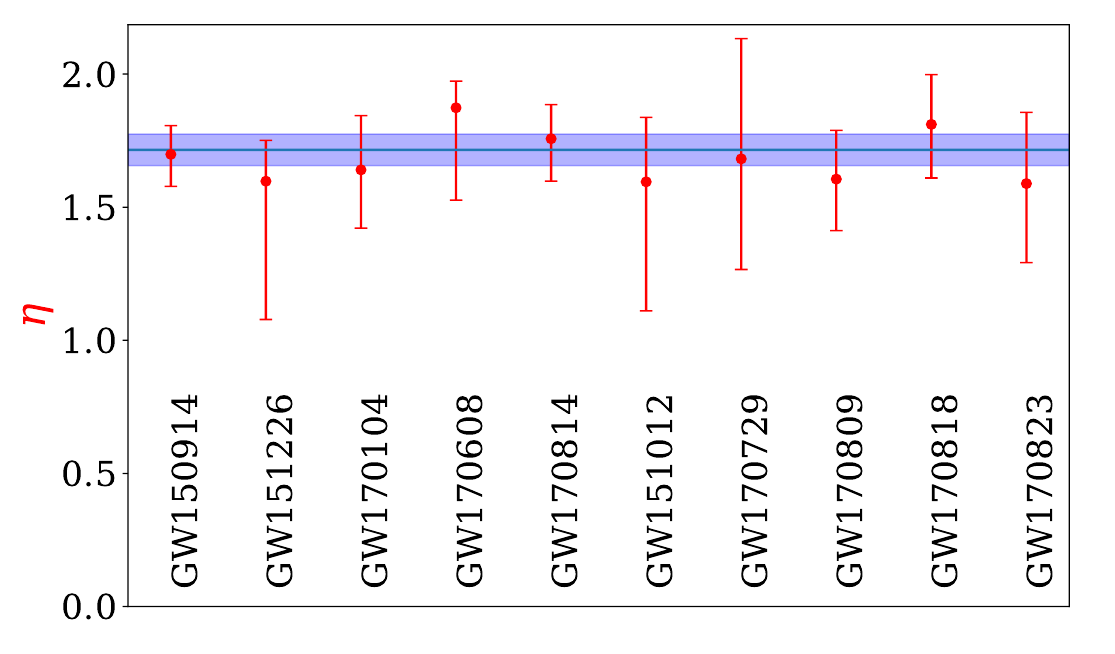}
\caption{Determination of $\eta$ for the events in O1 and O2 \cite{Holdom:2019bdv}.}
\label{Holdom1}
\end{center}
\end{figure}

Taking into account of errors in final mass, spin, and redshift it is realised that (see Fig \ref{Holdom1}) the echoes found are consistent with $\eta=1.7$ \cite{Conklin:2017lwb,Holdom:2019bdv}. Best fit properties of the peaks for O1 and O2 events also shown in Table \ref{table_6} \cite{Holdom:2019bdv}. Alternatively, this result can be interpreted as the energy scale for reflection from quantum horizons to be $6 \pm 2$ orders of magnitude below Planck energy \cite{Oshita:2016pbh}. 

\begin{table}
\begin{center}
\begin{tabular}{ |c|c|c|c| }
\hline
Event (method) & Best-fit $\Delta t_{\rm echo}$ (sec) & $N_{E}$ & $\Delta t_{\rm echo}/M$  \\
\hline
GW150914 (II) & 0.251 & 200 & 806  \\
\hline
GW151012 (II) & 0.145 & 160 & 826 \\
\hline
GW151226 (II) & 0.0791 & 783 & 270 \\
\hline
GW170104 (II) & 0.201 & 150 & 831 \\
\hline
GW170608 (II) & 0.0756 & 200 & 862 \\
\hline
GW170729 (II) & 0.489 & 180/170 & 1240 \\
\hline 
GW170809 (II) & 0.235 & 170 & 845 \\
\hline 
GW170814 (II) & 0.231 & 200 & 878 \\
\hline
GW170817 (II) & 0.00719 & 250 & 663 \\
\hline
GW170818 (II) & 0.275 & 140 & 933 \\
\hline
\end{tabular}
\caption{Echoes best fit time delays and corresponding $N_{E}$ using method II for O1 and O2 events \cite{Holdom:2019bdv}.}\label{table_6}
\end{center}
\begin{center}
\end{center}
\end{table}

\subsubsection{Comment on: "Gravitational wave echoes through new windows by Conklin et al. \cite{Conklin:2017lwb, Holdom:2019bdv}"}

Conklin et al. \cite{Conklin:2017lwb} remark that "We have not found signals for the two earlier events, GW150914 and GW151012, which play a significant role in ADA results in \cite{Abedi:2016hgu}". While in their updated search \cite{Holdom:2019bdv} they indicate existence of signals for both GW150914 and GW151012, although with no p-value estimation which is crucial in this search. Therefore, it is hard to truly evaluate the significance of echoes reported in \cite{Holdom:2019bdv}. Moreover, it is not clear how much the choices made in Method II in \cite{Conklin:2017lwb} might have been affected by {\it a posteriori} statistics.

\subsection{\label{Mixed}Mixed Results}

\subsubsection{\label{Results of Westerweck et al.}Results of Westerweck et al. and Nielsen et al. \cite{Westerweck:2017hus,Nielsen:2018lkf}}

Westerweck et al. \cite{Westerweck:2017hus} re-analysed the same model proposed by ADA \cite{Abedi:2016hgu}, using more background data and a modified procedure. They focused on the data analysis methods of ADA \cite{Abedi:2016hgu} and their significance estimation, namely the concerns presented in \cite{Ashton:2016xff} suggesting a different significance estimate using 4096 seconds of LOSC data.

The results of p-value estimation and comparison with ADA results are shown in Table \ref{table_7}. In addition, Nielsen et al. \cite{Nielsen:2018lkf} have searched for echoes signals in GW data via Bayesian model selection probabilities, comparing signal and no-signal hypotheses using ADA model \cite{Abedi:2016hgu}. Accordingly, calculation of Bayes factors for the ADA model in O1 events presented in Table \ref{table_8}.

\begin{table}
\begin{center}
\begin{tabular}{ |c|c|c|c| }
\hline
Event & \cite{Abedi:2016hgu} & original 16s (32s)& widened priors 16s (32s)  \\
\hline
GW150914 & 0.11 & 0.199 (0.238) & 0.705 (0.365)  \\
\hline
GW151012& - & 0.056 (0.063) & 0.124 \\
\hline 
GW151226& - & 0.414 (0.476) & 0.837 \\
\hline 
GW170104 & - & 0.725 & 0.757 \\
\hline
(1,2) & - & 0.004 & 0.36 \\
\hline
(1,3) & - & 0.159 & 0.801 \\
\hline 
(1,2,3) & 0.011 & 0.020 (0.032) & 0.18 (0.144)\\
\hline 
(1,3,4) & - & 0.199 (0.072) & 0.9 (0.32)\\
\hline
(1,2,3,4) & - & 0.044 (0.032) & 0.368 (0.112)\\
\hline
\end{tabular}
\caption{Comparison of p-values obtained in \cite{Abedi:2016hgu} and using larger portion of data (4096 seconds of LOSC data) \cite{Westerweck:2017hus}. This data is divided into segments of 16 or 32 seconds length. Here different combinations of the events are considered, denoted as (GW150914, GW151012, GW151226, GW170104) $\rightarrow (1, 2, 3, 4)$. Having the original priors, the Poisson errors (as suggested in \cite{Abedi:2018pst}): for GW150914 our p-values are $0.199 \pm 0.028$ ($0.238 \pm 0.043$), and for (1,2,3) our p-values are $0.02 \pm 0.009$ ($0.032 \pm 0.016$). The Poisson errors for the full combination (1,2,3,4) with original priors, are $0.044 \pm 0.013$ ($0.032 \pm 0.016$). The comparison of p-values using widened priors are also shown in this table.}\label{table_7}
\end{center}
\end{table}

\begin{table}
\begin{center}
\begin{tabular}{ |c|c|c| }
\hline
Event & Log Bayes factor & Max SNR  \\
\hline
GW150914 & -1.8056 & 2.86  \\
\hline
GW151012& 1.2499 & 5.5741 \\
\hline 
GW151226& 0.4186 & 4.07 \\
\hline 
\end{tabular}
\caption{Results of Bayes factor \cite{Nielsen:2018lkf} using ADA model. Gaussian noise hypothesis is preferred for negative values of Log Bayes factor. Echoes hypothesis is preferred for positive values of Log Bayes factor. Log Bayes values of $<$ 1 are ``not worth more than a bare mention'' \cite{Nielsen:2018lkf}.}\label{table_8}
\end{center}
\end{table}

In the following we explain the results given in different plots:
\begin{enumerate}

\item In Fig. \ref{Julian1} it is shown that depending on the overall amplitude of the injection, the signal either can be recovered or it would be difficult to recover.

\begin{figure}[!tbp]
\centering
    \includegraphics[width=0.5\textwidth]{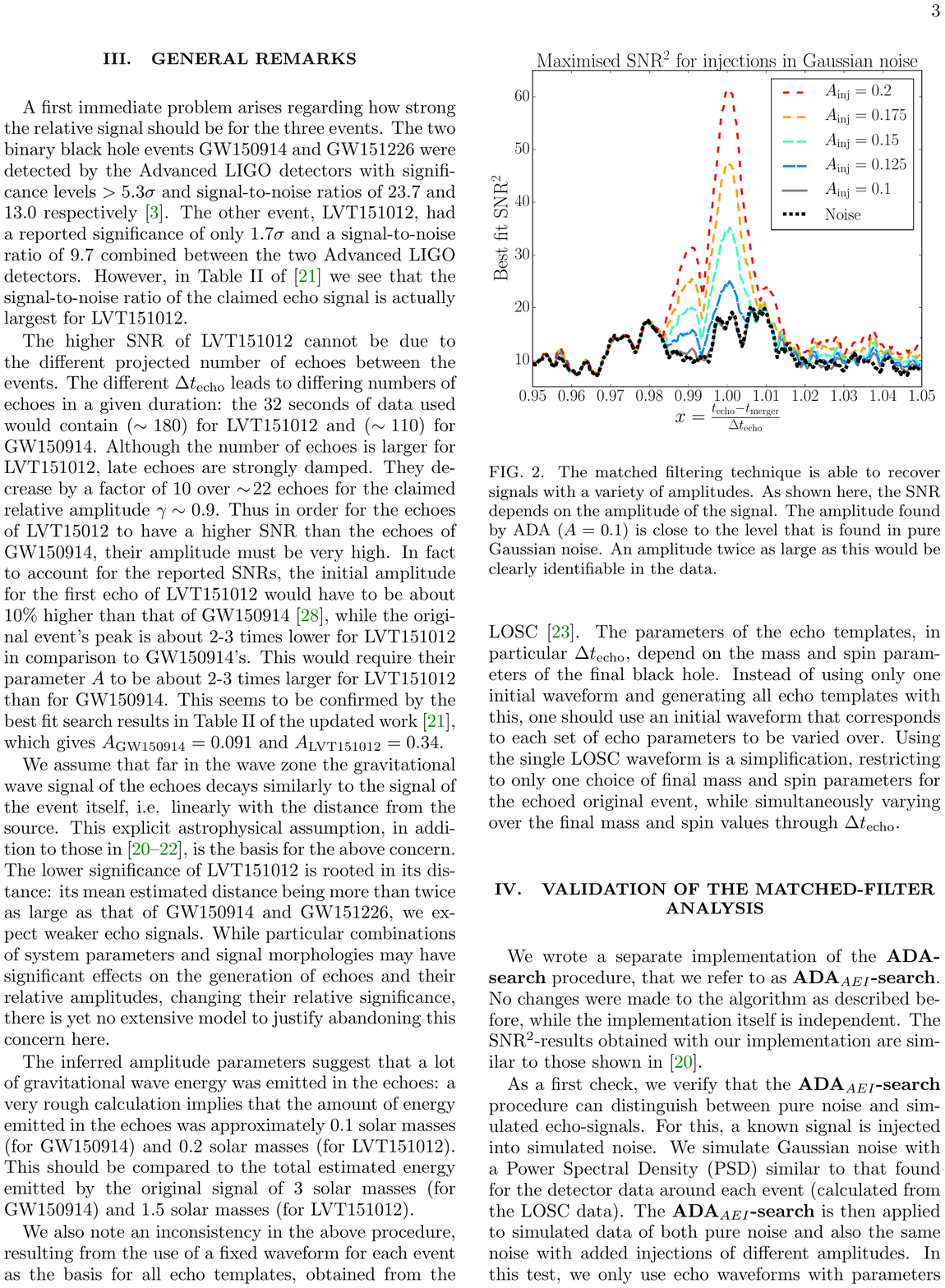}
 \caption{This plot shows whether it is possible to recover the potential signals with a variety of amplitudes in \cite{Westerweck:2017hus}. Here it can be seen that amplitudes less than A=0.1 in ADA \cite{Abedi:2016hgu} are difficult to be identified in data, while amplitude twice this value would be clearly identifiable.} \label{Julian1}
\end{figure}

\item Fig. \ref{Julian2} shows injected and recovered values for $\gamma$ having different overall amplitude $A$ used in \cite{Abedi:2016hgu}. Although this plot shows a preference for $\gamma = 1$, having high value of $\gamma$ can be recovered easily.

\begin{figure}[!tbp]
\centering
    \includegraphics[width=0.5\textwidth]{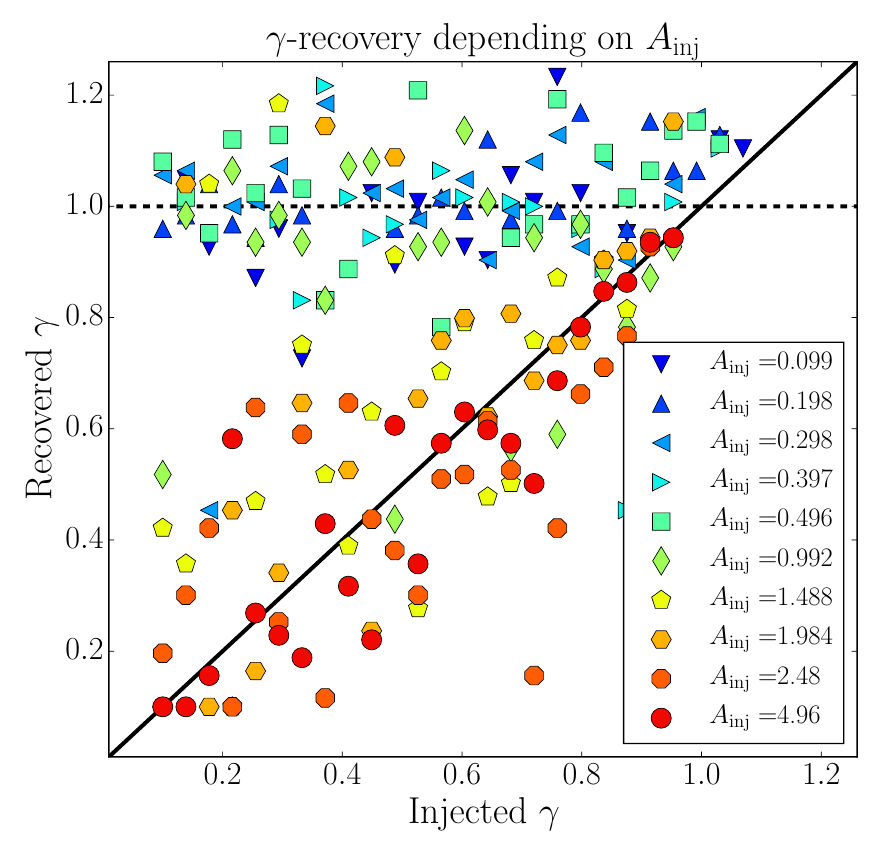}
 \caption{This plot shows injected and recovered values for $\gamma$. The diagonal line is accurate recovery \cite{Westerweck:2017hus}. The preference for $\gamma = 1$ (dashed line) at lower injection amplitudes can be clearly seen.}\label{Julian2}
\end{figure}

\item Fig. \ref{Julian3} shows injection of echoes signals into the Gaussian noise having different overall amplitudes A.

\begin{figure}[!tbp]
\centering
    \includegraphics[width=0.5\textwidth]{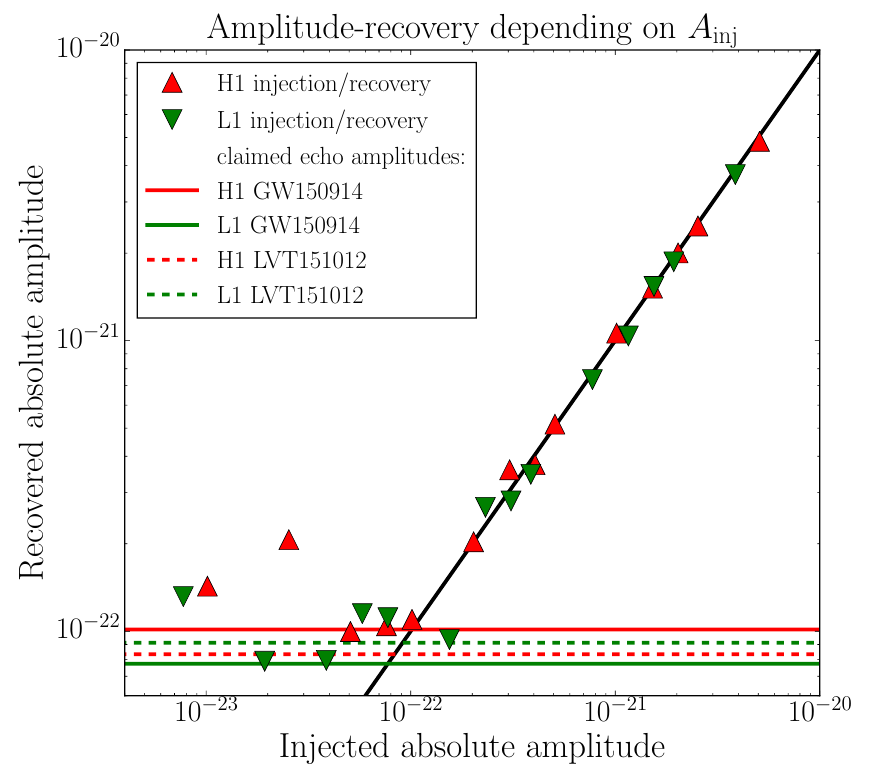}
 \caption{In this plot echoes signals are injected into Gaussian noise and are analysed for different overall amplitudes A \cite{Westerweck:2017hus}. Injection amplitudes above $10^{-22}$ of the peak strain, can be accurately recovered. For the injections with lower amplitudes no matter what value they take, recovered amplitudes are likely to be around $10^{-22}$.  Horizontal lines show the amplitudes reported in \cite{Abedi:2016hgu}. The injections are made with $\gamma = 0.8$.} \label{Julian3}
\end{figure}



\end{enumerate}

\subsubsection{Comment on: "Low significance of evidence for BH echoes in gravitational wave data" \cite{Westerweck:2017hus}}

\begin{enumerate}
\item \textit{Comment on search strategy}:

As described in former part in Fig. \ref{Julian2}, Westerweck et al. \cite{Westerweck:2017hus} demonstrated the preference of $\gamma=1$ at lower injection amplitudes for ADA model \cite{Abedi:2016hgu}. This is expected result as $\gamma\rightarrow 1$ extends the template to infinity. In other words, it extends the template range to infinite time, which is clearly dominated by noise.  However, $\gamma = 1$ is still far from the best-fit $\gamma = 0.9$ on edge of the prior where at least 90\% of energy goes to first 11 echoes. Besides, initial waveform must change significantly in subsequent echoes. Indeed, repeated template that does not damp ($\gamma=1$) is not physical. One solution to this problem might be to use a finite range of data (which is used by Westerweck et al. \cite{Westerweck:2017hus}) making the result strongly dependent on what portion of data has been taken.


It should be also noted that using 1-sigma range for errors in $\Delta \rm t_{echo}$, implies a 32\% chance for the signal to be missed that causes reduced significance by diluting $\rm SNR^{2}$ from some events.

Westerweck et al. \cite{Westerweck:2017hus} have found that the least significant event  LVT151012 which is now called GW151012 has the most contribution to tentative evidence for echoes. This peculiar finding does not disfavor echoes as there is no simply reasonable justification that significance of echoes should be directly related to the significance of main event. Additionally, Wang et al. \cite{Wang:2018gin} have shown that by changing only $\pm 20\%$ of frequency of initial condition of echoes the $\rm SNR^{2}$ for echoes can change by 3 orders of magnitude. As BBH events have different component spins and mass ratios we might expect a significant diversity in relative echo signal amplitude for each of them. Interestingly, as will be discussed in next part \ref{Results of Salemi et al.} mass ratio of BBH events appears to show correlation with the echo amplitude \cite{Abedi:2020sgg}.

\item \textit{Comment on abstract and conclusion}:

The most crucial comments for Westerweck et al. \cite{Westerweck:2017hus} goes to their abstract and conclusion (also provided in \cite{Abedi:2018pst}). Although ADA \cite{Abedi:2018pst} strongly acknowledge the analysis by Westerweck et al. \cite{Westerweck:2017hus}, which is a careful re-evaluation of ADA analysis, the Abstract/Conclusion of Westerweck et al. \cite{Westerweck:2017hus} misrepresents their finding. The most critical point of this misrepresentation is in Abstract claiming “a reduced statistical significance ... entirely consistent with noise”. Contrasted to this claim in their Table I (Table \ref{table_7} in this paper) they found p-value=0.020 for the noise hypothesis, with the same model and data as in ADA (as opposed to 0.011 in ADA \cite{Abedi:2016hgu}).
However, if one follows standard nomenclature (e.g., \href{https://en.wikipedia.org/wiki/P-value#Usage}{https://en.wikipedia.org/wiki/P-value\#Usage} or  \cite{Goodman}), p-values $< 0.05$ disfavour noise hypothesis, providing “moderate to strong” Bayesian evidence for echoes, which is contrary to what they state in their
Abstract.

To conclude, considering all the critiques of Westerweck et al. \cite{Westerweck:2017hus}, we see NO evidence that their improved analysis with p-value$=0.020\pm 0.009$ has reduced the significance of echoes, entirely consistent with p-value $=0.011$ of ADA \cite{Abedi:2016hgu}. The fact that completely independent events of O2 also show a low p-value$=0.039$ for ADA echoes (see Table \ref{table_4} \cite{Uchikata:2019frs}) further boosts the statistical evidence for ADA model in LIGO/Virgo data. 

\end{enumerate}

\subsubsection{\label{Results of Salemi et al.}Results of Salemi et al. \cite{Salemi:2019uea}}

Another independent group \cite{Salemi:2019uea} has found similar post-merger GW signals that can be attributed to GW echoes. However, the setup of this methodology, which is based on coherent WaveBurst (cWB) \cite{cWB} method, was not originally developed to search for echoes. This search, which is independent of the waveform models, has been developed based on coherent excess power in events from the GWTC-1 (catalog of compact binary coalescence). Here, loose bounds on the duration and bandwidth of the signal leads to evaluation of coherent response of independent detectors.

\begin{figure}[!tbp]
\centering
    \includegraphics[width=0.7\textwidth]{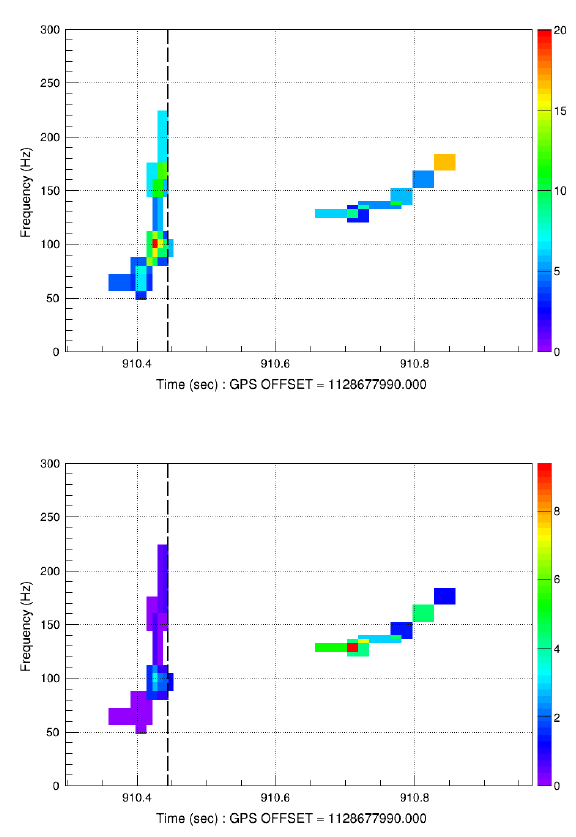}
 \caption{Reconstruction of cWB for the event GW151012 via color-coded time-frequency maps \cite{Salemi:2019uea}. The upper plot shows the squared coherent network SNR and the plot bellow shows the normalized residual noise energy. The residual plot is given after the reconstructed signal was subtracted from the data. In this plots a secondary cluster occurring 200 ms after the merger (consistent with echo times predicted and seen by ADA, Equation \ref{t_echo_meas}). The dashed vertical lines denote coalescence time for GW151012 (the network has used the Livingston detector time as a reference).\label{Salami1}}
\end{figure}

\begin{figure}[!tbp]
\centering
    \includegraphics[width=0.7\textwidth]{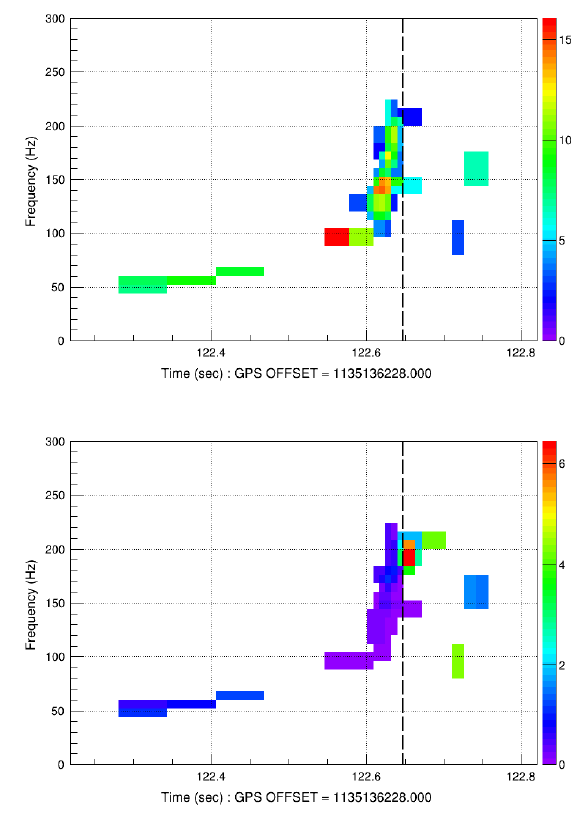}
 \caption{Reconstruction of cWB for the event GW151226, as color-coded time-frequency maps \cite{Salemi:2019uea}. The upper plot shows the squared coherent network SNR and the plot bellow shows the normalized residual noise energy. The residual plot is given after the reconstructed signal was subtracted from the data. In this plots a secondary cluster occurring 100 ms after the merger (consistent with echo times predicted and seen by ADA, Equation \ref{t_echo_meas}). The dashed vertical lines denote coalescence time for GW151226 (the network has used the Livingston detector time as a reference).\label{Salami2}}
\end{figure}

\begin{figure}[!tbp]
\centering
    \includegraphics[width=0.7\textwidth]{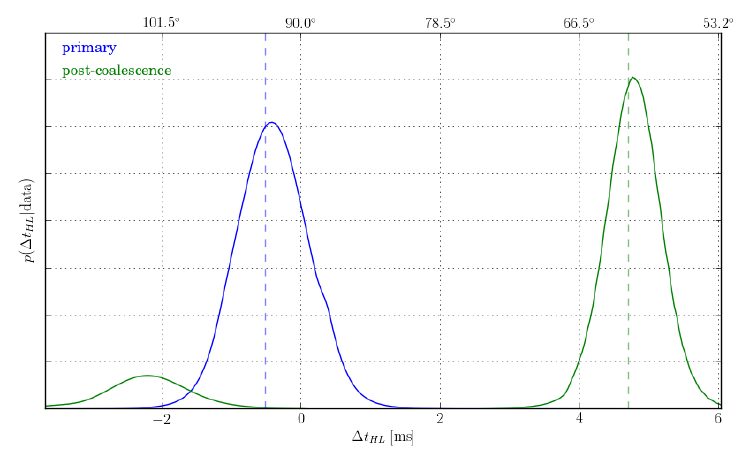}
 \caption{Maximum posteriori probability for time delay between Hanford (H) and Livingston (L) in line-of-sight frame for the main event GW151012 (blue contour) and the secondary signal (green contour) \cite{Salemi:2019uea}.\label{Salami3}}
\end{figure}

This search has focused on detected features as deviations from GR and has presented the method to obtain their significance. It appears that from eleven events reported in the GWTC-1, two of them (GW151012 and GW151226) in Figs. \ref{Salami1} and \ref{Salami2} respectively, show an excess of coherent energy after the merger ($\Delta t \simeq 0.2$ s and $\simeq 0.1$ s, respectively) with p-values (0.004 and 0.03, respectively). However, \cite{Salemi:2019uea} have shown that (Fig. \ref{Salami3}) the post-merger signal from GW151012 favours different sky location than that of the main event.

In TableI \ref{table_9}, $\rm SNR^{minR}_{pc}$, and its upper and lower bounds are presented. In the last column of this table, estimated p-values for postmerger features are reported. Along with, $2\sigma$ upper and lower bounds for SNR, upper and lower bounds for the p-values are reported. 

\begin{table*}
\begin{center}
\begin{tabular}{ |c|c|c|c|c|c|c|c| }
\hline
Event & source & SNR & SNR$_{\rm{pc}}^{\rm{minR}}\left\{^{\rm{SNR}^{\rm{sup}}_{\rm{pc}}}_{\rm{SNR}^{\rm{inf}}_{\rm{pc}}}\right\}$ & p-value$_{\rm{pc}}\left\{^{\rm{P}_{\rm{sup}}}_{\rm{P}_{\rm{inf}}}\right\}$ \\
\hline
GW150914 & BBH & 25.2 & $5.72^{6.92}_{5.64}$ & $0.94\pm 0.02^{0.95}_{0.71}$ \\
\hline
GW151012 & BBH & 10.5 & $6.60^{6.54}_{6.26}$ & $0.0037\pm 0.0014^{0.0068}_{0.0042}$ \\
\hline 
GW151226 & BBH & 11.9 & $4.40^{4.41}_{4.36}$ & $0.025\pm 0.005^{0.03}_{0.02}$ \\
\hline 
GW170104 & BBH & 13.0 & $5.29^{5.30}_{3.95}$ & $0.07\pm 0.01^{0.31}_{0.07}$ \\
\hline 
GW170608 & BBH & 14.1 & $1.69^{1.75}_{1.64}$ & $0.51\pm 0.02^{0.54}_{0.49}$ \\
\hline 
GW170729 & BBH & 10.2 & $4.81^{4.86}_{3.43}$ & $0.09\pm0.01^{0.35}_{0.08}$ \\
\hline
GW170809 & BBH & 11.9 & $3.89^{4.71}_{3.88}$ & $0.28\pm 0.01^{0.28}_{0.11}$ \\
\hline 
GW170814 & BBH & 17.2 & $5.98^{6.02}_{5.94}$ & $0.10\pm 0.01^{0.11}_{0.09}$ \\
\hline 
GW170817 & BNS& 29.3 & $0.21^{0.21}_{0.21}$ & $0.55\pm 0.01^{0.56}_{0.55}$ \\
\hline 
GW170818 & BBH & 8.6 & $1.97^{2.04}_{1.76}$ & $0.87\pm 0.02^{0.91}_{0.86}$ \\
\hline 
GW170823 & BBH & 10.8 & $3.11^{3.54}_{2.69}$ & $0.60\pm 0.02^{0.74}_{0.44}$ \\
\hline
\end{tabular}
\caption{p-values for post-coalescence deviations from GR obtained by cWB for the eleven GW events from GWTC-1 \cite{Salemi:2019uea}.  Post-coalescence SNR, SNR$^{minR}_{pc}$; and estimated probability such that SNR$^{minR}_{pc}$ produced by a noise fluctuation. Here $P_{sup}$ and $P_{inf}$ are refering to the probability of SNR$^{inf}_{pc}$ and SNR$^{sup}_{pc}$, respectively.}\label{table_9}
\end{center}
\end{table*}

\subsubsection{Hint of dependence of significance of echoes on binary BH mass ratio,\\ \textit{Comment on: Results of Salemi et al.} \cite{Salemi:2019uea}}
In this part, we first re-examine the interpretation of  Salemi et al. \cite{Salemi:2019uea} about the signals they found and then give a more conclusive support for echoes hypothesis.
\begin{itemize}
\item \textit{Comment on: Results of Salemi et al.} \cite{Salemi:2019uea}

Salemi et al. \cite{Salemi:2019uea} disfavoured echoes hypothesis pointing that post-merger signal of GW151012 has arrived from a different sky location than that of the main event. However, the p-value$\sim 0.004$ of this secondary signal, disfavours two signals being unrelated.  We see that all the secondary (post-merger) clusters they claim as signals in Figs. \ref{Salami1} and \ref{Salami2} are nearly monochromatic. That means the waveform of these signals are quasi-periodic, leading to degeneracies in inferred time-delays. So looking again to the secondary signal of GW151012, the null (residual) plot in Fig. \ref{Salami1} shows the peak of the cluster (which is mostly responsible to make a different sky localization) is at $\sim 130$ Hz which corresponds to 7.7 m sec time delay. This 7.7 m sec is the same time delay of first peak and second peak in Fig. \ref{Salami3} for post-coalescence signal (green),  confirming that the monochromatic degeneracy in Fig. \ref{Salami3} might have caused the different sky localization. Fig. \ref{Salami4} shows better interpretation of cause of this error.

\textbf{\begin{figure}[!tbp]
\centering
    \includegraphics[width=0.6\textwidth]{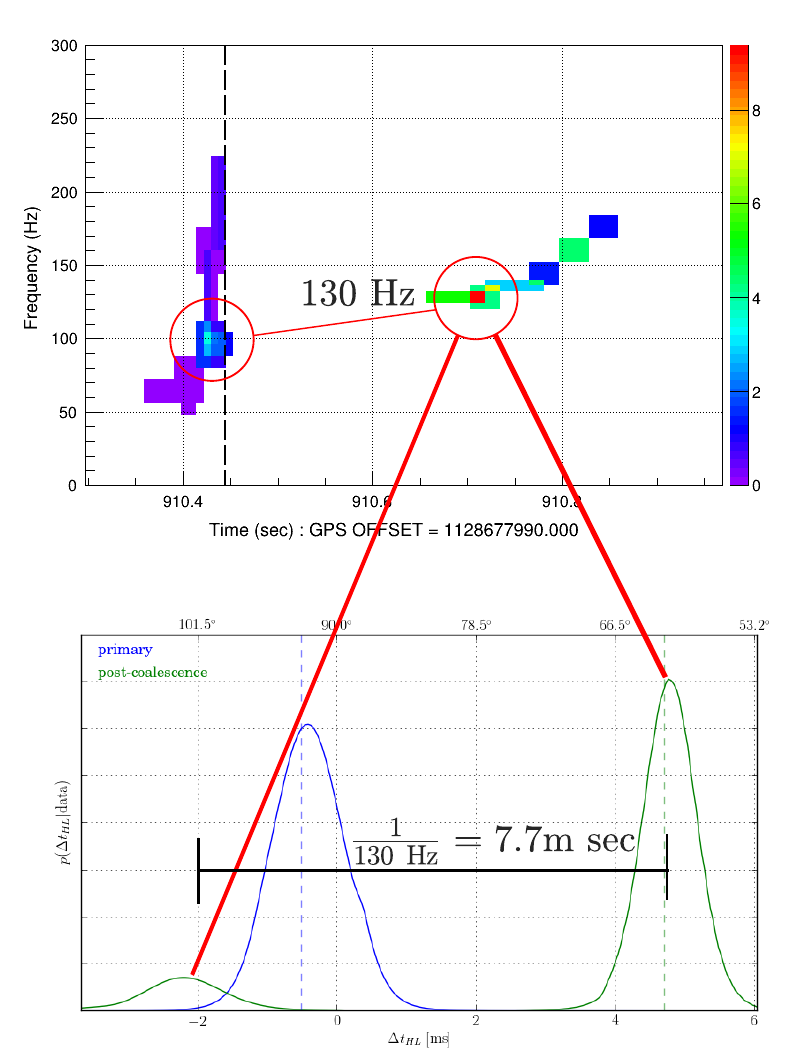}
 \caption{This plot shows how residual signal because of wavelength degeneracy can cause 7.7 m sec $=1/(130\ \rm Hz)$ shift in maximum poster probability for time delay. The upper plot is residual plot in Fig. \ref{Salami1} and the plot below is the poster probability in Fig. \ref{Salami3} for the event GW151012 \cite{Salemi:2019uea}.\label{Salami4}}
\end{figure}}



\item \textit{Dependence of significance of echoes on binary BH mass ratio}

Consider a system of binary BHs (BBH), with progenitor masses $m_{1}$ and $m_{2}$. It is known that these systems consist of two almost equal mass BHs $m_{1}\sim m_{2}$. However, the inevitable diversity in the initial conditions, specifically binary mass ratio, can lead to different echo properties. Here, we review the evidence for correlation between the significance of echoes and the progenitor BBH mass ratio (first presented in \cite{Abedi:2020sgg}): 
    \begin{enumerate}
        \item We have used LIGO parameter estimation samples for BBH events provided in \cite{GWTC-1}. Then obtained mass ratios by weighting all events as equal. We used full m1 vs m2 distribution samples for "Overall\_posterior". We fit these  posterior points to a straight line shown in Fig. \ref{mass ratio error}.
\\
        \item We take p-values reported in \cite{Salemi:2019uea} for each event post-coalescence signal and plot the best fit line of mass ratio vs $(-\log(p-value))^{0.5}$. We used least square method \cite{least-square-method} in fitting a straight line consisting all the posterior points provided by LIGO. Finally, we use the slope of best fit line as our primary parameter in determining significance.
\\
        \item Finally, we obtain the significance (Fig. \ref{histogram-slope}) assuming that there is no relation between p-value and mass ratio by taking random p-values for BBH Catalog events within the uniform range $0<(-\log(\rm{p-value}))^{0.5}<2.5$ ($0.0019<$ p-values$<1$). Indeed, having no relation between p-value and mass ratio of events shall end up with zero slope in large number of random selections. Therefore, in order to find a false detection rate for correlation, number of slopes higher than the actual measured slope is calculated (see Fig. \ref{histogram-slope}).  Accounting for the "look elsewhere" effect, we find tentative hint of mass-ratio dependence of echo significance reported in \cite{Salemi:2019uea} at false detection probability of 1\%.

    \end{enumerate}
    
    \begin{figure}
    \centering
    \includegraphics[width=0.7\textwidth]{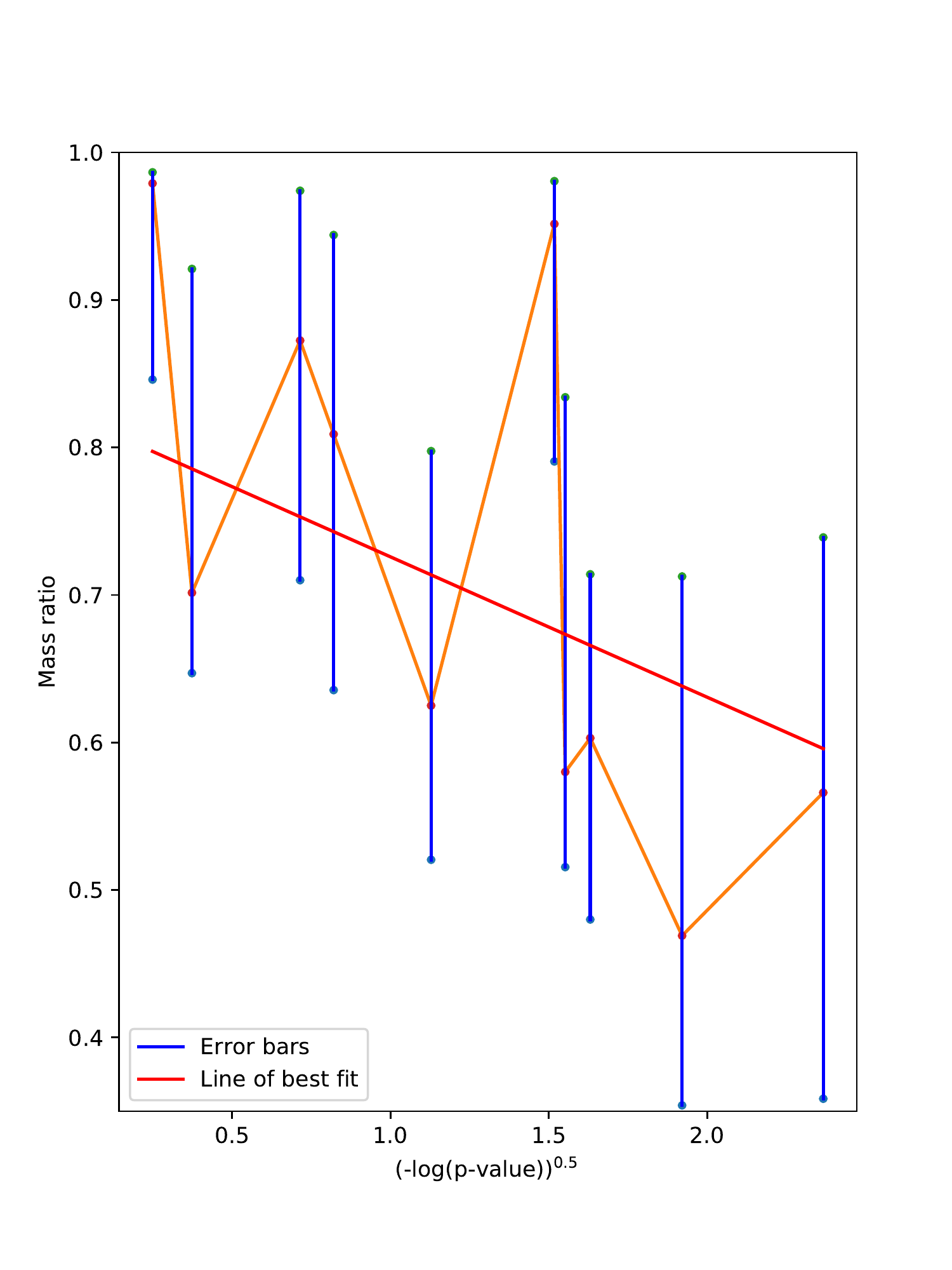}
    \caption{Plot of mass ratio dependence of p-values in \cite{Salemi:2019uea}. Vertical lines are error bars for 50\% credible region and central points are best value of mass ratio obtained from posteriors distribution. Because of relation of p-value to error function erf(SNR) we took roughly SNR $\sim \sqrt{-\log(\rm p-value)}$ as horizontal axis \cite{Abedi:2020sgg}.}
    \label{mass ratio error}
    \end{figure}

    \begin{figure}
    \centering
    \includegraphics[width=0.6\textwidth]{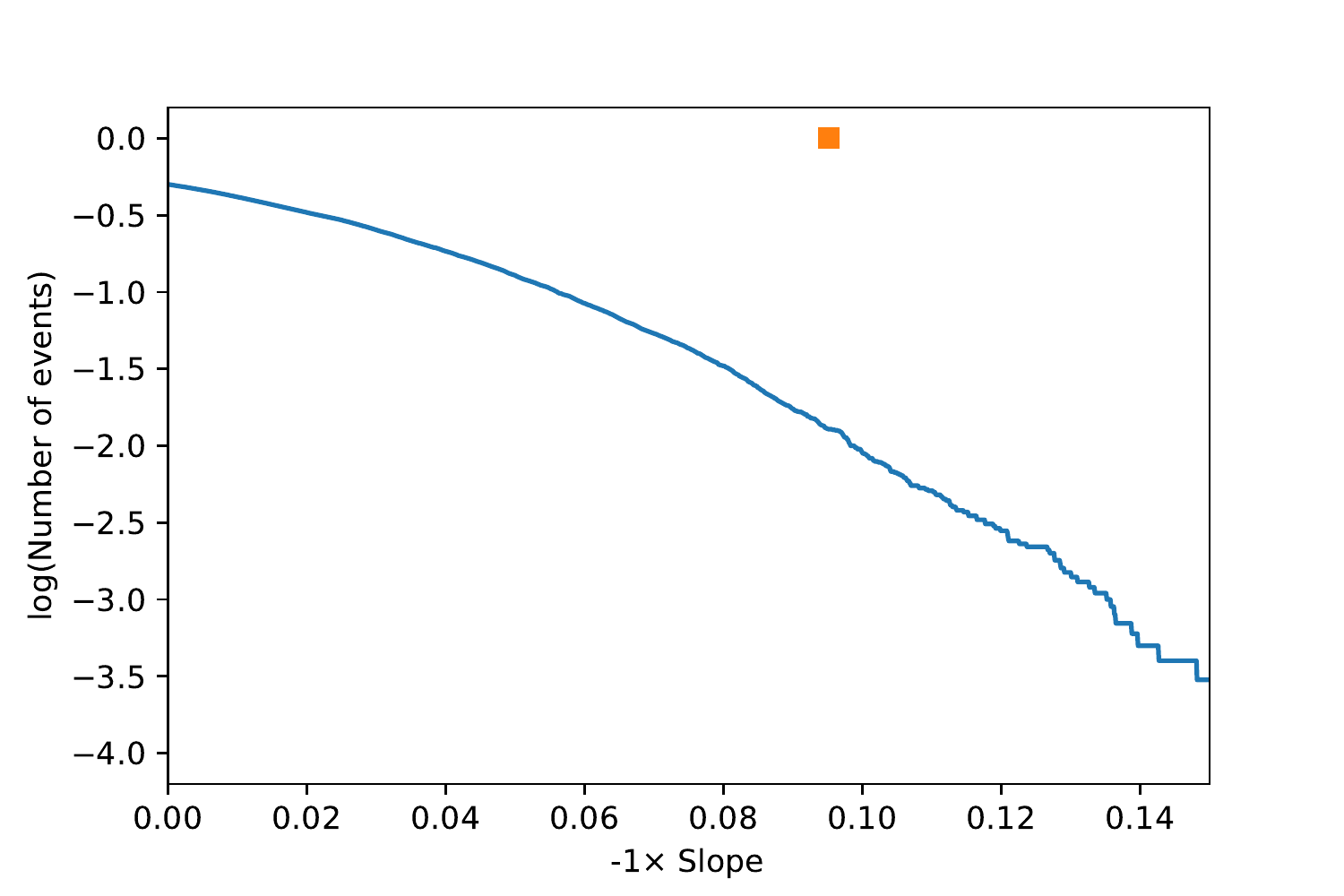}
    \caption{Histogram of slopes for uniform random choices of $0<(-\log(\rm{p-value}))^{0.5}<2.5$. We see that only 1.3\% of these random realizations, the slope can exceed the observed value \cite{Abedi:2020sgg}. }
    \label{histogram-slope}
    \end{figure}

\end{itemize}

Table \ref{table_20} shows events and p-value of their post-coalescence signal reported in \cite{Salemi:2019uea} versus expected Planckian echo time delays and average mass ratios. At a glance it is seen that smallest mass ratios go to smallest p-values. This can happen with 1/10 chance out of 10 BBH events. For two most significant events, being also the most extreme BBHs, the random chance becomes p-value$=\frac{1}{9}\times \frac{1}{10} = 0.011$ which is consistent with the statistics of regression analysis in Fig. \ref{histogram-slope}.

\begin{table*}
\begin{center}
\begin{tabular}{ |c|c|c|c|c| }
\hline
Event & p-value $\pm 2\sigma$& average mass ratio  \\
\hline
GW150914 & $0.94 \pm 0.02$& 0.86  \\
\cellcolor{blue!25} GW151012& \cellcolor{blue!25} $0.0037 \pm 0.0014$ & \cellcolor{blue!25} 0.58 \\
\cellcolor{blue!25} GW151226& \cellcolor{blue!25}$0.025 \pm 0.005$ & \cellcolor{blue!25}0.56 \\
GW170104& $0.07 \pm 0.01$& 0.65 \\
GW170608& $0.51 \pm 0.02$& 0.68 \\
GW170729& $0.09 \pm 0.01$& 0.68 \\
GW170809& $0.28 \pm 0.01$& 0.68\\
GW170814& $0.10 \pm 0.01$& 0.82\\
GW170818& $0.87 \pm 0.02$& 0.75\\
GW170823& $0.60 \pm 0.02$& 0.74\\

\hline
\end{tabular}
\caption{Events, p-values reported in \cite{Salemi:2019uea}, and average mass ratio of event. Highlighted rows are the most significant post-coalescence signals reported in  \cite{Salemi:2019uea}. }\label{table_20}
\end{center}
\begin{center}
\end{center}
\end{table*}

\subsection{\label{Negative}Negative Results}

\subsubsection{Template-based gravitational-wave echoes search using Bayesian model selection by Lo et al. \cite{Lo:2018sep}}
Lo et al. \cite{Lo:2018sep} found that using a wider range of priors (listed in Table \ref{table_16}) compared to that of ADA model \cite{Abedi:2016hgu}, including the main event template in the echo template, and keeping just three echoes, leads to lower significance on echo signals evaluated by the Bayes factor using Bayesian analysis.
\begin{table}
\begin{center}
\begin{tabular}{ |c|c| }
\hline
Parameter & Prior range \\
\hline
A & [0.0, 1.0] \\
\hline
$\gamma$ & [0.0, 1.0] \\
\hline
$t_{0}$ (sec) & [-0.1,0.01] \\
\hline
$t_{\rm echo}$ (sec) & [0.05, 0.5] \\
\hline
$\Delta t_{\rm echo}$ (sec) & [0.05, 0.5] \\
\hline
\end{tabular}
\caption{Prior range proposed by Lo et al. \cite{Lo:2018sep} of the echo parameters of ADA model \cite{Abedi:2016hgu}.}\label{table_16}
\end{center}
\end{table}
They have considered two hypotheses as the null hypothesis $\mathcal{H}_{0}$ and alternative hypothesis $\mathcal{H}_{1}$,
\begin{eqnarray}
\mathcal{H}_{0} \equiv \rm{No\ echoes\ in\ the\ data } \Rightarrow d=n+h_{\rm IMR} \nonumber \ \ \ \ \ \ \ \ \ \\
\mathcal{H}_{1} \equiv \rm{There\ are\ echoes\ in\ the\ data } \Rightarrow d=n+h_{\rm IMRE} \nonumber \\
\end{eqnarray}
where $d$ and $n$ indicate the GW data, and the instrumental noise respectively and $h_{\rm IMR}$, and $h_{\rm IMRE}$ are the inspiral-merger-ringdown (IMR) gravitational-wave signal and inspiral-merger-ringdown-echo (IMRE) gravitational-wave signal respectively.

The log Bayes factor $\ln \mathcal{B}$ in their search method gives the detection statistics which determines whether there is an IMRE signal or an IMR signal in data. From the Bayesian perspective if the log Bayes factor, is greater than 0, we can conclude that the data favor the alternative hypothesis.

The relation between p-value and null distribution of detection statistic $\ln \mathcal{B}$ is given by
\begin{eqnarray}
\rm p-value=Pr(\ln \mathcal{B} \geq \ln \mathcal{B}_{\rm detected}|\mathcal{H}_{0}) \nonumber \\
=1-\int_{-\infty}^{\ln \mathcal{B}_{\rm detected}} p(\ln \mathcal{B}|\mathcal{H}_{0}) d\ln \mathcal{B},
\end{eqnarray}
where $\ln \mathcal{B}_{\rm detected}$ is the detection statistic which is obtained using a segment of data in analysis, and $p(\ln \mathcal{B}|\mathcal{H}_{0})$ is called the null distribution of $\ln \mathcal{B}$, i.e., the distribution of $\ln \mathcal{B}$ assuming that $\mathcal{H}_{0}$ is true.

Lo et al. have injected an IMRE injection of template with echo parameters discussed earlier for the event GW150914 into simulated Gaussian noise. The detection statistic for this injection is given as follows,
\begin{eqnarray}
\ln \mathcal{B}_{\rm detected,Gaussian}=-0.2576<0.
\end{eqnarray}
Therefore, this finding indicates that the data slightly favor the null hypothesis from Bayesian analysis point of view. While the p-value and the corresponding statistical significance, for Gaussian noise, show that the data favors echo hypothesis,
\begin{eqnarray}
{\rm p-value} = 0.01275,\nonumber \\
{\rm statistical~ significance} = 2.234~  \sigma .
\end{eqnarray}

Table \ref{table_17} shows the values of the detection statistic $\ln \mathcal{B}$ vs corresponding statistical significance in Gaussian and O1 backgrounds. Therefore, for a detection of gravitational-wave echoes having statistical significance $\geq 5\sigma$, the detection threshold would be,
\begin{eqnarray}
\ln \mathcal{B}_{\rm threshold,Gaussian}=1.9,\nonumber \\
\ln \mathcal{B}_{\rm threshold,O1}=5.7,
\end{eqnarray}
for Gaussian noise and O1 noise respectively.
\begin{table}
\begin{center}
\begin{tabular}{ |c|c|c| }
\hline
Statistical significance & Detection statistic (Gaussian noise) &  Detection statistic (O1 noise) \\
\hline
$1\sigma$ & -0.9 & 0.1 \\
\hline
$2\sigma$ & -0.4 & 1.5 \\
\hline
$3\sigma$ & 1.1 & 4.0 \\
\hline
$4\sigma$ & 1.5 & 5.4 \\
\hline
$5\sigma$ & 1.9 & 5.7 \\
\hline 
\end{tabular}
\caption{ Detection statistic $\ln \mathcal{B}$ vs its corresponding statistical significances shown for both Gaussian and O1 backgrounds \cite{Lo:2018sep}.}\label{table_17}
\end{center}
\end{table}

Table \ref{table_18} shows the detection statistics and the corresponding statistical significance and p-value for the O1 events. This table also shows that the ordering of the events by their statistical significance is consistent with what has been reported by Nielsen et al.  \cite{Nielsen:2018lkf}.

\begin{eqnarray}
\ln \mathcal{B}_{\rm O1}^{\rm (cat)}=-1.1,
\end{eqnarray}

\begin{table}
\begin{center}
\begin{tabular}{ |c|c|c|c| }
\hline
Event & Detection statistic & p-value & Statistical significance $(\sigma)$ \\
\hline
GW150914 & -1.3 & 0.806 & $<1$\\
\hline
GW151012 & 0.4 & 0.0873 & 1.4 \\
\hline
GW151226 & -0.2 & 0.254 & $<1$ \\
\hline
\end{tabular}
\caption{ The detection statistic and its corresponding statistical significance and p-value for O1 events \cite{Lo:2018sep}. The ordering of events by their statistical significance is consistent with what reported by Nielsen et al.  \cite{Nielsen:2018lkf}}\label{table_18}
\end{center}
\end{table}

\subsubsection{Comments on Lo et al. \cite{Lo:2018sep}}

In their analysis, Lo et al. \cite{Lo:2018sep} have included both the main event, as well as the ADA echo waveform in  their template, but they used expanded priors in Table \ref{table_16}. Although expanding the ADA priors covers a larger space of possibilities, it tends to dilute marginal signals and bury them in the noise. For example, there is no good physical interpretation for repeating echoes that do not damp (with $\gamma=1$), as they violate energy conservation.

For these reasons, it is not surprising that Lo et al. \cite{Lo:2018sep} find a smaller Bayes factors, due to their expanded priors. However, it is well-known that expanding prior into nonphysical regimes will artificially lower Bayesian evidence for any model, especially since $\gamma=1$ is a (formal) singularity of the likelihood function.

\subsubsection{\label{Uchikatanegative}Results of Uchikata et al. \cite{Uchikata:2019frs}}

As discussed in Section \ref{Uchikata} above,  Uchikata et al. \cite{Uchikata:2019frs} can approximately reproduce the evidence for ADA echoes in both O1 and O2 events. Here, we present their search using an alternative template that failed to find any evidence for echoes. In order to build the latter, they considered Kerr spacetime, replacing the event horizon with a reflective membrane. They then used the transmissivity of the Kerr angular momentum barrier ${\cal T}_{\rm BH}(\omega)$ to filter the ADA template, which acts as a hi-pass filter (truncating the low frequency part of the ADA phenomenological waveform \ref{template}). Moreover, the overall phase shift of the waveform as a free parameter is taken into account contrary to ADA search. Using this template, they have found no significant echo signals in the binary BH merger events. The background estimation, has used the same method provided by Westerweck et al. \cite{Westerweck:2017hus}. 

\begin{enumerate}

\item The results in Table \ref{table_10} gives p-values for all events. The combined p-value is well above the critical p-value 0.05. In other words, echo signals using this model do not exist in the data, or their amplitudes are too small to be detected within the current detector sensitivity.

\begin{table}
\begin{center}
\begin{tabular}{ c }
\ \ \ \ \ \ \ \ \ \ \ \ \ \ \ Data version \\
\end{tabular}\\
\begin{tabular}{ |c|c|c| }
\hline
Event & C01 & C02 \\
\hline
GW150914 & 0.992 & 0.984 \\
\hline
GW151012 & 0.646 & 0.882 \\
\hline
GW151226 & 0.276 & - \\
\hline
GW170104 & 0.717 & 0.677 \\
\hline
GW170608 & - & 0.488 \\
\hline
GW170729 & - & 0.575 \\
\hline 
GW170814 & - & 0.472 \\
\hline 
GW170818 & - & 0.976 \\
\hline 
GW170823 & - & 0.315 \\
\hline 
Total & 0.976 & 0.921 \\
\hline 
\end{tabular}
\caption{Obtained P-values for each event along with total p-value \cite{Uchikata:2019frs}. A hyphen means that 4096-second of data are not available.
}\label{table_10}
\end{center}
\end{table}

\item Since the phase shift at the membrane (due to the reflection and boundary condition) is model dependent, it is physically reasonable to assume a total phase shift as a parameter (see Fig. \ref{Uchikata1} that has used extra phase parameter for GW150914). In contrast, former studies \cite{Abedi:2016hgu,Westerweck:2017hus} only considered phase inversion (or Dirichlet boundary conditions) at the reflective  membrane. Therefore, the results of two cases, when the phase shift is fixed to $\pi$ (result 1) and when it is a free parameter (result 2), respectively, in Table \ref{table_11} are compared. In this table we see that, p-values become slightly larger (taking GW151226 as an exception) when the phase shift due to the reflection has taken as a free parameter.

\begin{table}
\begin{center}
\begin{tabular}{ |c|c|c| }
\hline
Event & Result 1 & Result 2 \\
\hline
GW150914 & 0.638 & 0.992 \\
\hline
GW151012 & 0.417 & 0.646 \\
\hline
GW151226 & 0.953 & 0.276 \\
\hline
GW170104 & 0.213 & 0.717 \\
\hline
Total & 0.528 & 0.976 \\
\hline 
\end{tabular}
\caption{ P-value for each event and total p-value \cite{Uchikata:2019frs}. Result 1 is the case when the phase shift is fixed to $\pi$, and result 2 is the case when the total phase shift is also a parameter.}\label{table_11}
\end{center}
\end{table}

\end{enumerate}

\subsubsection{Comment on Negative results of Uchikata et al. \cite{Uchikata:2019frs}}

Model provided by Uchikata et al. \cite{Uchikata:2019frs} substantially truncates the low frequency part of the GR waveform (which is the basis of ADA template \ref{template}). However, one may argue that the GR waveform has already been filtered once by the transmissivity of the angular momentum barrier, ${\cal T}_{\rm BH}(\omega)$ as it is what is seen by observers at infinity. In fact, \cite{Wang:2019rcf}  have shown that both GR signal (main event) and echoes can be constructed from a superposition of the QNMs of the quantum BH, which are essentially the modes trapped between the angular momentum barrier and the quantum membrane (see Section \ref{subsec_ergoinst} above).  Therefore, both the GR signal and the echoes pass through same barrier and are thus truncated by the same ${\cal T}_{\rm BH}(\omega)$. This implies that ${\cal T}_{\rm BH}(\omega)$ cancels in the ratio of echo to main signal waveform, and in contrast to Uchikata et al. \cite{Uchikata:2019frs}, no truncation is needed.    

Indeed, as evidenced by their own analysis (Section \ref{Uchikata} above), the low-frequency part of the ADA template is necessary to obtain a significant signal. This is physically justified since the high frequencies leak out of the angular momentum barrier quickly, leading to a rapid decay, while echoes can last much longer at lower frequencies \cite{Wang:2018gin}. 

\subsubsection{\label{Results of Tsang et al.}Results of Tsang et al. \cite{Tsang:2019zra}}

Tsang et al. \cite{Tsang:2018uie,Tsang:2019zra}  proposed a morphology-independent search method which consists of a large number of free parameters for echoes compared to ADA model \cite{Abedi:2016hgu} (49 versus 5). They search for echoes in all the significant events in (GWTC-1), and found that for all the events, the ratios of evidences for signal versus noise and signal versus glitch do not rise above their respective background. Only the smallest p-value=3\% goes to the event GW170823. Hence they found no significant evidence for echoes in GWTC-1. The results of search are given in Table \ref{table_12} and \ref{table_13}.

\begin{table}
\begin{center}
\begin{tabular}{ |c|c|c|c|c| }
\hline
Event & Log $B_{S/N}$ & $p_{S/N}$ & Log $B_{S/G}$ & $p_{S/N}$ \\
\hline
GW150914 & 2.32 & 0.26 & 2.95 & 0.43  \\
\hline
GW151012 & -0.59 & 0.70 & 0.35 & 0.88 \\
\hline 
GW151226 & -0.67 & 0.72 & 2.48 & 0.53 \\
\hline 
GW170104 & 1.09 & 0.44 & 3.80 & 0.28 \\
\hline 
GW170608 & -0.90 & 0.75 & 0.90 & 0.82 \\
\hline 
GW170823 & 6.11 & 0.03 & 5.29 & 0.11 \\
\hline 
Combined & - & 0.34 & - & 0.57 \\
\hline 
\end{tabular}
\caption{Log Bayes factors for signal versus noise and signal versus glitch, and the corresponding p-values, for events seen in two detectors of GWTC-1 \cite{Tsang:2019zra}. The bottom row shows the combined p-values for all these events together.}\label{table_12}
\end{center}
\begin{center}
\end{center}
\end{table}

\begin{table}
\begin{center}
\begin{tabular}{ |c|c|c|c|c| }
\hline
Event & Log $B_{S/N}$ & $p_{S/N}$ & Log $B_{S/G}$ & $p_{S/N}$ \\
\hline
GW170729 & 4.24 & 0.67 & 5.64 & 0.62  \\
\hline
GW170809 & 9.05 & 0.31 & 12.69 & 0.09 \\
\hline 
GW170814 & 8.75 & 0.33 & 8.54 & 0.34 \\
\hline 
GW170817 & 11.05 & 0.19 & 10.30 & 0.20 \\
\hline 
GW170817$+1s$ & 6.19 & 0.52 & 9.39 & 0.27 \\
\hline 
GW170818 & 10.39 & 0.23 & 9.36 & 0.27 \\
\hline 
Combined & - & 0.47 & - & 0.22 \\
\hline 
\end{tabular}
\caption{Same as Table \ref{table_12}, while for the events that three detectors has been involved \cite{Tsang:2019zra}. In the case of GW170817, in order to cover 1.0 sec after the merger for echoes found by Abedi and Afshordi \cite{Abedi:2018npz}, additional search as first echo being from this time has been set. For this particular event, latter prior choice has been taken for combined p-values.}\label{table_13}
\end{center}
\begin{center}
\end{center}
\end{table}

\subsubsection{Comment on: Results of Tsang et al. \cite{Tsang:2019zra}}

Tsang et al. \cite{Tsang:2019zra} have developed  a model which consists of a large number of free parameters (49 by our count). Indeed, a larger space of possiblities leads to lower significance. So it is not surprising to get large p-values out of large free-parameter space. Indeed, all the SNRs for echoes reported in ADA would be below the detection threshold reported by \cite{Tsang:2018uie}, given their large number of parameters (see \cite{Abedi:2020sgg} for more detail).

\subsection{A concordant picture of Echoes}

In order to have an optimal search for echoes, one may want to take the following guidelines into consideration:
\begin{enumerate}

\item Have a good physical model (or you will not find them!)

\item Use a simple template (avoid too many arbitrary choices)

\item Avoid {\it a posteriori} statistics (don’t look at data to make your model)

\end{enumerate}

Based on the positive (Section \ref{Positive}),  mixed (Section \ref{Mixed}), and negative (Section \ref{Negative}) results, one may offer the following general observations:  
\begin{enumerate}

\item Coherent searches appear to give more significant evidence for echoes

\item Template searches can find evidence for echoes, if they include lower frequencies

\item Models with a large number of free parameters and/or wider priors can weaken the echoes below the detection threshold

\end{enumerate}

An executive summary of these observations is shown in Tables \ref{table_positive} and \ref{table_fail} as positive evidence (p-value$\leq 0.05$) and failed results, respectively.
\begin{table*}[htbp]
\begin{center}
\begin{tabular}{ c c c c c }
  & \cellcolor{yellow}Authors & \cellcolor{yellow}Method & \cellcolor{yellow}Data & \cellcolor{yellow}p-value \\
\hline
1 & \cellcolor{green}{Abedi, Dykaar, Afshordi (ADA) 2017} \cite{Abedi:2016hgu} & \textcolor{blue}{ADA template} & \textcolor{blue}{O1} & \textcolor{blue}{1.1\%} \\
\hline
2 & \cellcolor{green}{Conklin, Holdom, Ren 2018} \cite{Conklin:2017lwb} & \textcolor{blue}{spectral comb} & \textcolor{blue}{O1+O2} & \textcolor{blue}{0.2\% - 0.8\%} \\
\hline
3 & \cellcolor{green}{Westerweck, et al. 2018} \cite{Westerweck:2017hus} & \textcolor{blue}{ADA template} & \textcolor{blue}{O1} & \textcolor{blue}{2.0\%} \\
\hline
4 & \cellcolor{green}{Nielsen, et al. 2019} \cite{Nielsen:2018lkf} & \textcolor{blue}{ADA+Bayes} & \textcolor{blue}{GW151012, GW151226} & \textcolor{blue}{2\%} \\
\hline
5 & \cellcolor{green}{Uchikata, et al. 2019} \cite{Uchikata:2019frs} & \textcolor{blue}{ADA template} & \textcolor{blue}{O1} & \textcolor{blue}{5.5\%} \\
\hline
6 & \cellcolor{green}{Uchikata, et al. 2019} \cite{Uchikata:2019frs} & \textcolor{blue}{ADA template} & \textcolor{blue}{O2} & \textcolor{blue}{3.9\%} \\
\hline
7 & \cellcolor{green}{Salemi, et al. 2019} \cite{Salemi:2019uea} & \textcolor{blue}{coherent WaveBurst} & \textcolor{blue}{GW151012, GW151226} & \textcolor{blue}{0.4\%,3\%} \\
\hline
8 & \cellcolor{green}{Abedi, Afshordi 2019} \cite{Abedi:2018npz} & \textcolor{blue}{spectral comb} & \textcolor{blue}{BNS} & \textcolor{blue}{0.0016\%} \\
\hline
9 & \cellcolor{green}{Gill, Nathanail, Rezolla 2019}  \cite{Gill:2019bvq} & \textcolor{blue}{Astro Modelling} & \textcolor{blue}{BNS EM} & \textcolor{blue}{$t_{\rm coll}=t_{\rm echo}$} \\
\hline
\end{tabular}
\caption{Table of positive results (p-value$\leq 0.05$) by different groups (The p-value for Nielsen et al. above \cite{Nielsen:2018lkf}  is a rough estimate, based on the log-Bayes $= 1.66$).}\label{table_positive}
\end{center}
\end{table*}

\begin{table*}[htbp]
\begin{center}
\begin{tabular}{ c c c c c }
  & \cellcolor{yellow}Authors & \cellcolor{yellow}Method & \cellcolor{yellow}Data & \cellcolor{yellow}possible caveat \\
\hline
1 & \cellcolor{green}{Westerweck, et al. 2018} \cite{Westerweck:2017hus} & \textcolor{blue}{ADA template} & \textcolor{blue}{O1} & \textcolor{blue}{“Infinite” prior} \\
\hline
2 & \cellcolor{green}{Nielsen, et al. 2019} \cite{Nielsen:2018lkf} & \textcolor{blue}{ADA+Bayes} & \textcolor{blue}{GW150914} & \textcolor{blue}{mass-ratio dependence} \\
\hline
3 & \cellcolor{green}{Uchikata, et al. 2019} \cite{Uchikata:2019frs} & \textcolor{blue}{ADA, hi-pass} & \textcolor{blue}{O1,O2} & \textcolor{blue}{no low-frequencies} \\
\hline
4 & \cellcolor{green}{Salemi, et al. 2019} \cite{Salemi:2019uea} & \textcolor{blue}{coherent WaveBurst} & \textcolor{blue}{O1,O2} & \textcolor{blue}{mass-ratio dependence,} \\
 &\cellcolor{green} & & & \textcolor{blue}{only 1st echo} \\
\hline
5 & \cellcolor{green}{Lo, et al. 2019} \cite{Lo:2018sep}  & \textcolor{blue}{ADA+Bayes} & \textcolor{blue}{O1} & \textcolor{blue}{“Infinite” prior} \\
\hline
6 & \cellcolor{green}{Tsang, et al. 2019} \cite{Tsang:2018uie}  & \textcolor{blue}{BayesWave} & \textcolor{blue}{O1+O2} & \textcolor{blue}{needs very loud echoes} \\
 &\cellcolor{green} & & & \textcolor{blue}{(49 free parameters!)} \\
\hline
\end{tabular}
\caption{Table of failed searches and their possible caveat.}\label{table_fail}
\end{center}
\end{table*}

The study of post-merger GW observations with the above-mentioned motivations have lead to tentative signals, at varying levels of significance, by different groups. We shall outline several similarities amongst these findings below. However, it is also important to note that these similarities do not mean that the signals found are the same, but it does provide a {\it preponderance of corroborating evidence} for GW echoes in current observations. 

\subsubsection{\label{Five independent groups, Five independent methods, identical results}Five independent groups, Five independent methods, identical results!}

\begin{enumerate}
\item In \cite{Abedi:2016hgu} (Table II), using the reported masses and spins of LIGO O1 events, the time delays of 0.1 sec and 0.2 sec for GW151226 and GW151012 were predicted respectively for Planckian echoes. These happen to be exactly the same as the times for post-merger signals found in \cite{Salemi:2019uea}.

\item Results of \cite{Abedi:2016hgu,Uchikata:2019frs,Westerweck:2017hus,Nielsen:2018lkf,Salemi:2019uea,Abedi:2018npz} all are consistent with Planckian echoes at p-values of ${\cal O}(\%)$.

\item Furthermore, the reconstructed detector responses for GW151226 and GW151012 \cite{SalemiGW151012, SalemiGW151226} in \cite{Salemi:2019uea} give consistent amplitudes (0.33, 0.34)$\times$(maximum amplitude of main event) comparing with \cite{Abedi:2016hgu} (Table II).
Energy reported in \cite{Abedi:2016hgu} (Appendix A) is also consistent with strength of signals found in \cite{Salemi:2019uea}. Finally, SNR reported for GW151012 in \cite{Abedi:2016hgu} (Table II and Fig. 6) has highest value which is also consistent with highest significance event in \cite{Salemi:2019uea}.

\item  Log Base factor values in Table II of \cite{Nielsen:2018lkf}, where they found positive evidence for ADA echoes in GW151012 and GW151226 (where GW151012 is more significant) is consistent with significance of signals found in \cite{Salemi:2019uea}.

\item Also note that the echo signal of GW150914 \cite{Abedi:2016hgu} at time delay 0.3 sec had narrowest time window ($\pm 3\%$ in Table II) and smallest energy (Table II) compared to GW151226 and GW151012, which could explain its absence in \cite{Salemi:2019uea}, and no evidence (negative Log Base factor value) in \cite{Nielsen:2018lkf} Table II.

\item Nevertheless, the residual signal in \cite{SalemiGW150914} which is a supporting results for \cite{Salemi:2019uea} is consistent with 300 m sec echo signal time delay in \cite{Abedi:2016hgu} (table II).

\item The percent-level evidence of Uchikata et al. \cite{Uchikata:2019frs} for ADA echoes in O1 events (shown in Table \ref{table_3}) are consistent with the results of other groups \cite{Abedi:2016hgu,Westerweck:2017hus,Nielsen:2018lkf,Salemi:2019uea}.

\item The results of Uchikata et al. \cite{Uchikata:2019frs} for O2 events shown in Table \ref{table_4} are given $\sim 4\%$ overall p-value which are small as O1 events. 

\item Lo et al. \cite{Lo:2018sep} Table \ref{table_18} by adding main event to the ADA waveform and keeping only three echoes with larger prior ranges also found similar ordering of events by their statistical significance with what reported by Nielsen et al.  \cite{Nielsen:2018lkf}.

\end{enumerate}

\subsubsection{Other findings}
Here we present similarities found for echoes in binary neutron star merger GW170817:
\begin{enumerate}
\item We note that the time-scale of 1.0 sec after merger for collapse into BH (first reported by Abedi and Afshordi \cite{Abedi:2018npz}) is now also independently found from purely Astrophysical considerations by \cite{Gill:2019bvq}, who found $t_{\rm coll}=0.98^{+0.31}_{-0.26}$ second.

\item Along with echo signal found by Abedi and Afshordi \cite{Abedi:2018npz} another group \cite{Conklin:2017lwb} claimed evidence for an echo frequency of $f'_{\rm echo} \simeq (0.00719 ~{\rm sec}) ^{-1}= 139$ Hz for GW170817, with a p-value of $1/300$. Noting the proximity of this value to the second harmonic of Abedi and Afshordi finding with echo frequency $2\times f_{\rm echo} = 144$ Hz, it is feasible that the two different methods are seeing (different harmonics of) the same echo signal. However, the method applied in  \cite{Conklin:2017lwb} is sub-optimal, as they whiten (rather than Wiener filter in \cite{Abedi:2018npz}) the data, and thus could underestimate the significance of the correlation peak they found (see \cite{Abedi:2018pst} for further discussion).

\end{enumerate}

To contrast, let us point out two apparent inconsistencies:
\begin{enumerate}

\item The results of Conklin et al. \cite{Conklin:2017lwb,Holdom:2019bdv} are finding non-Planckian echoes signal which might not be consistent with the results of other groups \cite{Abedi:2016hgu,Westerweck:2017hus,Nielsen:2018lkf,Salemi:2019uea,Abedi:2018npz}. However, it may not be appropriate to do a one-to-one comparison of  \cite{Conklin:2017lwb,Holdom:2019bdv} to other studies, as the employed method is significantly different. 

\item Results of Salemi et al. \cite{Salemi:2019uea} show that using their reconstruction of cWB, the post merger signal of GW151012 seen in Fig. \ref{Salami1} appear to come from a different sky location, compared to the main event signal. However, as we discussed above (Section \ref{Results of Salemi et al.}), this might be due to time-delay degeneracy in the cWB monochromatic signals. 

\end{enumerate}

\subsubsection{Independent confirmations of model predictions}
Here, we outline model predictions that have been confirmed using independent data by independent groups
\begin{enumerate}

\item{\textit{Binary BH mergers:}}

Uchikata et al. \cite{Uchikata:2019frs} have used ADA model \cite{Abedi:2016hgu} to search for echoes for both O1 (first observing run) and O2 (second observing run) while the original search of ADA only covers O1. The results for O2 (with p-value=0.039) in Tables \ref{table_3} and \ref{table_4} show similar evidence as O1 (with p-value=0.055).

\item{\textit{Binary neutron star merger:}}

After detection of echoes signal with $4.2\sigma$ significance around 1 sec after BNS merger GW170817 \cite{Abedi:2018npz} where Abedi and Afshordi claim that it has collapsed to BH at this time, Gill et al. \cite{Gill:2019bvq} with independent Astrophysical consideration have also determined that the remnant of GW170817 must have collapsed to a BH after $t_{\rm coll}=0.98^{+0.31}_{-0.26}$ sec. Error-bar for this observation compared to the detected signal of echoes by Abedi and Afshordi as a consequence of BH collapse is shown in Fig. \ref{NS-NS_9}.

\end{enumerate}

\subsubsection{Concerns about ADA searches}

\begin{enumerate}

\item \textit{Concerns about errors in $\Delta t_{echo}$}: The original ADA search \cite{Abedi:2016hgu} had used an ad-hoc method for finding symmetric 1-sigma errors for $\Delta t_{echo}$, which would miss $\sim 1/3$ of Planckian echoes.  Furthermore, the actual LIGO posteriors for these parameters (which are now publicly available, even though they were not at the time), are not Gaussian. A fully Bayesian Bayesian search, using actual mass and spin posteriors would avoid these short-comings.


\item \textit{Concerns about keeping $t_{0}=-0.1$ and $\gamma=0.9$ fixed}: Uchikata et al. \cite{Uchikata:2019frs} have fixed $t_{0}$ at its best fit value of O1 in order to search in O2. It might be a good idea to keep these parameters at their best-fit values (from O1 or O2), to make perform more efficient searches in e.g., O3 events. 

\item \textit{Concerns about mass ratio dependence of echoes overall amplitude}: As pointed out in Section \ref{Results of Salemi et al.} (Fig. \ref{mass ratio error}), it appears that the amplitudes of echo signals may depend on the BBH mass ratio, which should be taken into account if one wants to optimally combine echo signals in different events.

\end{enumerate}

\subsubsection{On negative GW Echo searches}
For attempts that fail to yield any evidence for echoes \cite{Tsang:2018uie,Uchikata:2019frs}, we again point out that an optimal echo search should us a simple model with minimum number of free parameters. The current positive results turn out to be weak signals with SNR$\sim 4$, which is below the threshold for those searches that consist of many free parameters ($SNR>8$ for 49 free parameters).
In addition, negative results of Uchikata et al. \cite{Uchikata:2019frs} indicate that the echo signals found in \cite{Abedi:2016hgu} mostly consist of low frequency modes, which is independently confirmed in  \cite{Abedi:2018npz}, and Uchikata et al's own search in both O1 and O2, using the original ADA template.

\subsubsection{Non-Gaussianity of backgrounds}

In order to search for signals of a given echo model, we need a proper understanding of background behaviour. Only then we might be able to determine the best statistical methodology. Since ADA \cite{Abedi:2016hgu} used different parts of LIGO data to get a combined significance, one may think about what would be the best method of combination of separate sets of data with different background behaviour. Then, we can also ask how the search changes by including three or higher number of detectors.

\begin{figure}
  \centering
    \includegraphics[width=0.5\textwidth]{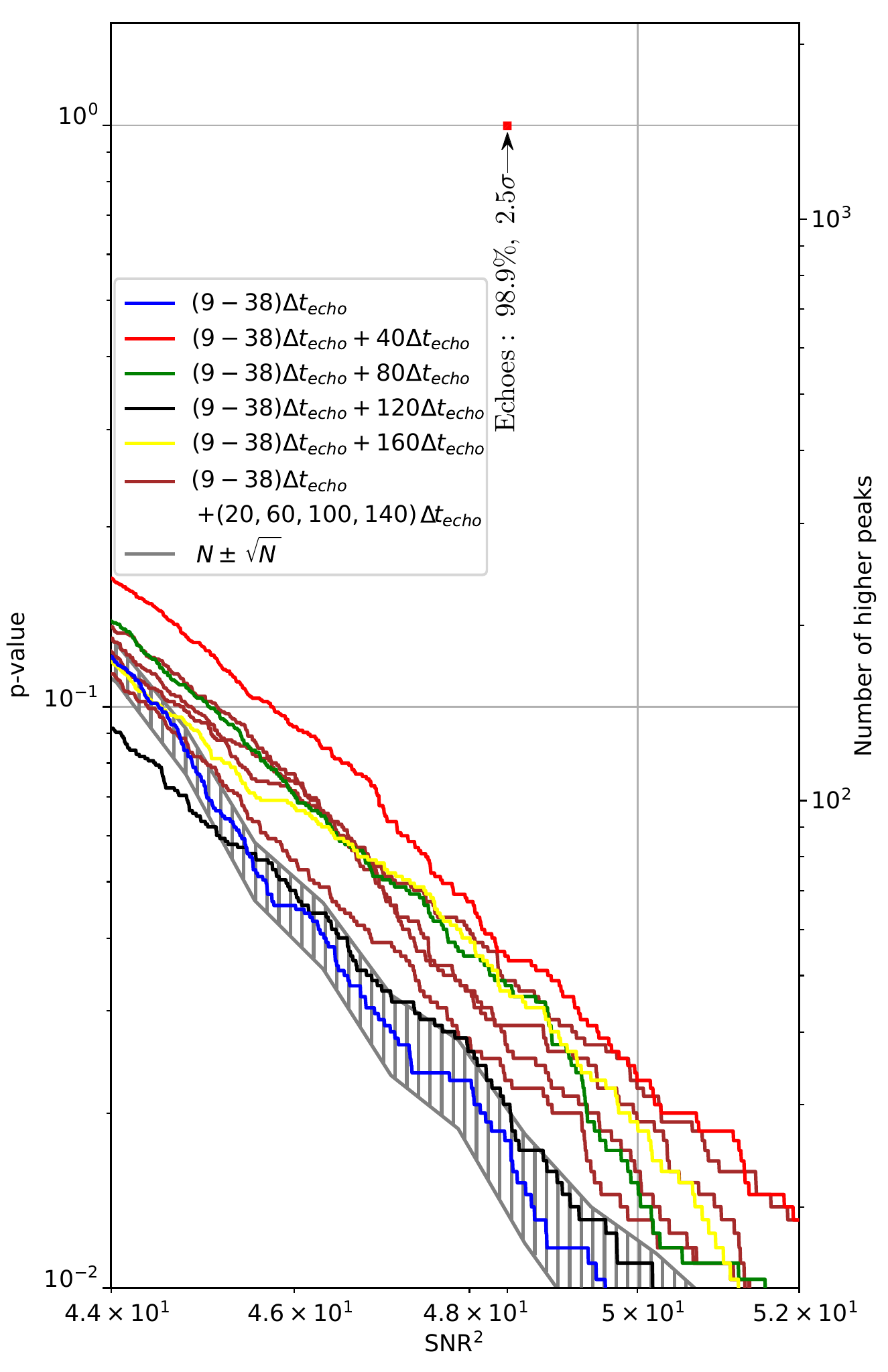}
    \caption{p-value distribution for combined events of different stretches of data within 1 minute of the main events. Surprisingly, the blue line which is closest to the main event, and has used to define p-value in \cite{Abedi:2016hgu} (Fig.  \ref{Histogramloglog}), happens to give the smallest p-value. The shaded region represents the Poisson error range for blue histogram. This shows that the variation in p-values is clearly much larger. This behaviour is interpreted as non-gaussianity and/or non-stationarity of the LIGO noise. In this plot the y-axis on the left (right) shows p-value (number of higher peaks) within the mentioned range of data. In each histogram the total number of ``peaks'' is $(38-9)/0.02= 1450$.}
 \label{Histogram-non-gaussian}
\end{figure}

It was already observed that LIGO noise vary significantly and is very non-gaussian over long time-scales (see Fig's 14-15 in \cite{TheLIGOScientific:2016src}). This either non-stationary or non-Gaussian background makes the interpretation of p-value ambiguous, particularly in finding marginal echo signals which are often near the detection threshold. Therefore, one must examine how much this varying background affects the inferred significance of a detection. This is studied by looking at other different minute-long stretches of data within a minute of the main events \cite{Abedi:2016hgu}. As can be seen in Fig. \ref{Histogram-non-gaussian}, the variation of p-values at the tail of the distribution is much higher than what is expected from Poisson statistics of the SNR peaks. This becomes more interesting when we see that the smallest p-value is coming from the range which is closest to the main event. This might be because  the marginal LVT151012 (now called GW151012) detection is over a minimum of the LIGO (combined) detector noise.
\newpage
\section{Future Prospects}
\label{sec:future_prospects}
\subsection{Towards Synergistic Statistical Methodologies}

As we summarized in the previous sections, the past fours years have witnessed hundreds of theoretical studies focusing on model-building for echoes, as well as dozens of observational searches and statistical methodologies. However, in spite of remarkable progress on both fronts, the theoretical and observational tracks have largely developed independently. However, it appears that both tracks have become mature enough, so much so that the time is ripe for a synergistic convergence. For example, Bayesian methods developed in \cite{Nielsen:2018lkf, Lo:2018sep} applied to a superposition of QNMs of quantum BHs (as outlined in \cite{Wang:2019rcf}) would put coherent methods developed by \cite{Conklin:2017lwb,Abedi:2018pst} on more sound statistical {\it and} physical footings. The analogy will be with helio- or astro-seismology, where modeling a dense spectrum of QNM frequencies can be used to infer the intrenal structure of the compact objects \cite{Oshita:2019seis1}.

The real challenge will be in allowing enough freedom in our best physical models, in order to capture all the remaining theoretical uncertainties, {\it but not any more!}

\subsection{\label{numerical} Echoes in Numerical Relativity}

Most studies of echoes have so far focused on the linear perturbation theory around the final BH for simplicity, but in reality the mergers start with the highly nonlinear binary BH inspiral. Hence, we need a covariant numerical implementation of binary quantum BHs within a highly-nonlinear dynamical spacetime to fully address the entire dynamics, especially the initial conditions. There are several possible approaches borrowed from numerical relativity which can be modified to either include the quantum boundary condition or the full dynamics of binary quantum BHs.

For instance, the effective one body (EOB)  formalism \cite{Buonanno:1998gg, Damour:2008yg} is a concrete strategy which only needs to solve ordinary differential equations rather than to perform the costly 3d numerical relativity simulations. It uses higher-order post-Newtonian expansion in a resummed form (different from the usual the Taylor-expansion), to include the non-perturbative result using a conservative description of binary BHs dynamics, radiation-reaction and emitted GW waveform. One possible approach, that is currently underway, is to capture the nonlinear effects in echoes by modifying the boundary condition in the EOB codes to implement the quantum BH dynamics.

Another route is to directly modify numerical relativity codes that have successfully produced waveforms for BBH merger events.  A concrete strategy could be incorporating the mock fuzzball energy-momentum tensor (Section \ref{sec3}) as a source for Einstein equations, directly into the numerical relativity codes. If the fuzzball ``fluid'' manages to stay just outside the apparent BH horizons in a dynamical setting, then it can potentially generate echoes in a fully nonlinear numerical simulation of quantum BBH merger. 

Recently, \cite{Okounkova:2019dfo,Okounkova:2019zjf} presented the first numerical simulation of BBH mergers in Chern-Simon gravity. They start with the modified action and predict the dynamics order by order. It is possible that a similar iterative approach can be applied to model boundary conditions at apparent horizons, or evolution of mock fuzzballs. 

\subsection{Quantum Gravity, Holography, and Echoes}

As we discussed in Section \ref{sec:echo_predictions} above, any modification of event horizons that could lead to echoes should be a non-perturbative modification of general relativity, and can only be fully captured by a non-perturbative description of quantum gravity. A possible example of this is the fuzzball program in string theory (Section \ref{sec:QBHs} above). But more generally, what can non-perturbative approaches to quantum gravity tell us about BH echoes? 

One of our greatest insights into the dynamics of quantum gravity has come from the Holographic Principle, that extending Bekenstein-Hawking area law for entropy of BHs \cite{Bekenstein:1973ur}, suggests the entire dynamics of a quantum gravitational system should be captured on its boundary. The most concrete realization of this principle was proposed by Juan Maldacena \cite{Maldacena:1997re}, in the form a conjectured duality between quantum gravity in  
Anti-de Sitter (AdS) spacetime and a Conformal field theory (CFT), commonly known as AdS/CFT correspondence or conjecture.  It proposes that CFT in spacetime of d-1 dimension, at the asymptotic boundary of an AdS spacetime is mathematically equivalent to string theory (or quantum gravity) within the bulk AdS in d dimension. This topic has been extremely fruitful over the past two decades, offering many synergies between seemingly disparate notions in geometry and quantum information. For example, the Ryu-Takayanagi conjecture \cite{Ryu:2006ef} relates the entanglement entropy of boundary CFTs with the areas of extremal surfaces in the bulk AdS, generalizing the notion of Bekenstein-Hawking BH enetropy to arbitrary geometries. 

An intriguing connection between AdS/CFT and echoes is the appearance of echo times:
\begin{equation}
\Delta t_{\rm echo} = t_{\rm scrambling} = \frac{\ln(S_{\rm BH})}{2\pi T_{\rm H}},    
\end{equation}
as ``scrambling time'', in the AdS/CFT literature \cite{Sekino:2008he}. Here, $S_{\rm BH}$ and $T_{\rm H}$ are the entropy and temperature of the BH respectively.  The scrambling time refers to the time it takes to destroy quantum entanglements in a chaotic system, while BHs (and their CFT duals) are conjectured to be fast scramblers, i.e. the most efficient in destroying entanglement (e.g., \cite{Maldacena:2015waa}). Interestingly, Saraswat and Afshordi \cite{Saraswat:2019npa} have recently shown that the scrambling time (computed using Ryu-Takayanagi conjecture in a dynamical setting) is identical to the Planckian echo times, for generic charged AdS BHs. Could this imply that echoes could be a generic property of (possibly a certain class of) quantum chaotic systems?

Another possible connection could come in the form of the fluid-gravity correspondence, e.g., in the context of membrane paradigm discussed in Section \ref{sec:membrane}. For example, in \cite{Oshita:2019sat}, we have argued that Boltzmann reflectivity of GW echoes, implies that viscosity of the boundary fluid should vanish at small frequencies $\hbar \omega \ll k T$. One may also speculate that other holographic manifestations of BH echoes may appear in the Kerr/CFT conjecture \cite{Castro:2010fd}, Braneworld BHs \cite{Dey:2020lhq}, or as Regge poles of the boundary plasma in AdS/CFT.

\subsection{Einstein Telescope, Cosmic Explorer}
\begin{figure}[h]
  \centering
    \includegraphics[width=1\textwidth]{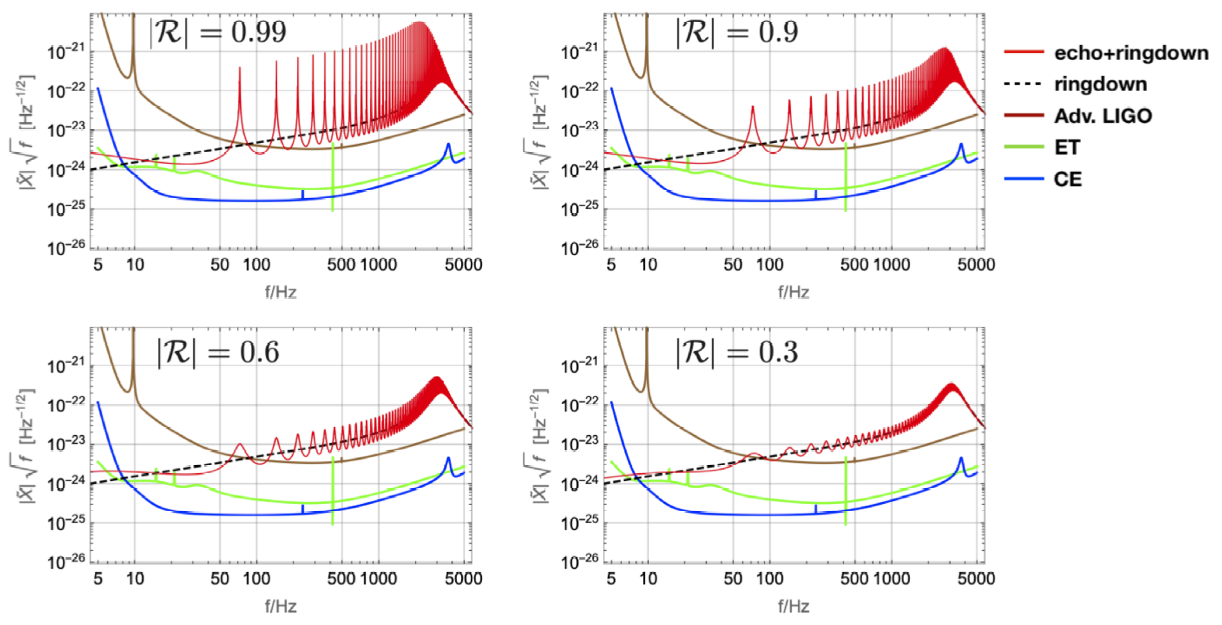}
    \caption{Spectra of ringdown and echo phases with the reflectivity of $|{\cal R}| = 0.99$, $0.9$, $0.6$, and $0.3$. We set $D_o = 40$ Mpc, $\bar{a} =0.1$, $\ell = m= 2$, $M = 4 M_{\odot}$, $\theta = 20^{\circ}$, and $\epsilon_{\rm rd} = 0.1 \%$.}
 \label{third_gene}
\end{figure}

The Einstein Telescope (ET) \cite{Punturo:2010zz} and Cosmic Explore (CE) \cite{PhysRevD.91.082001} are the third-generation ground based GW detectors. The ET consists of three underground detectors with three arms $10$ kilometers long and CE will be realized with two arms 40 kilometers long, which are 10 times longer than Advanced LIGO's. These next-generation GW detectors might allow us to observe some Planckian signatures from quantum BHs such as GW echoes from merger events leading to a remnant BH. We plot the spectra of GW echoes and ringdown with the sensitivity curves of Advanced LIGO, ET, and CE in FIG. \ref{third_gene}. The detection of GW echoes with the third generation GW observatories are discussed in \cite{Testa:2018bzd,Maggio:2019zyv}, and it may be possible to distinguish ECOs with $|{\cal R}| \lesssim 0.3$ from BHs with at $2\sigma$ level when SNR $\sim 100$ in ringdown, which would be possible for the third-generation GW detectors. The relative error on the reflectivity of would-be horizon is also investigated in \cite{Testa:2018bzd,Maggio:2019zyv}, and the relative error for measurement of relectivity in ground-based detectors is approximately given by
\begin{equation}
\left| \frac{\Delta {\cal R}}{1-\cal R} \right| \simeq 0.5 \times \left(\frac{8}{\rho_{\text{ringdown}}} \right),
\end{equation}
where $M = 30 M_{\odot}$, $\rho_{\text{ringdown}}$ is the SNR in the ringdown phase, while the distance between the top of the angular momentum barrier and the would-be horizon is assumed to be longer than $50 M$ in the tortoise coordinate. For comparison, we note that the loudest detected BBH event, GW150914, has $\rho_{\text{ringdown}} \simeq 8$.

\begin{figure}[h]
  \centering
    \includegraphics[width=1\textwidth]{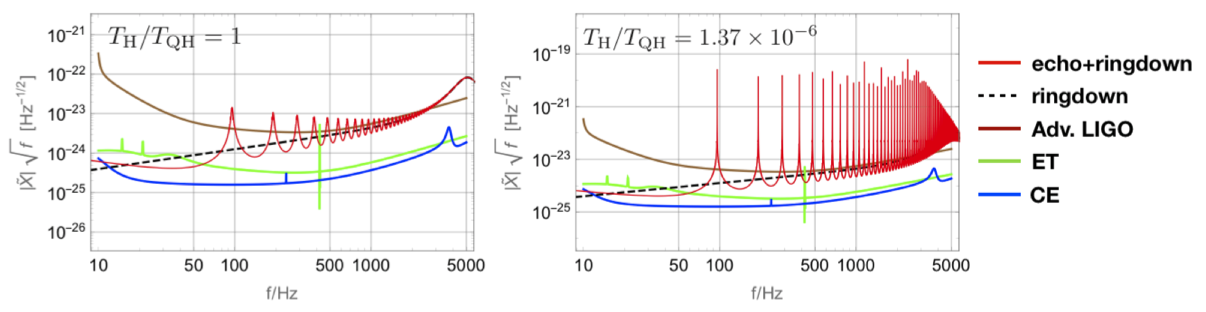}
    \caption{Spectra of ringdown and echo phases in the Boltzmann reflectivity model with $\bar{a} = 0.1$, $\epsilon_{\rm rd} = 6 \times 10^{-7}$, $M = 2.4 M_{\odot}$, $\theta = 90^\circ$, and $D_o = 1$ Mpc. Here we also assume $\gamma = 10^{-10}$, $T_{\rm H} / T_{\rm QH} = 1$ (left) and $T_{\rm H} / T_{\rm QH} = 1.37 \times 10^{-6}$ (right).}
 \label{third_gene_boltz}
\end{figure}

The detectability of GW echoes from failed supernovae, leading to the formation of BHs, with the third-generation GW observatories is also discussed in \cite{Oshita:2019seis1}. Calculating the SNR of GW spectrum consisting of echo and ringdown, $\rho_{\rm ringdown + echo}$, in the Boltzmann reflectivity model, the horizon distance $D_{\rm h}$, defined as the distance where $\rho_{\rm ringdown + echo} = 8$, is estimated. Given the optimistic case in the Boltzmann reflectivity model, $T_{\rm H}/ T_{\rm QH} = e^{15 (\bar{a}-1)}$, the horizon distance can be estimated as $D_{\rm h} \sim 10$ Mpc for the Advanced LIGO at design sensitybity and $D_{\rm h} \sim 100$ Mpc for the third-generation detectors such as ET and CE. Therefore, the authors in \cite{Oshita:2019seis1} argue that the searching for GW echoes, sourced by failed supernovae within our Galaxy and nearby galaxies, may be possible. However, in the case of $T_{\rm QH} = T_{\rm H}$, the horizon distance is less than or comparable with $10$ Mpc and so the echo search with failed supernovae would be restricted to within the Local Group. For the comparison, the strain amplitude of GW echoes in $T_{\rm QH}/ T_{\rm H} = 1$ and $T_{\rm QH}/ T_{\rm H} = e^{15 (\bar{a} -1)}$ are shown\footnote{We here assume that the energy fraction of ringdown phase is $\epsilon_{\rm rd} = 6 \times 10^{-7}$ although it highly depends on the detail of nonlinear gravitational collapse.} in FIG. \ref{third_gene_boltz}.

\subsection{LISA}
The Laser Interferometer Space Antena (LISA) is planned to be the first GW observatory in space. It will have three satellites separated by millions of kilometers and their orbits maintain near-equilateral triangular formation. LISA might enable us to reach high-precision detection of ringdown in SNR $\sim {\cal O} (10^3)$, which puts stronger constraints on the reflectivity of BHs \cite{Testa:2018bzd,Maggio:2019zyv}.

Recently, a novel proposal to discriminate BH horizons based on the tidal heating was proposed in \cite{Datta:2019epe}. One of the main targets of the LISA mission is precision measurements of  extreme-mass-ratio inspirals (EMRIs), in which the tidal heating could be important. The (partial) absorption of ECOs or BHs plays the role of dissipation at the surface, by which tides back-react on the orbital trajectory. It is argued that this tidal heating is responsible for a large dephasing between the orbits of a BH and ECO. This dephasing accumulates over the timescale of months and the accumulation speed is faster for a higher spin. The authors in \cite{Datta:2019epe} also found a proportionality relation between the dephasing $\delta \phi$ and energy reflectivity $|{\cal R}|^2$.

In order to make use of this scheme to put strong constraints on the reflectivity of ECOs, one has to obtain accurate EMRI waveforms by properly taking into account the tidal heating for orbiting objects, which may decrease systematic errors in data analysis.

Not only the tidal heating, but also the tidal deformability contributes to the GW Fourier phase and it can be characterized by the tidal Love number $k$. 
The Love number of ECO of mass $M$ may scale as $1/|\log{\delta}|$, where $\delta \equiv r_0 - r_{\rm h}$, with $r_{\rm h}$ is the BH horizon radius of mass $M$ and $r_0$ is the radius of the ECO. So the $k-\delta$ relation is
\begin{equation}
\delta = r_{\rm h} e^{-1/k},
\end{equation}
and assuming this relation, one can infer the near-horizon structure characterized by $\delta$ from the measurement of the Love number $k$. For instance, if the Love number of the order of $k \sim 10^{-2}$ is measured by LISA from a supermassive BH binary signal, leading to the formation of a BH of $M \sim 10^6 M_{\odot}$, it yields the resolution of $\delta \sim l_{\rm Pl}$.

However, the authors in \cite{Addazi:2018uhd} point out that the statistical and quantum mechanical uncertainties in measurements of near-horizon lead to some difficulty to measure $\delta$ precisely. The former one comes from the fact that the statistical uncertainty in $\delta$ is proportional to $1/k$, and the inferred value of $k$, where the inferred value of $\delta$ is comparable with its statistical uncertainty, is around $k \sim 0.2$. Therefore, any inferred value of $\delta$, derived from $k$ that is smaller than $\sim 0.2$, would be dominated by the statistical uncertainty. The latter one comes from the uncertainty principle in quantum mechanics. Once precisely measuring $\delta \sim l_{\rm Pl}$, it may lead to the uncertainty in the mass of the ECO, which then leads to the uncertainty in the binding energy. This results in the uncertainty in the orbital and GW frequencies, which means that one cannot measure $\delta$ precisely if it is much shorter than $l_{\rm Pl}$.

\subsection{Pulsar Timing Arrays}

\begin{figure}[h]
  \centering
    \includegraphics[width=0.8\textwidth]{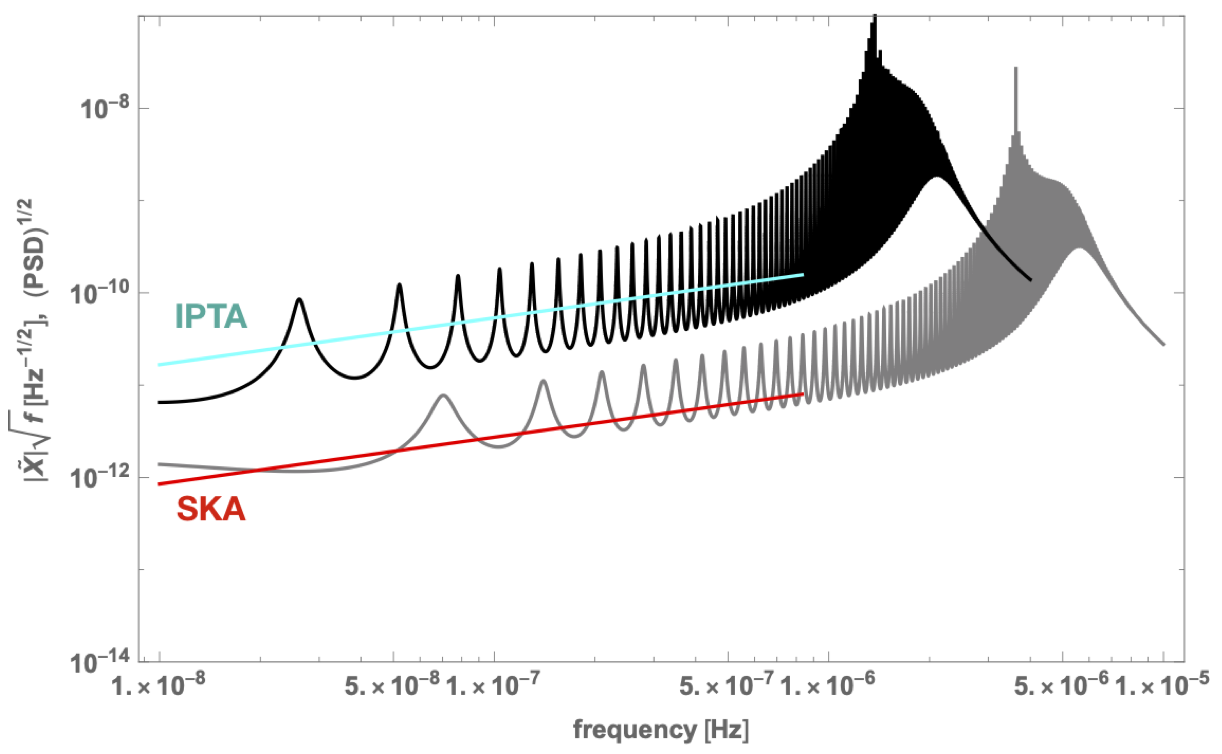}
    \caption{Spectra of GW echoes in the Boltzmann reflectivity model with $\bar{a}=0.6$, $\ell=m=2$, $D_o= 1$ Gpc, and $\gamma =1$. The gray line shows the case of $M=3 \times 10^{9} M_{\odot}$, $\epsilon_{\rm rd} = 0.005$, $T_{\rm H}/T_{\rm QH} = 0.1$ and the black line shows one for $M= 8 \times 10^{9} M_{\odot}$, $\epsilon_{\rm rd} = 0.01$, $T_{\rm H}/T_{\rm QH} = 0.05$. We also plot the PSD for the IPTA (blue) and SKA (red).}
 \label{IPTA_SKA}
\end{figure}

Following their first discovery in 1968 \cite{Hewish:1968bj}, over 2000 pulsars have now been detected by radio telescopes across the world.  The pulsars' intrinsic properties, as well as propagation effects in the interstellar medium, can influence the arrival times of pulsar pulses. Therefore, pulsar timing arrays (PTA) can be used as a detection tool for BH binaries \cite{Hobbs:2009yy}, and thus, might be used to detect singatures of echoes from quantum BHs. In particular, millisecond pulsars stand out for their unparalleled stability (comparable to atomic clocks!) without being subject to starquakes and accretion. To give an explicit example, we show the spectrum of GW echoes predicted by the Boltzmann reflectivity model \cite{Oshita:2019sat,Wang:2019rcf} with the sensitivity curve of International Pulsar Timing Array (IPTA) and Square Kilometre Array (SKA) (FIG. \ref{IPTA_SKA}). The lower curve in Fig. \cite{Hewish:1968bj} is for a $3 \time 10^9 M_{\odot}$ BH merger at $D_o = 1$ Gpc. Given that this mass is comparable to that of M87 supermassive BH, located at 16 Mpc, we expect $\sim 2 \times 10^5$ of such BHs at < Gpc. Assuming that each BH merges once every Hubble time $\sim 10^{10}$ years, and that echoes last for 20 years (from simple mass scaling), the chances of detecting such a loud event with PTAs at any time is 0.1\%. However, fainter events will be more prevalent as their number increases as SNR$^{-3/2}$ from volume scaling. Furthermore,  increase in supermassive BH merger activity observed at high redshifts shall boost this statistics.

PTAs are anticipated to detect the low frequency GW signal from supermassive BBH  within the next few years \cite{Hobbs:2009yy}. We expect that the first GW detection will be a stochastic background of supermassive BH binaries. With any luck, this shall lead to new insights into the nature of quantum BHs and gravity.

\section{Final Word}
\label{sec:final_words}

In this review article, we provided a comprehensive overview of the theoretical motivations for why quantum black holes in our universe may have different observable properties, in contrast to their classical counterparts in Einstein's theory of general relativity. The most prominent and potentially observable smoking gun for these quantum black holes comes in the form of gravitational wave echoes, which have been the subject of intense theoretical and observational scrutiny over the past few years. We provided a concise account of theoretical predictions, as well as the exciting and confusing state of observational searches for echoes in LIGO/Virgo observations. We closed by article by our vision of the future of ``Quantum Black Holes in the Sky'', via a synergy of statistical methodology, quantum gravity, and numerical relativity, and in light of the next generation of gravitational wave observatories. 

While this review article focuses on the gravitational wave echoes, as arguably the most concrete and promising signature of quantum black holes, other possible observable signatures can be (and should be) explored. For example, interactions of photons or neutrinos with near-horizon quantum structure could lead to signatures in radio images in Event Horizon Telescope observations \cite{Rummel:2019ads}, or ultra high energy neutrinos in Ice Cube observatory \cite{Afshordi:2015foa}, respectively. However, these signals will be suppressed if Boltzmann reflectivity is assumed, as they have $\hbar \omega \gg k T_{\rm H}$. Another alternative to echoes may come through non-localities in non-violent unitarization, which would be observable far from the horizon (see Section \ref{sec:non-violent}). However, it is arguably difficult to pin down concrete predictions in this scenario.  

To conclude, the world of Quantum Black Holes remains a wide open and largely uncharted territory, spanning from the dark corners of obscure mathematical structures to the nitty-gritty details of gravitational wave detector noise. It also holds the promise to crack the century-old puzzle of quantum gravity, and yet be imminently testable in the next few years. Therefore, the study of ``Quantum Black Holes in the Sky'' remains extremely exciting, active, and confusing, and is bound to provide us with new surprises in the new decade, and beyond.   

\section*{Acknowledgments} \label{sec::acknowledgments}

We would like to thank Michael Balogh, Ofek Birnholtz, Avery Broderick, Vitor Cardoso, Ramit Dey,  Hannah Dykaar, Will East, Steve Giddings, Bob Holdom, Badri Krishnan, Lam Hui, Luis Lehner, Luis Longo, Samir Mathur, Emil Mottola, Shinji Mukohyama, Rob Myers, Ramesh Narayan, Alex Nielsen, Paolo Pani, Joe Polchinski (RIP), Chanda Prescod-Weinstein, Jing Ren, Krishan Saraswat, Rafael Sorkin, Daichi Tsuna, Yasaman Yazdi, Huan Yang, Aaron Zimmerman, and many others for discussions and/or collaborations about Quantum Black Holes over the past decade. N. O. is supported by JSPS Overseas Research Fellowships and the Perimeter Institute for Theoretical Physics.

Research at the Perimeter Institute is supported by the Government of Canada through Industry Canada, and by the Province of Ontario through the Ministry of Research and Innovation. This research has made use of data, software and/or web tools obtained from the Gravitational Wave Open Science Center (https://www.gw- openscience.org), a service of LIGO Laboratory, the LIGO Scientific Collaboration and the Virgo Collaboration. LIGO is funded by the U.S. National Science Foundation. Virgo is funded by the French Centre National de Recherche Scientifique (CNRS), the Italian Instituto Nazionale della Fisica Nucleare (INFN) and the Dutch Nikhef, with contributions by Polish and Hungarian institutes.

 \appendixtitles{yes} 
\appendix
\section{Energy density and flux seen by a radially free falling observer in mock fuzzball\label{fuzzballflux}}

We start with the geodesic of a radially falling observer in the mock fuzzball. 
\begin{align}
\label{line}u^{\mu}u_{\mu}=-1&=-(1-\frac{2M}{r}+b)(\frac{dt}{d\tau})^2-\frac{1}{1-\frac{2M}{r}}(\frac{dr}{d\tau})^2,\\
\label{lineinitial}-1&=-(1-\frac{2M}{r}+b)(\frac{dt}{d\tau})^2,\\
\label{conservation}(1-\frac{2M}{r}+b)\frac{dt}{d\tau}&=\text{constant}.
\end{align}
Where $\tau$ is the proper time of the observer. $\theta$ and $\varphi$ vanish with a radially observer. Eq. \ref{line} is from the line element, Eq. \ref{lineinitial} is initial condition of Eq. \ref{line} : the observer rest at infinite. Eq. \ref{conservation} is energy conservation. Eq. \ref{line}-\ref{conservation} give four-velocity:
\begin{align}
u^{\mu}=(\frac{r}{b r-2M+r},- \sqrt{\frac{2M (r-2 M)}{r (b r-2 M+r)}},0,0).
\end{align}
Then we consider area vector $a_{\mu}=A(-u^1,u^0,0,0)$ which is normal to four-velocity $a^{\mu}u_{\mu}=0$. We determine $A=\sqrt{\frac{1-2M/r+b}{1-2M/r}}$ by normalization of the area vector $a^{\mu}a_{\mu}=1$. In the end, the energy density and flux are defined as:
\begin{align}
 \mathcal{U}&= T_{\mu \nu} u^{\mu}  u^{\nu}=-\frac{4 b M^2}{r^2 (b r-2M+r)^2},\\
 \mathcal{F}&= -T_{\mu \nu} u^{\mu}  a^{\nu}=-\frac{ b \left(\frac{2M}{r}\right)^{3/2}}{(b r-2 M+r)^2}.
\end{align}

\bibliography{template}
\end{document}